\documentclass[aps,prb,amsmath,superscriptaddress,twocolumn,longbibliography]{revtex4-2}

\usepackage{graphicx}
\usepackage{amsmath}
\usepackage{amsfonts}
\usepackage{amssymb}
\usepackage{braket}
\usepackage{bm}
\usepackage{multirow}
\usepackage{color}
\usepackage[normalem]{ulem}
\usepackage{mathrsfs}
\usepackage{mathtools}
\usepackage{float}

\makeatletter

\newcommand{\Rmnum}[1]{\expandafter\@slowromancap\romannumeral #1@}
\makeatother

\usepackage[breaklinks]{hyperref}
\hypersetup{colorlinks=true, linkcolor=blue, citecolor=blue, filecolor=blue, urlcolor=blue}

\AtBeginDocument{%
    \newwrite\bibnotes
    \def\bibnotesext{Notes.bib}
    \immediate\openout\bibnotes=\jobname\bibnotesext
    \immediate\write\bibnotes{@CONTROL{REVTEX41Control}}
    \immediate\write\bibnotes{@CONTROL{%
    apsrev41Control,author="08",editor="1",pages="1",title="0",year="1"}}
     \if@filesw
     \immediate\write\@auxout{\string\citation{apsrev41Control}}%
    \fi
}%

\begin{document}

\title{Coexistence of Topological Dirac and Dirac Nodal line semimetal in SrCaP belonging to Nodal line semimetal family SrCaX(X= Bi, Sb, As, P)}
\author{Shivendra Kumar Gupta}
\email{shivendrakumarg900@gmail.com}
\affiliation{Department of Physics, Visvesvaraya National Institute of Technology, Nagpur, 440010, India \\}
\author{Ashish Kore}%
\affiliation{Physics Division, National Center for Theoretical Sciences, National Taiwan University, Taipei 106319, Taiwan \\}
\author{Saurabh Kumar Sen}%
 
\author{Poorva Singh}
\email{poorvasingh@phy.vnit.ac.in}
\affiliation{Department of Physics, Visvesvaraya National Institute of Technology, Nagpur, 440010, India \\}

\date{\today}

\begin{abstract}
 Abstract.---
 Nodal line semimetals represent precursor states for various topological phases, exhibiting intrinsic topological characteristics and intriguing properties. These materials host rare and distinctive topological features, which can give rise to exotic phenomena, thereby garnering significant attention in both fundamental research and technological applications. In this study, we conduct ab-initio calculations to explore the properties of SrCaX (X = Bi, Sb, As, P), identifying these as multiple Dirac nodal line semimetals protected by $Z_2$ quantized Berry phases and manifesting multiple drum-head-like surface states. The nodal lines in these compounds are situated at the M point when $k_z$ = 0 and at the A point when $k_z$ = $\pi$. Notably, SrCaX family exhibits a unique characteristic wherein they host both type II Dirac point and topological nodal line semimetal within a single crystal structure, hence providing an excellent platform for studying the interplay between different topological properties. Additionally, in SrCaP topological Dirac semimetal, Type II Dirac point and topological nodal line semimetal features coexist in a single crystal. These special features in this series of materials make them ideal candidates for further investigation by experimental means.

\end{abstract}

\maketitle

{\it Introduction.---}Topological materials have the unique properties of dissipation-less spin transport through symmetry-protected surface states. These materials are classified into topological insulators (TIs) and topological semimetals (TSMs). TIs were first discovered to have an exchange of orbital bands characteristic of induced band inversion and have the properties of surface band crossing at the Fermi level associated with spin-momentum locking due to protection from time-reversal symmetry\cite{PhysRevLett.115.136801, PhysRevB.94.201104, PhysRevB.94.125152, RevModPhys.82.3045, PhysRevB.107.075143, mukherjee2020fermi}. Later TSMs have been discovered to have similar kind of symmetry-protected surface states resulting in the absence of a hybridized gap opening in the bulk band crossing near the Fermi level i.e. the crossing is non-accidental and it cannot be avoided without breaking any symmetry\cite{fu2019dirac,fang2016topological,weng2017new}. TSMs are mainly classified into Dirac semimetal, Weyl semimetal, and nodal line semimetal based on the degeneracy of bands, dimensionality of band crossing, and symmetry protection.  The Dirac semimetals have a four-fold degenerate point along with the linear dispersion in the band crossing of valence and conduction band near the Fermi level in conjunction with the protection of both time-reversal symmetry (TRS) and inversion symmetry (IS) analogous to relativistic Dirac fermions that are theoretically predicted\cite{wang2012dirac, liu2014discovery, wang2013three, wan2011topological, peng2018predicting,shende2023first,gupta2023pressure} and experimentally confirmed\cite{quintela2017epitaxial, xiong2015evidence, wang2012multiband, yen2020tunable, song2021coexistence}. If any one or both among TRS and IS break, the degeneracy of the crossing bands also breaks resulting in the conversion of Dirac semimetal into Weyl semimetal(WSMs). WSMs have a unique characteristic of open Fermi arc connected Weyl points of opposite chirality and transport of the Weyl fermions which have long been theoretically predicted in 1929\cite{weng2015weyl, yan2017topological,mccormick2017minimal}. 
The material realization of Weyl semimetal has been revealed theoretically and experimentally in special types of non-centrosymmetric tetragonal crystal symmetry groups of AB-type materials (where A=Ta, Nb and B= As, P)\cite{osterhoudt2019colossal, weng2015weyl, liu2016evolution, yuan2020discovery, chang2016signatures}and many others\cite{sun2015prediction, alidoust, li2021type, jin2020fully, shende2023pressure}. 
Moreover, distinct topological materials including the three-fold\cite{cheung2018, barik2018, lv2017, fang2020ideal}, six-fold\cite{sun2020direct, yang2020observation, thiru, jin2021}, and eight-fold\cite{rong2023, guo2021eightfold} fermionic systems have been theoretically predicted. Among these, the three-fold and six-fold fermionic systems which have no counterpart in high-energy physics have been experimentally confirmed\cite{lv2017observation,sun2020direct}. 

A more fundamental topological state of matter, nodal-line semimetals (NLSMs) are a class of materials characterized by the presence of one-dimensional topological nodal lines that are protected by the additional mirror or glide or rotational symmetries in their electronic band structure along with the formation of closed loops or lines in three-dimensional momentum space. The surface states of NLSMs have drumhead-like surface bands\cite{he2018type}. In contrast to isolated zero-dimensional points (Dirac and Weyl nodes), nodal lines possess richer topological properties and can form one-dimensional nodal rings, nodal chains, nodal links, or even nodal knots\cite{fu2019dirac}. The distinct characteristics exhibited by nodal-line semimetal present an ideal environment for the investigation of physical phenomena stemming from interactions among the massless quasiparticles\cite{bian2016drumhead}. Furthermore, the presence of a torus-shaped Fermi surface in a doped nodal-line semimetal can give rise to extraordinary transport properties\cite{bian2016drumhead}. Theoretical band calculations have successfully predicted numerous nodal-line semimetals with drumhead-like surface states in the absence of spin-orbit coupling(SOC), however in the presence of SOC the crossing is avoided and the nodal line converts to TI’s\cite{yu2015topological}. In the realm of synthesized materials, SOC is notable for its universality and extensive presence, so the material that loses its nodal line properties in the presence of SOC cannot be experimentally suitable for the NLSM realization, therefore robust NLSM phase is required in the presence of SOC\cite{bian2016drumhead, fang2015topological}. Some of the materials acquiring the above properties and confirmed experimentally remain limited \cite{201800897, wang2021spectroscopic, wang2017topological, takane2018observation, xu2018trivial, lou2018experimental, feng2017experimental, sato2018observation, yi2018observation, hosen2017tunability, bian2016topological, chen2017dirac}. %

\begin{figure}[!t]
 \includegraphics[width=.48\textwidth]{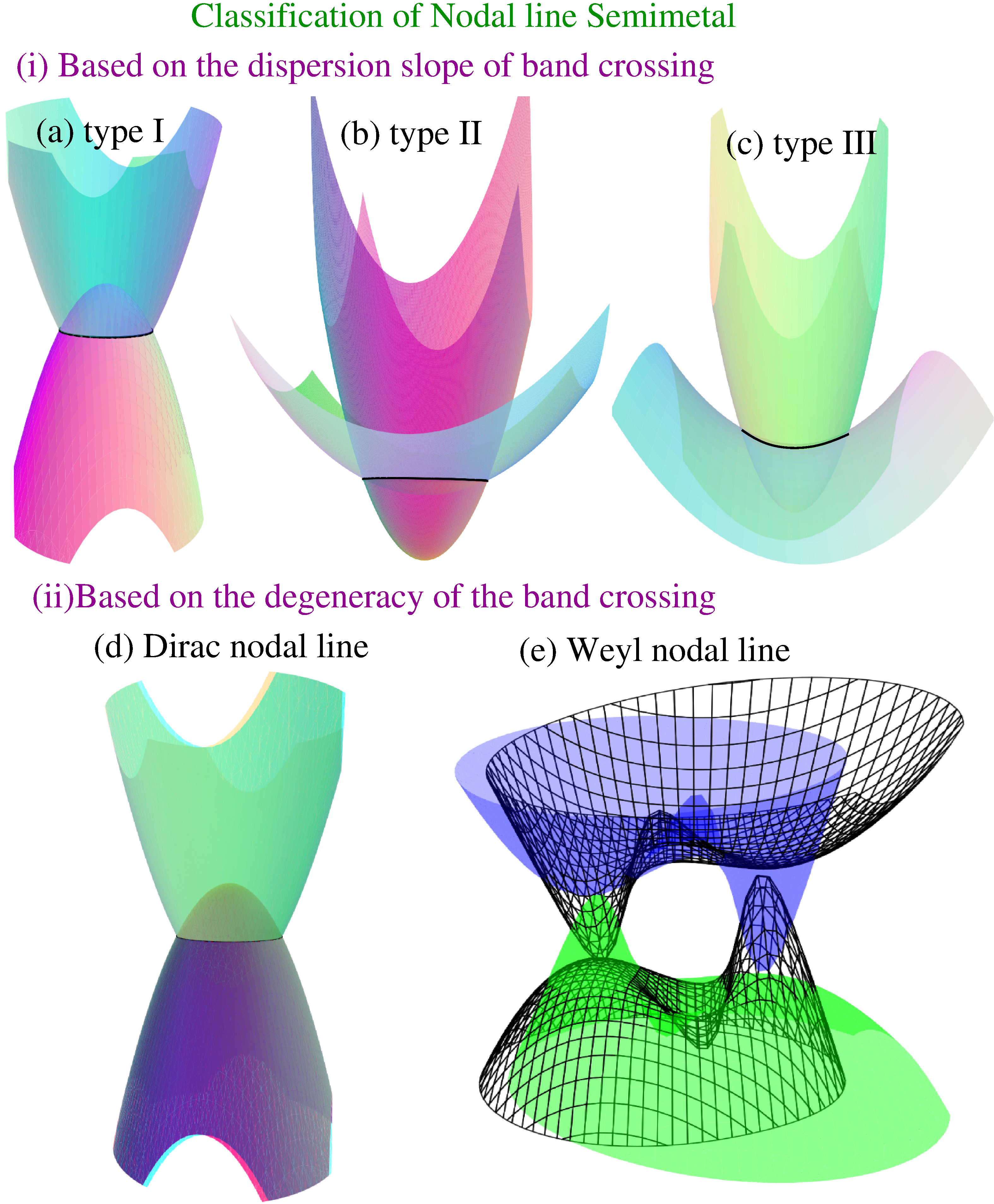}
 \caption{\label{fig:epsart} Scematics of classification of Topological nodal line semimetals (i) based on the dispersion slope of the band crossing include (a) type I (b) type II (c) type III and (ii) based on the Degeneracy of the band crossing (d) Dirac nodal line semimetals (e) Weyl nodal line semimetals.} 
 \label{fig:figure1} 
 \end{figure}

 NLSMs have been classified in different perspectives relative to the characteristic of the nodal line; (i)the degeneracy of the band crossing includes the Dirac and Weyl NLSMs\cite{hu2016evidence, liu2018experimental, hirayama2017topological, cheng2019visualizing, jin2020ferromagnetic, cui2020three, jin2020two}; (ii)topological invariants include the Z and Z$_2$ Berry phase NLSMs\cite{huang2016topological, suzumura2018berry, yu2017topological, gao2019topological}; (iii)the dispersion slope of the band crossing includes the type I, type II  and type III NLSMs\cite{li2017type, xu2020centrosymmetric, wang2018type, he2018type}; (iv)geometry of the Fermi-surface which may menifest as nodal line, nodal rings, and nodal surface\cite{bzduvsek2016nodal,song2022spectroscopic}; (v)multiple nodal lines and its connections include nodal rings, nodal chains, nodal links, and nodal knots\cite{PhysRevB.96.201305, PhysRevB.96.041102, PhysRevB.96.081114}. The broad classification of NLSMs is shown in Figure \ref{fig:figure1}. 

Dirac nodal line semimetals (DNLSMs) host four-fold degeneracy of nodal line in the momentum space protected by the Z$_2$ quantized Berry phase. In this type of nodal line semimetals, the time-reversal symmetry and inversion symmetry are present, and doubly degenerate Kramers' band crosses at the Fermi level which are responsible for the four-fold degenerate Dirac nodal line\cite{ji2023observation}. When the bands are inverted such that the parity eigenvalues exchange at a TRIM (time reversal invariant momentum) point and the presence of drum-head-like surface band guarantees the presence of a nodal line\cite{gao2019topological}. Strongly correlated flat bands localized at the surface can be obtained by tuning band inversion and the size of the nodal line via impurity doping and strain to the system\cite{kim2015dirac, zhao2016topological}. DNLSMs are obtained in $CaAgAs$\cite{xie3apl,okamoto2016low}, $Ca_3P_2$\cite{yu2017topological}, and many others systems.   

In the surge of new quantum phenomena, characteristic features, exotic properties and to gain more understanding about the intrinsic nature of quantum materials, the researchers intend to find materials that inherit two or more distinct properties simultaneously and examine their mutual interaction. In this perspective, some materials have been investigated like the Kagome compound $Mg_3 Bi_2$ possesses a 3D Dirac point and type II nodal line that are independent of each other \cite{zhang2017topological}. Also, $LaAgSb_2$ is defined as topological NLSMs as well as Dirac semimetal in different terminations and different studies\cite{rosmus2022electronic,observation}. Theoretical proposal of the coexistence of Weyl semimetal and Weyl nodal loop were reported in anti-ferromagnetic transition metal oxide $RuO_2$ and ferromagnetic double perovskite $Ba_2 CdReO_6$\cite{zhan2023coexistence,zhao2021coexistence}. Also, in the absence of SOC, TaS possesses two nodal rings and a triply degenerate point whereas after applying SOC one of the nodal lines gets converted into six pairs of Weyl points while the other one remains unchanged\cite{sun2017coexistence}.
 \begin{figure}[!t]
 \includegraphics[width=8cm,height=5cm]{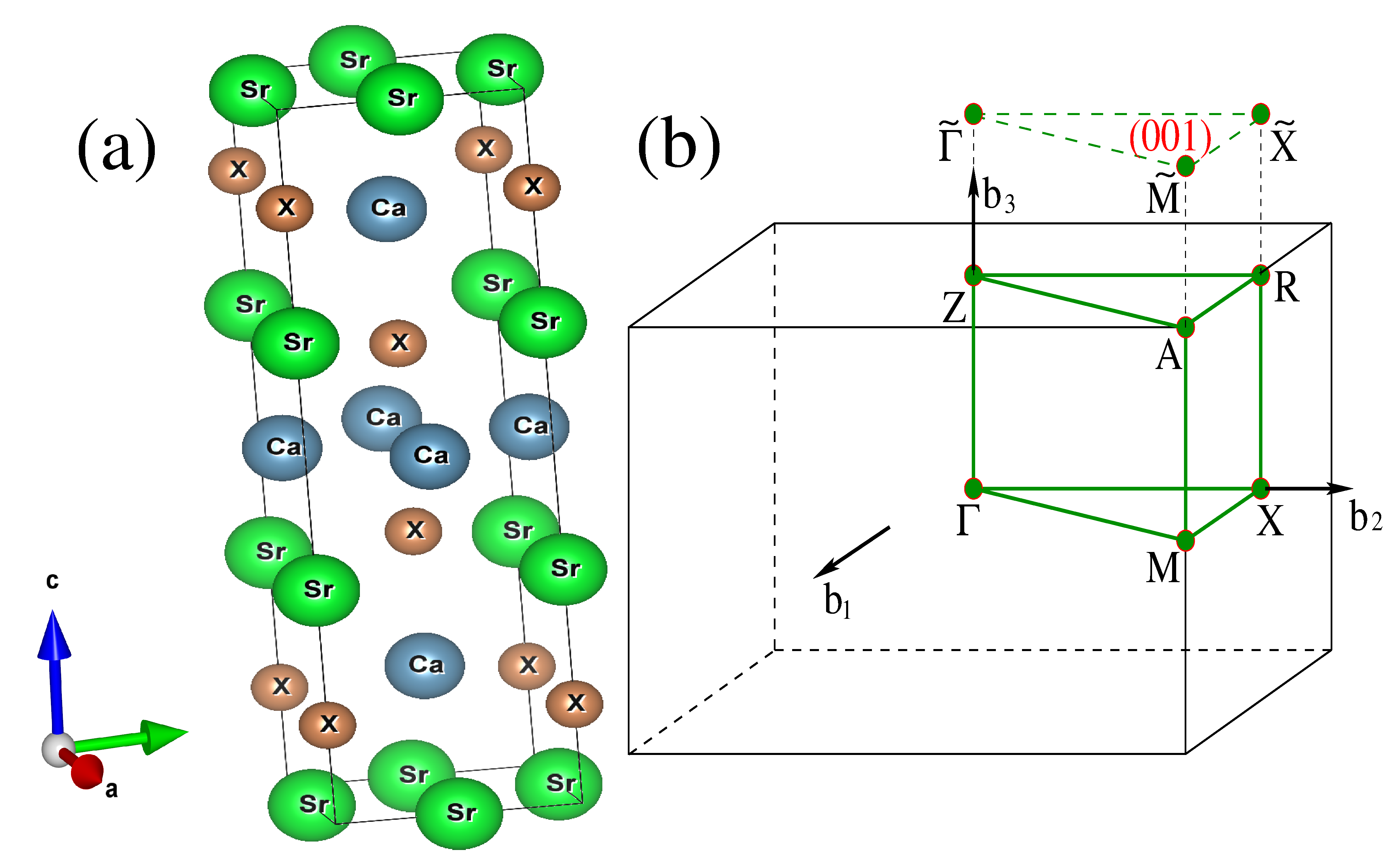}
 \caption{\label{fig:epsart} (a) Crystal structure of SrCaX (X= Bi, Sb, As, P) (b) Bulk and surface Brillouin zone (001) of tetragonal crystal structure.} 
 \label{fig:figure2} 
 \end{figure}
\begin{table}
\begin{center}
\begin{tabular}{||c c c c||} 
 \hline
 S. No.  &  Material  &   a(in \AA)   &     c(in \AA)    \\ [0.5ex] 
 \hline\hline
 1 & SrCaBi & 5.01 & 17.63 \\ 
 \hline
 2 & SrCaSb & 4.92 & 17.33 \\
 \hline
 3 & SrCaAs & 4.72 & 16.43 \\
 \hline
 4 & SrCaP & 4.63 & 16.03 \\
 \hline
\end{tabular}
\caption{\label{demo-table}Optimized lattice parameter of SrCaX(X = Bi, Sb, As, P).}
\end{center}
\end{table}

 \begin{figure*}[!t]
\includegraphics[width=18cm,height=11cm]{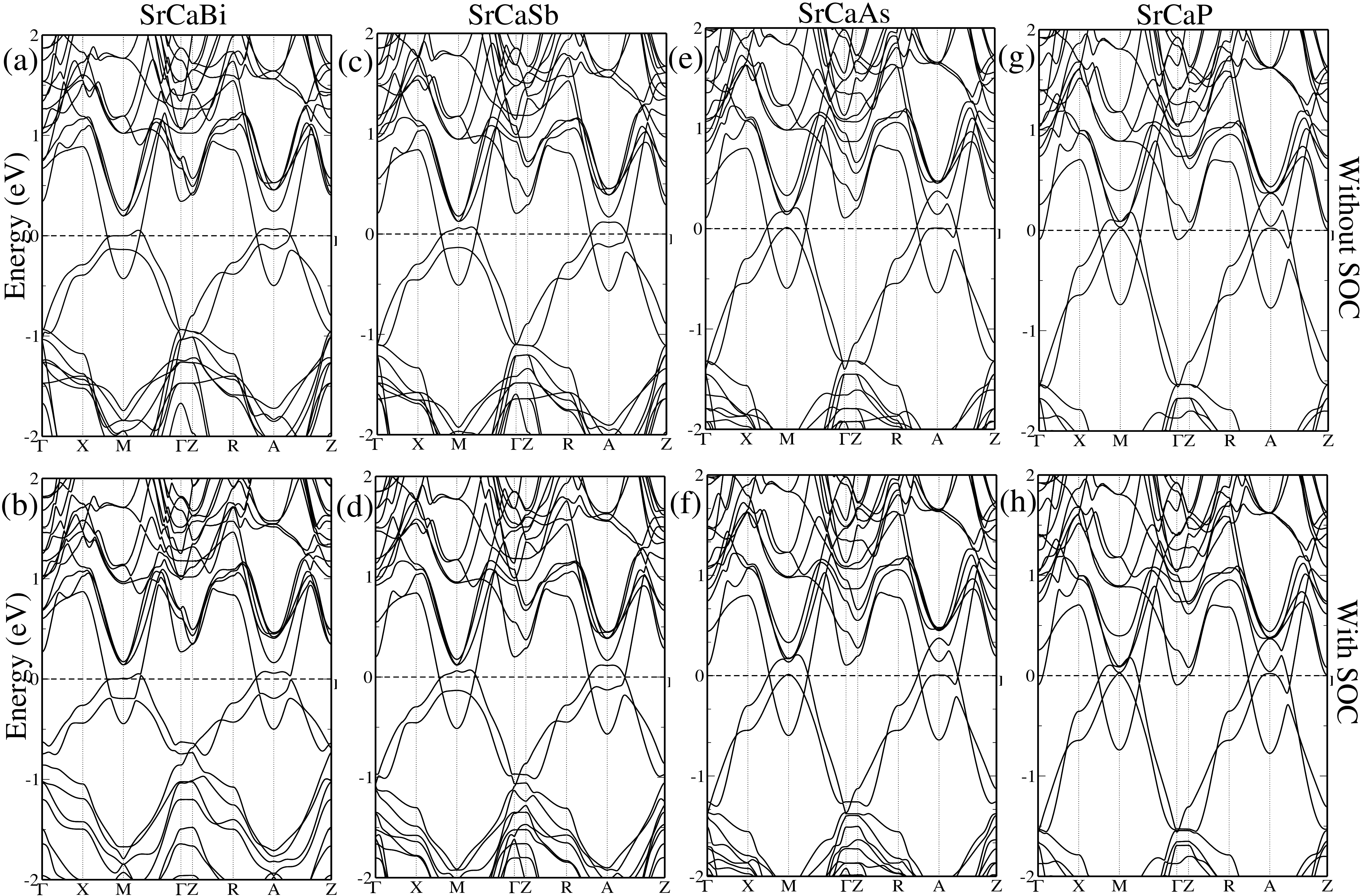}
 \caption{\label{fig:epsart} Electronic band structure of SrCaBi (a) without SOC (b) with SOC. Electronic band structure of SrCaSb (c) without SOC (d) with SOC. Electronic band structure of SrCaAs (e) without SOC (f) with SOC. Electronic band structure of SrCaP (g) without SOC (h) with SOC.}
 \label{fig:figure3} 
 \end{figure*}

In the present study, the electronic properties of a series of materials SrCaX ( X = Bi, Sb, As, P) have been investigated via first-principles calculations. The TRS and IS are present in the crystal making all the bands doubly degenerate.  The study reveals that all four materials in the series are topological DNLSMs and host type II dirac points without considering the SOC. Applying SOC does not change the band topology of the system and they remain nodal-line semimetal along with the presence of type II Dirac points. The topological nature of the DNLSMs are confirmed by the drum-head-like surface states and the $Z_2$ Berry phase quantized into $\pm\pi$. Moreover, in SrCaP topological Dirac and topological Dirac nodal line features coexist along with presence of type II Dirac point in a single crystal. Nodal loops lie around the high symmetry points M and A which coincide when observed along the c-axis.

{\it Result and Discussion.---}
The crystal structure of materials SrCaX ( X = Bi, Sb, As, P) is tetragonal and belongs to the symmorphic space group P4/mmm (123) as shown in Figure\ref{fig:figure2}(a). These materials have been structurally optimized to find the most stable structure and the optimized lattice parameters are shown in Table \ref{demo-table}. The three-dimensional (3D) and associated two-dimensional (2D) Brillioun zone are shown in Figure \ref{fig:figure2}(b). To determine the stability of the material the phonon band structure calculation of SrCaBi has been performed and is shown in the supplementary Figure S1. The absence of imaginary phonon frequency in the phonon dispersion band structure indicates the dynamical stability of the material. The electronic band structure of SrCaBi, SrCaSb, SrCaAs, and SrCaP without and with SOC is shown in Figure\ref{fig:figure3}(a), \ref{fig:figure3}(b); \ref{fig:figure3}(c), \ref{fig:figure3}(d); \ref{fig:figure3}(e), \ref{fig:figure3}(f); and \ref{fig:figure3}(g), \ref{fig:figure3}(h), respectively. The crossing of bands near the Fermi level indicates the semimetallic nature of these materials. 
 \begin{figure*}[!t]
 \includegraphics[width= 1 \textwidth]{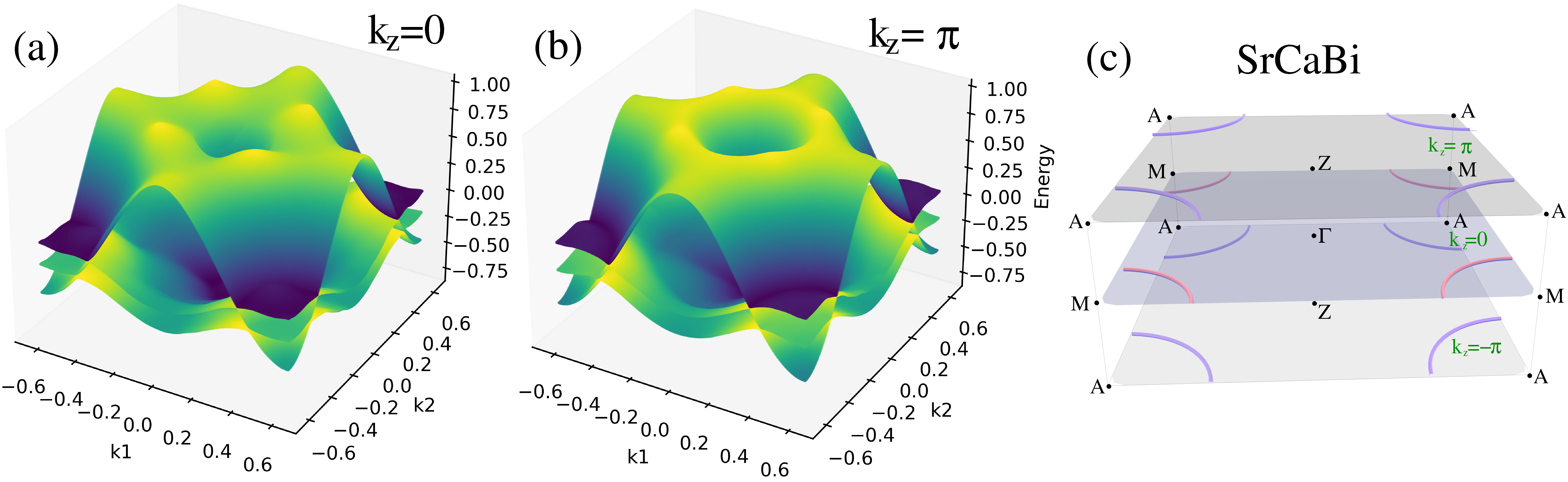}
 \caption{\label{fig:epsart} 3D band structure of (a) SrCaBi $k_z =0$ (b) SrCaBi $k_z =\pi$ (c) nodal loop in first BZ in SrCaBi.}
 \label{fig:figure4} 
 \end{figure*}

As the TRS and IS is present in these system, all the bands after inclusion of SOC are doubly degenerate due to Kramers' degeneracy (Figure \ref{fig:figure3}). Hence, if there is a crossing in the bulk band at the Fermi level, the material may possess either topological Dirac or Dirac nodal line semi-metal based on whether the conduction and valence band intersect  at discrete points or along a continuous line. For SrCaBi and SrCaSb, two valence bands and one conduction band cross each other at the Fermi level around M point at $k_z =0$ and around A point at $k_z =\pi$ in Figure \ref{fig:figure3}(b)and \ref{fig:figure3}(d). In order to ascertain the band gap at band crossing points, the individual band structures of SrCaBi and SrCaSb are shown in supplementary Figure S2 and Figure S3, respectively.  For SrCaBi and SrCaSb the calculated band gap at band crossing is found to be  minimal as compared to previously reported nodal line semimetals\cite{li2020insight,xu2020hexagonal}. In SrCaAs due to the contraction of the lattice parameter, more bands from valence and conduction near the Fermi level interact with each other, increasing the number of crossing in the system at both $k_z =0$ and $k_z =\pi$ (Figure \ref{fig:figure3}(f)). In SrCaP even more number of crossing can be seen due to further contraction of the lattice parameter (Figure \ref{fig:figure3}(h)). The individual band structure of SrCaAs and SrCaP are shown in supplementary Figure S4 and Figure S5, respectively, along with the calculated gap at band crossing. Additionally, for SrCaP, at the Fermi level at the M point on the $k_z =0$ plane, there is a Dirac-like cone as shown in supplementary Figure S5.

Now to confirm whether the crossing is a zero-dimensional Dirac point or a one-dimensional nodal line we calculated the three-dimensional band structure at $k_z =0$ and $k_z =\pi$ as shown in Figure\ref{fig:figure4}(a) and \ref{fig:figure4}(b), along with the location of the nodal loop in first Brillouin zone for SrCaBi. The investigation reveals that there is one nodal line at $k_z =0$ plane around M point and another nodal line at $k_z = \pi$  plane around the A point as shown in Figure \ref{fig:figure4}(c). For SrCaSb, SrCaAs, and SrCaP the 3D band structure for $k_z =0$ and $k_z =\pi$ are shown in supplementary Figure S6 along with the position of nodal loop in first Brillioun zone. Except for SrCaBi that possesses one nodal loop at $k_z =0$, SrCaSb, SrCaAs and SrCaP possess two nodal loops at $k_z =0$. All the materials  possess one nodal loop at $k_z =\pi$. All these nodal loops are protected by $Z_2$ quantized Berry phase. For better visualization of the nodal line in SrCaX the 3D band structures for $k_z =0$ taking M as the center point have been shown in supplementary Figure S7.
 \begin{figure}[b]
 \includegraphics[width=0.5 \textwidth]{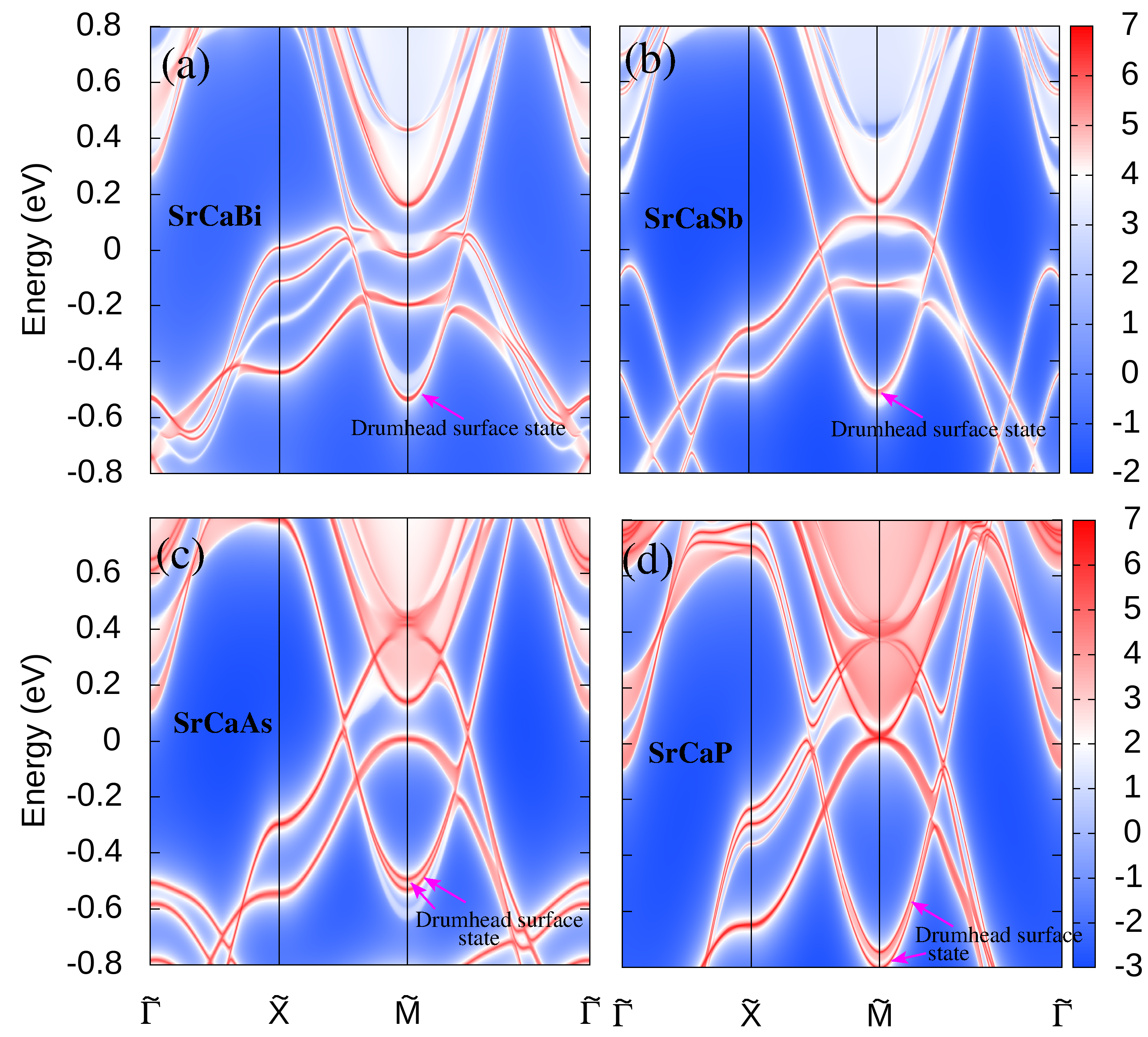}
 \caption{\label{fig:epsart} Surface band structure having drum-head-like surface state for (a) SrCaBi (b) SrCaSb (c) SrCaAs (d) SrCaP }
 \label{fig:figure5} 
 \end{figure}
To further confirm the one-dimensional nodal line, we performed gap plane calculations (supplementary Figure S8). The figure demonstrates that the band closing occurs only between two bands at the Fermi plane, indicating that the band crossing is one-dimensional and forms a loop, thereby confirming the material as a topological nodal line semimetal. In SrCaP, alongside the nodal line, a Type I Dirac point at the M point is also observed, as shown in supplementary Figure S8(d), in agreement with Dirac-like cone feature (supplementary Figure S5). The individual band structures plotted in supplementary Figure S2, S3, S4, and S5 for SrCaX reveal Type II Dirac points around the Z point in the $k_z =\pi$ plane. There are four Type II Dirac points around the M points due to this crossing. These findings have been confirmed by Fermi arc calculations at energies near the crossing, as shown in supplementary Figure S9.

Nodal line semi-metals have the unique feature of having drumhead-like surface states that are popping out from the bulk region and originate from the nodal points. The drumhead surface state guarantees the existence of a topological nodal line. The number of drumhead-like surface state can be up to the number of nodal lines present in the system. More than one drumhead-like surface state indicates the existence of multiple nodal lines in the system. In SrCaX, to further confirm the topological nodal line semi-metal, we have calculated the surface band structure of these materials and identified the drumhead-like surface structure in these materials as shown in Figure \ref{fig:figure5}. In SrCaBi we identify one whereas in SrCaSb, SrCaAs, and SrCaP two drumhead-like surface states are present, which indicate multiple nodal lines at $k_z =0$ plane in these materials.

\begin{figure}[!t]
 \includegraphics[width= 0.5 \textwidth]{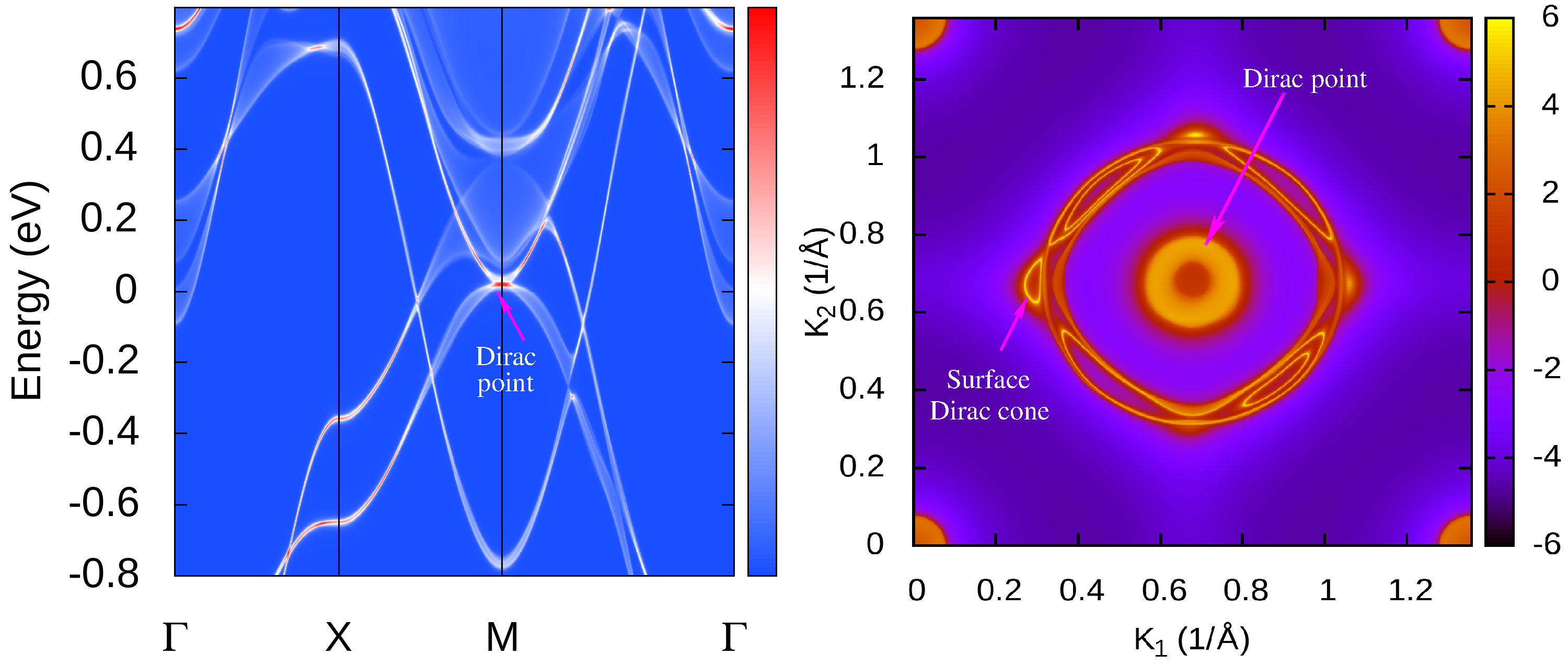}
 \caption{\label{fig:epsart} (a) Representation of bulk band showing type I Dirac point. (b) Fermi arc at energy around the type I crossing for SrCaP, that highlights the Dirac point and surface Dirac cone.}
 \label{fig:figure6} 
 \end{figure}

To confirm the existence of the type I Dirac point in SrCaP at M, the bulk surface band has been calculated that shows the bright point at the touching of two bands just above the Fermi level (shown in Figure \ref{fig:figure6}(a)), indicating the presence of Dirac fermions. We have also calculated the Fermi arc, which shows the bright point at M high symmetry point indicating the existence of Dirac crossing in Figure \ref{fig:figure6}(b). These results confirm the coexistence of multiple Dirac nodal line semimetal and Dirac semimetal properties together in SrCaP.


{\it Computational Methods.---}Electronic properties of SrCaX compounds have been investigated by the first-principles calculation based on standard density functional theory with the full potential linearized augmented plane-wave method provided by the WIEN2K package \cite{blaha2001wien2k}. Generalized gradient approximation (GGA) with Projector augmented wave (PAW) potentials have been utilized to incorporate the exchange-correlation function.  $\mathrm {R_{MT}K_{max}}$ was set to 6.5 to acquire the plane wave cut-off. The 10 ×10× 3 k-mesh is used to perform self-consistent calculations (SCF). Maximally localized Wannier functions (MLWF) are used to develop the tight-binding model, which is used to calculate the surface states of the materials using Wannier90 code \cite{mostofi2008wannier90}. WannierTools is used to obtain topological characteristics such as topological surface state, Fermi arc, etc\cite{wu2018wanniertools}. To confirm the dynamical stability, the phonon band structure for SrCaBi has been calculated for $3\times 3\times 1$ supercell using phonopy code\cite{phonopy,phonopy1}.

{\it Conclusion.---} We employ first-principles calculation to search a series of materials that have a topologically intrinsic nature  and are suitable for topological nodal line semimetals, and that host multiple coexisting topological phases. In this work, we have investigated a series of materials  SrCaX (X=Bi, Sb, As, P) that are semimetalic in nature with band crossing in both at $k_z =0$ and $k_z =\pi$. The TRS and IS present in the system makes all the bands doubly degenerate. The investigation indicates that  type II Dirac point and topological Dirac nodal line semimetal features coexist, protected by $Z_2$ Berry phase, for all the materials. The Topological nodal line semimatic phase of materials is confirmed by the drum-head-like surface states present in the surface band structure. The multiple drum-head-like surface states guarantee the multiple nodal line semimetal feature present in the system. SrCaBi possesses one nodal line at $k_z =0$ and another at $k_z =\pi$, whereas all other materials SrCaSb, SrCaAs, and SrCaP possess two-nodal lines at $k_z =0$ and one at $k_z =\pi$. Additionally, SrCaP possesses topological Dirac semimetal and topological Dirac nodal line semimetal properties together in a single crystal. Due to the coexistance of Multiple Toplogical Dirac nodal lines, Dirac semimetal, and type II Dirac Point features, these systems can host several exotic phenomena that can be implemented for various applications. We anticipate that the findings presented here will encourage physicists to investigate these materials experimentally.

{\it Acknowledgments.---}
{This work is supported by the Science and Engineering Research Board (SERB), Govt. of India, through Grant No. CRG/2022/006419 and Council of Scientific and Industrial Research-Human Resource Development Group (CSIR-HRDG) through ASPIRE grant No. 03WS (006)/2023-24/EMR-II/ASPIRE. Author PS would like to thank NPSF C-DAC Pune for providing the HPC facility. Author AK would like to thank National Center for High-performance Computing (NCHC) for providing computational and storage resources. This work is also supported by the National Science and Technology Council, the Ministry of Education (Higher Education Sprout Project NTU-113L104022-1), and the National Center for Theoretical Sciences of Taiwan.}

Appendix A.--- Supplementary material
Supplementary data to this article can be found online at \cite{supplementary2024}

\bibliography{biblo}

\begin{thebibliography}{98}%
\makeatletter
\providecommand \@ifxundefined [1]{%
 \@ifx{#1\undefined}
}%
\providecommand \@ifnum [1]{%
 \ifnum #1\expandafter \@firstoftwo
 \else \expandafter \@secondoftwo
 \fi
}%
\providecommand \@ifx [1]{%
 \ifx #1\expandafter \@firstoftwo
 \else \expandafter \@secondoftwo
 \fi
}%
\providecommand \natexlab [1]{#1}%
\providecommand \enquote  [1]{``#1''}%
\providecommand \bibnamefont  [1]{#1}%
\providecommand \bibfnamefont [1]{#1}%
\providecommand \citenamefont [1]{#1}%
\providecommand \href@noop [0]{\@secondoftwo}%
\providecommand \href [0]{\begingroup \@sanitize@url \@href}%
\providecommand \@href[1]{\@@startlink{#1}\@@href}%
\providecommand \@@href[1]{\endgroup#1\@@endlink}%
\providecommand \@sanitize@url [0]{\catcode `\\12\catcode `\$12\catcode
  `\&12\catcode `\#12\catcode `\^12\catcode `\_12\catcode `\%12\relax}%
\providecommand \@@startlink[1]{}%
\providecommand \@@endlink[0]{}%
\providecommand \url  [0]{\begingroup\@sanitize@url \@url }%
\providecommand \@url [1]{\endgroup\@href {#1}{\urlprefix }}%
\providecommand \urlprefix  [0]{URL }%
\providecommand \Eprint [0]{\href }%
\providecommand \doibase [0]{https://doi.org/}%
\providecommand \selectlanguage [0]{\@gobble}%
\providecommand \bibinfo  [0]{\@secondoftwo}%
\providecommand \bibfield  [0]{\@secondoftwo}%
\providecommand \translation [1]{[#1]}%
\providecommand \BibitemOpen [0]{}%
\providecommand \bibitemStop [0]{}%
\providecommand \bibitemNoStop [0]{.\EOS\space}%
\providecommand \EOS [0]{\spacefactor3000\relax}%
\providecommand \BibitemShut  [1]{\csname bibitem#1\endcsname}%
\let\auto@bib@innerbib\@empty
\bibitem [{\citenamefont {Chang}\ \emph {et~al.}(2015)\citenamefont {Chang},
  \citenamefont {Tang}, \citenamefont {Feng}, \citenamefont {Li}, \citenamefont
  {Ma}, \citenamefont {Duan}, \citenamefont {He},\ and\ \citenamefont
  {Xue}}]{PhysRevLett.115.136801}%
  \BibitemOpen
  \bibfield  {author} {\bibinfo {author} {\bibfnamefont {C.-Z.}\ \bibnamefont
  {Chang}}, \bibinfo {author} {\bibfnamefont {P.}~\bibnamefont {Tang}},
  \bibinfo {author} {\bibfnamefont {X.}~\bibnamefont {Feng}}, \bibinfo {author}
  {\bibfnamefont {K.}~\bibnamefont {Li}}, \bibinfo {author} {\bibfnamefont
  {X.-C.}\ \bibnamefont {Ma}}, \bibinfo {author} {\bibfnamefont
  {W.}~\bibnamefont {Duan}}, \bibinfo {author} {\bibfnamefont {K.}~\bibnamefont
  {He}},\ and\ \bibinfo {author} {\bibfnamefont {Q.-K.}\ \bibnamefont {Xue}},\
  }\href {https://doi.org/10.1103/PhysRevLett.115.136801} {\bibfield  {journal}
  {\bibinfo  {journal} {Phys. Rev. Lett.}\ }\textbf {\bibinfo {volume} {115}},\
  \bibinfo {pages} {136801} (\bibinfo {year} {2015})}\BibitemShut {NoStop}%
\bibitem [{\citenamefont {Ma}\ \emph {et~al.}(2016)\citenamefont {Ma},
  \citenamefont {Kou}, \citenamefont {Dai},\ and\ \citenamefont
  {Heine}}]{PhysRevB.94.201104}%
  \BibitemOpen
  \bibfield  {author} {\bibinfo {author} {\bibfnamefont {Y.}~\bibnamefont
  {Ma}}, \bibinfo {author} {\bibfnamefont {L.}~\bibnamefont {Kou}}, \bibinfo
  {author} {\bibfnamefont {Y.}~\bibnamefont {Dai}},\ and\ \bibinfo {author}
  {\bibfnamefont {T.}~\bibnamefont {Heine}},\ }\href
  {https://doi.org/10.1103/PhysRevB.94.201104} {\bibfield  {journal} {\bibinfo
  {journal} {Phys. Rev. B}\ }\textbf {\bibinfo {volume} {94}},\ \bibinfo
  {pages} {201104} (\bibinfo {year} {2016})}\BibitemShut {NoStop}%
\bibitem [{\citenamefont {Khazaei}\ \emph {et~al.}(2016)\citenamefont
  {Khazaei}, \citenamefont {Ranjbar}, \citenamefont {Arai},\ and\ \citenamefont
  {Yunoki}}]{PhysRevB.94.125152}%
  \BibitemOpen
  \bibfield  {author} {\bibinfo {author} {\bibfnamefont {M.}~\bibnamefont
  {Khazaei}}, \bibinfo {author} {\bibfnamefont {A.}~\bibnamefont {Ranjbar}},
  \bibinfo {author} {\bibfnamefont {M.}~\bibnamefont {Arai}},\ and\ \bibinfo
  {author} {\bibfnamefont {S.}~\bibnamefont {Yunoki}},\ }\href
  {https://doi.org/10.1103/PhysRevB.94.125152} {\bibfield  {journal} {\bibinfo
  {journal} {Phys. Rev. B}\ }\textbf {\bibinfo {volume} {94}},\ \bibinfo
  {pages} {125152} (\bibinfo {year} {2016})}\BibitemShut {NoStop}%
\bibitem [{\citenamefont {Hasan}\ and\ \citenamefont
  {Kane}(2010)}]{RevModPhys.82.3045}%
  \BibitemOpen
  \bibfield  {author} {\bibinfo {author} {\bibfnamefont {M.~Z.}\ \bibnamefont
  {Hasan}}\ and\ \bibinfo {author} {\bibfnamefont {C.~L.}\ \bibnamefont
  {Kane}},\ }\href {https://doi.org/10.1103/RevModPhys.82.3045} {\bibfield
  {journal} {\bibinfo  {journal} {Rev. Mod. Phys.}\ }\textbf {\bibinfo {volume}
  {82}},\ \bibinfo {pages} {3045} (\bibinfo {year} {2010})}\BibitemShut
  {NoStop}%
\bibitem [{\citenamefont {Gupta}\ \emph
  {et~al.}(2023{\natexlab{a}})\citenamefont {Gupta}, \citenamefont {Kore},
  \citenamefont {Sen},\ and\ \citenamefont {Singh}}]{PhysRevB.107.075143}%
  \BibitemOpen
  \bibfield  {author} {\bibinfo {author} {\bibfnamefont {S.~K.}\ \bibnamefont
  {Gupta}}, \bibinfo {author} {\bibfnamefont {A.}~\bibnamefont {Kore}},
  \bibinfo {author} {\bibfnamefont {S.~K.}\ \bibnamefont {Sen}},\ and\ \bibinfo
  {author} {\bibfnamefont {P.}~\bibnamefont {Singh}},\ }\href
  {https://doi.org/10.1103/PhysRevB.107.075143} {\bibfield  {journal} {\bibinfo
   {journal} {Phys. Rev. B}\ }\textbf {\bibinfo {volume} {107}},\ \bibinfo
  {pages} {075143} (\bibinfo {year} {2023}{\natexlab{a}})}\BibitemShut
  {NoStop}%
\bibitem [{\citenamefont {Mukherjee}\ \emph {et~al.}(2020)\citenamefont
  {Mukherjee}, \citenamefont {Jung}, \citenamefont {Weber}, \citenamefont {Xu},
  \citenamefont {Qian}, \citenamefont {Xu}, \citenamefont {Biswas},
  \citenamefont {Kim}, \citenamefont {Chapon}, \citenamefont {Watson} \emph
  {et~al.}}]{mukherjee2020fermi}%
  \BibitemOpen
  \bibfield  {author} {\bibinfo {author} {\bibfnamefont {S.}~\bibnamefont
  {Mukherjee}}, \bibinfo {author} {\bibfnamefont {S.~W.}\ \bibnamefont {Jung}},
  \bibinfo {author} {\bibfnamefont {S.~F.}\ \bibnamefont {Weber}}, \bibinfo
  {author} {\bibfnamefont {C.}~\bibnamefont {Xu}}, \bibinfo {author}
  {\bibfnamefont {D.}~\bibnamefont {Qian}}, \bibinfo {author} {\bibfnamefont
  {X.}~\bibnamefont {Xu}}, \bibinfo {author} {\bibfnamefont {P.~K.}\
  \bibnamefont {Biswas}}, \bibinfo {author} {\bibfnamefont {T.~K.}\
  \bibnamefont {Kim}}, \bibinfo {author} {\bibfnamefont {L.~C.}\ \bibnamefont
  {Chapon}}, \bibinfo {author} {\bibfnamefont {M.~D.}\ \bibnamefont {Watson}},
  \emph {et~al.},\ }\href {https://doi.org/10.1038/s41598-020-69926-8}
  {\bibfield  {journal} {\bibinfo  {journal} {Scientific reports}\ }\textbf
  {\bibinfo {volume} {10}},\ \bibinfo {pages} {12957} (\bibinfo {year}
  {2020})}\BibitemShut {NoStop}%
\bibitem [{\citenamefont {Fu}\ \emph {et~al.}(2019)\citenamefont {Fu},
  \citenamefont {Yi}, \citenamefont {Zhang}, \citenamefont {Caputo},
  \citenamefont {Ma}, \citenamefont {Gao}, \citenamefont {Lv}, \citenamefont
  {Kong}, \citenamefont {Huang}, \citenamefont {Richard} \emph
  {et~al.}}]{fu2019dirac}%
  \BibitemOpen
  \bibfield  {author} {\bibinfo {author} {\bibfnamefont {B.-B.}\ \bibnamefont
  {Fu}}, \bibinfo {author} {\bibfnamefont {C.-J.}\ \bibnamefont {Yi}}, \bibinfo
  {author} {\bibfnamefont {T.-T.}\ \bibnamefont {Zhang}}, \bibinfo {author}
  {\bibfnamefont {M.}~\bibnamefont {Caputo}}, \bibinfo {author} {\bibfnamefont
  {J.-Z.}\ \bibnamefont {Ma}}, \bibinfo {author} {\bibfnamefont
  {X.}~\bibnamefont {Gao}}, \bibinfo {author} {\bibfnamefont {B.}~\bibnamefont
  {Lv}}, \bibinfo {author} {\bibfnamefont {L.-Y.}\ \bibnamefont {Kong}},
  \bibinfo {author} {\bibfnamefont {Y.-B.}\ \bibnamefont {Huang}}, \bibinfo
  {author} {\bibfnamefont {P.}~\bibnamefont {Richard}}, \emph {et~al.},\ }\href
  {https://doi.org/10.1126/sciadv.aau6459} {\bibfield  {journal} {\bibinfo
  {journal} {Science advances}\ }\textbf {\bibinfo {volume} {5}},\ \bibinfo
  {pages} {eaau6459} (\bibinfo {year} {2019})}\BibitemShut {NoStop}%
\bibitem [{\citenamefont {Fang}\ \emph {et~al.}(2016)\citenamefont {Fang},
  \citenamefont {Weng}, \citenamefont {Dai},\ and\ \citenamefont
  {Fang}}]{fang2016topological}%
  \BibitemOpen
  \bibfield  {author} {\bibinfo {author} {\bibfnamefont {C.}~\bibnamefont
  {Fang}}, \bibinfo {author} {\bibfnamefont {H.}~\bibnamefont {Weng}}, \bibinfo
  {author} {\bibfnamefont {X.}~\bibnamefont {Dai}},\ and\ \bibinfo {author}
  {\bibfnamefont {Z.}~\bibnamefont {Fang}},\ }\href
  {https://doi.org/10.1088/1674-1056/25/11/117106} {\bibfield  {journal}
  {\bibinfo  {journal} {Chinese Physics B}\ }\textbf {\bibinfo {volume} {25}},\
  \bibinfo {pages} {117106} (\bibinfo {year} {2016})}\BibitemShut {NoStop}%
\bibitem [{\citenamefont {Weng}\ \emph {et~al.}(2017)\citenamefont {Weng},
  \citenamefont {Fang}, \citenamefont {Fang},\ and\ \citenamefont
  {Dai}}]{weng2017new}%
  \BibitemOpen
  \bibfield  {author} {\bibinfo {author} {\bibfnamefont {H.}~\bibnamefont
  {Weng}}, \bibinfo {author} {\bibfnamefont {C.}~\bibnamefont {Fang}}, \bibinfo
  {author} {\bibfnamefont {Z.}~\bibnamefont {Fang}},\ and\ \bibinfo {author}
  {\bibfnamefont {X.}~\bibnamefont {Dai}},\ }\href
  {https://doi.org/10.1093/nsr/nwx066} {\bibfield  {journal} {\bibinfo
  {journal} {National Science Review}\ }\textbf {\bibinfo {volume} {4}},\
  \bibinfo {pages} {798} (\bibinfo {year} {2017})}\BibitemShut {NoStop}%
\bibitem [{\citenamefont {Wang}\ \emph
  {et~al.}(2012{\natexlab{a}})\citenamefont {Wang}, \citenamefont {Sun},
  \citenamefont {Chen}, \citenamefont {Franchini}, \citenamefont {Xu},
  \citenamefont {Weng}, \citenamefont {Dai},\ and\ \citenamefont
  {Fang}}]{wang2012dirac}%
  \BibitemOpen
  \bibfield  {author} {\bibinfo {author} {\bibfnamefont {Z.}~\bibnamefont
  {Wang}}, \bibinfo {author} {\bibfnamefont {Y.}~\bibnamefont {Sun}}, \bibinfo
  {author} {\bibfnamefont {X.-Q.}\ \bibnamefont {Chen}}, \bibinfo {author}
  {\bibfnamefont {C.}~\bibnamefont {Franchini}}, \bibinfo {author}
  {\bibfnamefont {G.}~\bibnamefont {Xu}}, \bibinfo {author} {\bibfnamefont
  {H.}~\bibnamefont {Weng}}, \bibinfo {author} {\bibfnamefont {X.}~\bibnamefont
  {Dai}},\ and\ \bibinfo {author} {\bibfnamefont {Z.}~\bibnamefont {Fang}},\
  }\href {https://doi.org/10.1103/PhysRevB.85.195320} {\bibfield  {journal}
  {\bibinfo  {journal} {Physical Review B}\ }\textbf {\bibinfo {volume} {85}},\
  \bibinfo {pages} {195320} (\bibinfo {year} {2012}{\natexlab{a}})}\BibitemShut
  {NoStop}%
\bibitem [{\citenamefont {Liu}\ \emph {et~al.}(2014)\citenamefont {Liu},
  \citenamefont {Zhou}, \citenamefont {Zhang}, \citenamefont {Wang},
  \citenamefont {Weng}, \citenamefont {Prabhakaran}, \citenamefont {Mo},
  \citenamefont {Shen}, \citenamefont {Fang}, \citenamefont {Dai} \emph
  {et~al.}}]{liu2014discovery}%
  \BibitemOpen
  \bibfield  {author} {\bibinfo {author} {\bibfnamefont {Z.}~\bibnamefont
  {Liu}}, \bibinfo {author} {\bibfnamefont {B.}~\bibnamefont {Zhou}}, \bibinfo
  {author} {\bibfnamefont {Y.}~\bibnamefont {Zhang}}, \bibinfo {author}
  {\bibfnamefont {Z.}~\bibnamefont {Wang}}, \bibinfo {author} {\bibfnamefont
  {H.}~\bibnamefont {Weng}}, \bibinfo {author} {\bibfnamefont {D.}~\bibnamefont
  {Prabhakaran}}, \bibinfo {author} {\bibfnamefont {S.-K.}\ \bibnamefont {Mo}},
  \bibinfo {author} {\bibfnamefont {Z.}~\bibnamefont {Shen}}, \bibinfo {author}
  {\bibfnamefont {Z.}~\bibnamefont {Fang}}, \bibinfo {author} {\bibfnamefont
  {X.}~\bibnamefont {Dai}}, \emph {et~al.},\ }\href
  {https://doi.org/10.1126/science.1245085} {\bibfield  {journal} {\bibinfo
  {journal} {Science}\ }\textbf {\bibinfo {volume} {343}},\ \bibinfo {pages}
  {864} (\bibinfo {year} {2014})}\BibitemShut {NoStop}%
\bibitem [{\citenamefont {Wang}\ \emph {et~al.}(2013)\citenamefont {Wang},
  \citenamefont {Weng}, \citenamefont {Wu}, \citenamefont {Dai},\ and\
  \citenamefont {Fang}}]{wang2013three}%
  \BibitemOpen
  \bibfield  {author} {\bibinfo {author} {\bibfnamefont {Z.}~\bibnamefont
  {Wang}}, \bibinfo {author} {\bibfnamefont {H.}~\bibnamefont {Weng}}, \bibinfo
  {author} {\bibfnamefont {Q.}~\bibnamefont {Wu}}, \bibinfo {author}
  {\bibfnamefont {X.}~\bibnamefont {Dai}},\ and\ \bibinfo {author}
  {\bibfnamefont {Z.}~\bibnamefont {Fang}},\ }\href
  {https://doi.org/10.1103/PhysRevB.88.125427} {\bibfield  {journal} {\bibinfo
  {journal} {Physical Review B}\ }\textbf {\bibinfo {volume} {88}},\ \bibinfo
  {pages} {125427} (\bibinfo {year} {2013})}\BibitemShut {NoStop}%
\bibitem [{\citenamefont {Wan}\ \emph {et~al.}(2011)\citenamefont {Wan},
  \citenamefont {Turner}, \citenamefont {Vishwanath},\ and\ \citenamefont
  {Savrasov}}]{wan2011topological}%
  \BibitemOpen
  \bibfield  {author} {\bibinfo {author} {\bibfnamefont {X.}~\bibnamefont
  {Wan}}, \bibinfo {author} {\bibfnamefont {A.~M.}\ \bibnamefont {Turner}},
  \bibinfo {author} {\bibfnamefont {A.}~\bibnamefont {Vishwanath}},\ and\
  \bibinfo {author} {\bibfnamefont {S.~Y.}\ \bibnamefont {Savrasov}},\ }\href
  {https://doi.org/10.1103/PhysRevB.83.205101} {\bibfield  {journal} {\bibinfo
  {journal} {Physical Review B}\ }\textbf {\bibinfo {volume} {83}},\ \bibinfo
  {pages} {205101} (\bibinfo {year} {2011})}\BibitemShut {NoStop}%
\bibitem [{\citenamefont {Peng}\ \emph {et~al.}(2018)\citenamefont {Peng},
  \citenamefont {Yue}, \citenamefont {Zhang}, \citenamefont {Fang},\ and\
  \citenamefont {Weng}}]{peng2018predicting}%
  \BibitemOpen
  \bibfield  {author} {\bibinfo {author} {\bibfnamefont {B.}~\bibnamefont
  {Peng}}, \bibinfo {author} {\bibfnamefont {C.}~\bibnamefont {Yue}}, \bibinfo
  {author} {\bibfnamefont {H.}~\bibnamefont {Zhang}}, \bibinfo {author}
  {\bibfnamefont {Z.}~\bibnamefont {Fang}},\ and\ \bibinfo {author}
  {\bibfnamefont {H.}~\bibnamefont {Weng}},\ }\href
  {https://doi.org/10.1038/s41524-018-0124-5} {\bibfield  {journal} {\bibinfo
  {journal} {npj Computational Materials}\ }\textbf {\bibinfo {volume} {4}},\
  \bibinfo {pages} {68} (\bibinfo {year} {2018})}\BibitemShut {NoStop}%
\bibitem [{\citenamefont {Shende}\ \emph
  {et~al.}(2023{\natexlab{a}})\citenamefont {Shende}, \citenamefont {Gupta},
  \citenamefont {Kale},\ and\ \citenamefont {Singh}}]{shende2023first}%
  \BibitemOpen
  \bibfield  {author} {\bibinfo {author} {\bibfnamefont {A.}~\bibnamefont
  {Shende}}, \bibinfo {author} {\bibfnamefont {S.~K.}\ \bibnamefont {Gupta}},
  \bibinfo {author} {\bibfnamefont {D.}~\bibnamefont {Kale}},\ and\ \bibinfo
  {author} {\bibfnamefont {P.}~\bibnamefont {Singh}},\ }\href
  {https://doi.org/10.1016/j.physleta.2023.128937} {\bibfield  {journal}
  {\bibinfo  {journal} {Physics Letters A}\ }\textbf {\bibinfo {volume}
  {480}},\ \bibinfo {pages} {128937} (\bibinfo {year}
  {2023}{\natexlab{a}})}\BibitemShut {NoStop}%
\bibitem [{\citenamefont {Gupta}\ \emph
  {et~al.}(2023{\natexlab{b}})\citenamefont {Gupta}, \citenamefont {Singh},
  \citenamefont {Sen},\ and\ \citenamefont {Singh}}]{gupta2023pressure}%
  \BibitemOpen
  \bibfield  {author} {\bibinfo {author} {\bibfnamefont {S.~K.}\ \bibnamefont
  {Gupta}}, \bibinfo {author} {\bibfnamefont {N.}~\bibnamefont {Singh}},
  \bibinfo {author} {\bibfnamefont {S.~K.}\ \bibnamefont {Sen}},\ and\ \bibinfo
  {author} {\bibfnamefont {P.}~\bibnamefont {Singh}},\ }\bibfield  {journal}
  {\bibinfo  {journal} {arXiv preprint arXiv:2312.05636}\ }\href
  {https://doi.org/10.48550/arXiv.2312.05636} {10.48550/arXiv.2312.05636}
  (\bibinfo {year} {2023}{\natexlab{b}})\BibitemShut {NoStop}%
\bibitem [{\citenamefont {Quintela}\ \emph {et~al.}(2017)\citenamefont
  {Quintela}, \citenamefont {Campbell}, \citenamefont {Shao}, \citenamefont
  {Irwin}, \citenamefont {Harris}, \citenamefont {Xie}, \citenamefont
  {Anderson}, \citenamefont {Reiser}, \citenamefont {Pan}, \citenamefont
  {Tsymbal} \emph {et~al.}}]{quintela2017epitaxial}%
  \BibitemOpen
  \bibfield  {author} {\bibinfo {author} {\bibfnamefont {C.}~\bibnamefont
  {Quintela}}, \bibinfo {author} {\bibfnamefont {N.}~\bibnamefont {Campbell}},
  \bibinfo {author} {\bibfnamefont {D.}~\bibnamefont {Shao}}, \bibinfo {author}
  {\bibfnamefont {J.}~\bibnamefont {Irwin}}, \bibinfo {author} {\bibfnamefont
  {D.}~\bibnamefont {Harris}}, \bibinfo {author} {\bibfnamefont
  {L.}~\bibnamefont {Xie}}, \bibinfo {author} {\bibfnamefont {T.}~\bibnamefont
  {Anderson}}, \bibinfo {author} {\bibfnamefont {N.}~\bibnamefont {Reiser}},
  \bibinfo {author} {\bibfnamefont {X.}~\bibnamefont {Pan}}, \bibinfo {author}
  {\bibfnamefont {E.~Y.}\ \bibnamefont {Tsymbal}}, \emph {et~al.},\ }\bibfield
  {journal} {\bibinfo  {journal} {APL Materials}\ }\textbf {\bibinfo {volume}
  {5}},\ \href {https://doi.org/10.1063/1.4992006} {10.1063/1.4992006}
  (\bibinfo {year} {2017})\BibitemShut {NoStop}%
\bibitem [{\citenamefont {Xiong}\ \emph {et~al.}(2015)\citenamefont {Xiong},
  \citenamefont {Kushwaha}, \citenamefont {Liang}, \citenamefont {Krizan},
  \citenamefont {Hirschberger}, \citenamefont {Wang}, \citenamefont {Cava},\
  and\ \citenamefont {Ong}}]{xiong2015evidence}%
  \BibitemOpen
  \bibfield  {author} {\bibinfo {author} {\bibfnamefont {J.}~\bibnamefont
  {Xiong}}, \bibinfo {author} {\bibfnamefont {S.~K.}\ \bibnamefont {Kushwaha}},
  \bibinfo {author} {\bibfnamefont {T.}~\bibnamefont {Liang}}, \bibinfo
  {author} {\bibfnamefont {J.~W.}\ \bibnamefont {Krizan}}, \bibinfo {author}
  {\bibfnamefont {M.}~\bibnamefont {Hirschberger}}, \bibinfo {author}
  {\bibfnamefont {W.}~\bibnamefont {Wang}}, \bibinfo {author} {\bibfnamefont
  {R.~J.}\ \bibnamefont {Cava}},\ and\ \bibinfo {author} {\bibfnamefont
  {N.~P.}\ \bibnamefont {Ong}},\ }\href
  {https://doi.org/https:10.1126/science.aac6089} {\bibfield  {journal}
  {\bibinfo  {journal} {Science}\ }\textbf {\bibinfo {volume} {350}},\ \bibinfo
  {pages} {413} (\bibinfo {year} {2015})}\BibitemShut {NoStop}%
\bibitem [{\citenamefont {Wang}\ \emph
  {et~al.}(2012{\natexlab{b}})\citenamefont {Wang}, \citenamefont {Petrovic}
  \emph {et~al.}}]{wang2012multiband}%
  \BibitemOpen
  \bibfield  {author} {\bibinfo {author} {\bibfnamefont {K.}~\bibnamefont
  {Wang}}, \bibinfo {author} {\bibfnamefont {C.}~\bibnamefont {Petrovic}},
  \emph {et~al.},\ }\href {https://doi.org/10.1103/PhysRevB.86.155213}
  {\bibfield  {journal} {\bibinfo  {journal} {Physical Review B}\ }\textbf
  {\bibinfo {volume} {86}},\ \bibinfo {pages} {155213} (\bibinfo {year}
  {2012}{\natexlab{b}})}\BibitemShut {NoStop}%
\bibitem [{\citenamefont {Yen}\ and\ \citenamefont
  {Guo}(2020)}]{yen2020tunable}%
  \BibitemOpen
  \bibfield  {author} {\bibinfo {author} {\bibfnamefont {Y.}~\bibnamefont
  {Yen}}\ and\ \bibinfo {author} {\bibfnamefont {G.-Y.}\ \bibnamefont {Guo}},\
  }\href {https://doi.org/10.1103/PhysRevB.101.064430} {\bibfield  {journal}
  {\bibinfo  {journal} {Physical Review B}\ }\textbf {\bibinfo {volume}
  {101}},\ \bibinfo {pages} {064430} (\bibinfo {year} {2020})}\BibitemShut
  {NoStop}%
\bibitem [{\citenamefont {Song}\ \emph {et~al.}(2021)\citenamefont {Song},
  \citenamefont {Kim}, \citenamefont {Kim}, \citenamefont {Fu}, \citenamefont
  {Koo}, \citenamefont {Wang}, \citenamefont {Lee}, \citenamefont {Lee},
  \citenamefont {Oh}, \citenamefont {Bang} \emph
  {et~al.}}]{song2021coexistence}%
  \BibitemOpen
  \bibfield  {author} {\bibinfo {author} {\bibfnamefont {J.}~\bibnamefont
  {Song}}, \bibinfo {author} {\bibfnamefont {S.}~\bibnamefont {Kim}}, \bibinfo
  {author} {\bibfnamefont {Y.}~\bibnamefont {Kim}}, \bibinfo {author}
  {\bibfnamefont {H.}~\bibnamefont {Fu}}, \bibinfo {author} {\bibfnamefont
  {J.}~\bibnamefont {Koo}}, \bibinfo {author} {\bibfnamefont {Z.}~\bibnamefont
  {Wang}}, \bibinfo {author} {\bibfnamefont {G.}~\bibnamefont {Lee}}, \bibinfo
  {author} {\bibfnamefont {J.}~\bibnamefont {Lee}}, \bibinfo {author}
  {\bibfnamefont {S.~H.}\ \bibnamefont {Oh}}, \bibinfo {author} {\bibfnamefont
  {J.}~\bibnamefont {Bang}}, \emph {et~al.},\ }\href
  {https://doi.org/10.1103/PhysRevX.11.021065} {\bibfield  {journal} {\bibinfo
  {journal} {Physical Review X}\ }\textbf {\bibinfo {volume} {11}},\ \bibinfo
  {pages} {021065} (\bibinfo {year} {2021})}\BibitemShut {NoStop}%
\bibitem [{\citenamefont {Weng}\ \emph {et~al.}(2015)\citenamefont {Weng},
  \citenamefont {Fang}, \citenamefont {Fang}, \citenamefont {Bernevig},\ and\
  \citenamefont {Dai}}]{weng2015weyl}%
  \BibitemOpen
  \bibfield  {author} {\bibinfo {author} {\bibfnamefont {H.}~\bibnamefont
  {Weng}}, \bibinfo {author} {\bibfnamefont {C.}~\bibnamefont {Fang}}, \bibinfo
  {author} {\bibfnamefont {Z.}~\bibnamefont {Fang}}, \bibinfo {author}
  {\bibfnamefont {B.~A.}\ \bibnamefont {Bernevig}},\ and\ \bibinfo {author}
  {\bibfnamefont {X.}~\bibnamefont {Dai}},\ }\href
  {https://doi.org/10.1103/PhysRevX.5.011029} {\bibfield  {journal} {\bibinfo
  {journal} {Physical Review X}\ }\textbf {\bibinfo {volume} {5}},\ \bibinfo
  {pages} {011029} (\bibinfo {year} {2015})}\BibitemShut {NoStop}%
\bibitem [{\citenamefont {Yan}\ and\ \citenamefont
  {Felser}(2017)}]{yan2017topological}%
  \BibitemOpen
  \bibfield  {author} {\bibinfo {author} {\bibfnamefont {B.}~\bibnamefont
  {Yan}}\ and\ \bibinfo {author} {\bibfnamefont {C.}~\bibnamefont {Felser}},\
  }\href {https://doi.org/10.1146/annurev-conmatphys-031016-025458} {\bibfield
  {journal} {\bibinfo  {journal} {Annual Review of Condensed Matter Physics}\
  }\textbf {\bibinfo {volume} {8}},\ \bibinfo {pages} {337} (\bibinfo {year}
  {2017})}\BibitemShut {NoStop}%
\bibitem [{\citenamefont {McCormick}\ \emph {et~al.}(2017)\citenamefont
  {McCormick}, \citenamefont {Kimchi},\ and\ \citenamefont
  {Trivedi}}]{mccormick2017minimal}%
  \BibitemOpen
  \bibfield  {author} {\bibinfo {author} {\bibfnamefont {T.~M.}\ \bibnamefont
  {McCormick}}, \bibinfo {author} {\bibfnamefont {I.}~\bibnamefont {Kimchi}},\
  and\ \bibinfo {author} {\bibfnamefont {N.}~\bibnamefont {Trivedi}},\ }\href
  {https://doi.org/10.1103/PhysRevB.95.075133} {\bibfield  {journal} {\bibinfo
  {journal} {Physical Review B}\ }\textbf {\bibinfo {volume} {95}},\ \bibinfo
  {pages} {075133} (\bibinfo {year} {2017})}\BibitemShut {NoStop}%
\bibitem [{\citenamefont {Osterhoudt}\ \emph {et~al.}(2019)\citenamefont
  {Osterhoudt}, \citenamefont {Diebel}, \citenamefont {Gray}, \citenamefont
  {Yang}, \citenamefont {Stanco}, \citenamefont {Huang}, \citenamefont {Shen},
  \citenamefont {Ni}, \citenamefont {Moll}, \citenamefont {Ran} \emph
  {et~al.}}]{osterhoudt2019colossal}%
  \BibitemOpen
  \bibfield  {author} {\bibinfo {author} {\bibfnamefont {G.~B.}\ \bibnamefont
  {Osterhoudt}}, \bibinfo {author} {\bibfnamefont {L.~K.}\ \bibnamefont
  {Diebel}}, \bibinfo {author} {\bibfnamefont {M.~J.}\ \bibnamefont {Gray}},
  \bibinfo {author} {\bibfnamefont {X.}~\bibnamefont {Yang}}, \bibinfo {author}
  {\bibfnamefont {J.}~\bibnamefont {Stanco}}, \bibinfo {author} {\bibfnamefont
  {X.}~\bibnamefont {Huang}}, \bibinfo {author} {\bibfnamefont
  {B.}~\bibnamefont {Shen}}, \bibinfo {author} {\bibfnamefont {N.}~\bibnamefont
  {Ni}}, \bibinfo {author} {\bibfnamefont {P.~J.}\ \bibnamefont {Moll}},
  \bibinfo {author} {\bibfnamefont {Y.}~\bibnamefont {Ran}}, \emph {et~al.},\
  }\href {https://doi.org/10.1038/s41563-019-0297-4} {\bibfield  {journal}
  {\bibinfo  {journal} {Nature materials}\ }\textbf {\bibinfo {volume} {18}},\
  \bibinfo {pages} {471} (\bibinfo {year} {2019})}\BibitemShut {NoStop}%
\bibitem [{\citenamefont {Liu}\ \emph {et~al.}(2016)\citenamefont {Liu},
  \citenamefont {Yang}, \citenamefont {Sun}, \citenamefont {Zhang},
  \citenamefont {Peng}, \citenamefont {Yang}, \citenamefont {Chen},
  \citenamefont {Zhang}, \citenamefont {Guo}, \citenamefont {Prabhakaran} \emph
  {et~al.}}]{liu2016evolution}%
  \BibitemOpen
  \bibfield  {author} {\bibinfo {author} {\bibfnamefont {Z.}~\bibnamefont
  {Liu}}, \bibinfo {author} {\bibfnamefont {L.}~\bibnamefont {Yang}}, \bibinfo
  {author} {\bibfnamefont {Y.}~\bibnamefont {Sun}}, \bibinfo {author}
  {\bibfnamefont {T.}~\bibnamefont {Zhang}}, \bibinfo {author} {\bibfnamefont
  {H.}~\bibnamefont {Peng}}, \bibinfo {author} {\bibfnamefont {H.}~\bibnamefont
  {Yang}}, \bibinfo {author} {\bibfnamefont {C.}~\bibnamefont {Chen}}, \bibinfo
  {author} {\bibfnamefont {Y.~f.}\ \bibnamefont {Zhang}}, \bibinfo {author}
  {\bibfnamefont {Y.}~\bibnamefont {Guo}}, \bibinfo {author} {\bibfnamefont
  {D.}~\bibnamefont {Prabhakaran}}, \emph {et~al.},\ }\href
  {https://doi.org/doi.org/10.1038/nmat4457} {\bibfield  {journal} {\bibinfo
  {journal} {Nature materials}\ }\textbf {\bibinfo {volume} {15}},\ \bibinfo
  {pages} {27} (\bibinfo {year} {2016})}\BibitemShut {NoStop}%
\bibitem [{\citenamefont {Yuan}\ \emph {et~al.}(2020)\citenamefont {Yuan},
  \citenamefont {Zhang}, \citenamefont {Zhang}, \citenamefont {Yan},
  \citenamefont {Lyu}, \citenamefont {Zhang}, \citenamefont {Li}, \citenamefont
  {Song}, \citenamefont {Zhao}, \citenamefont {Leng} \emph
  {et~al.}}]{yuan2020discovery}%
  \BibitemOpen
  \bibfield  {author} {\bibinfo {author} {\bibfnamefont {X.}~\bibnamefont
  {Yuan}}, \bibinfo {author} {\bibfnamefont {C.}~\bibnamefont {Zhang}},
  \bibinfo {author} {\bibfnamefont {Y.}~\bibnamefont {Zhang}}, \bibinfo
  {author} {\bibfnamefont {Z.}~\bibnamefont {Yan}}, \bibinfo {author}
  {\bibfnamefont {T.}~\bibnamefont {Lyu}}, \bibinfo {author} {\bibfnamefont
  {M.}~\bibnamefont {Zhang}}, \bibinfo {author} {\bibfnamefont
  {Z.}~\bibnamefont {Li}}, \bibinfo {author} {\bibfnamefont {C.}~\bibnamefont
  {Song}}, \bibinfo {author} {\bibfnamefont {M.}~\bibnamefont {Zhao}}, \bibinfo
  {author} {\bibfnamefont {P.}~\bibnamefont {Leng}}, \emph {et~al.},\ }\href
  {https://doi.org/10.1038/s41467-020-14749-4} {\bibfield  {journal} {\bibinfo
  {journal} {Nature communications}\ }\textbf {\bibinfo {volume} {11}},\
  \bibinfo {pages} {1259} (\bibinfo {year} {2020})}\BibitemShut {NoStop}%
\bibitem [{\citenamefont {Chang}\ \emph {et~al.}(2016)\citenamefont {Chang},
  \citenamefont {Xu}, \citenamefont {Zheng}, \citenamefont {Lee}, \citenamefont
  {Huang}, \citenamefont {Belopolski}, \citenamefont {Sanchez}, \citenamefont
  {Bian}, \citenamefont {Alidoust}, \citenamefont {Chang} \emph
  {et~al.}}]{chang2016signatures}%
  \BibitemOpen
  \bibfield  {author} {\bibinfo {author} {\bibfnamefont {G.}~\bibnamefont
  {Chang}}, \bibinfo {author} {\bibfnamefont {S.-Y.}\ \bibnamefont {Xu}},
  \bibinfo {author} {\bibfnamefont {H.}~\bibnamefont {Zheng}}, \bibinfo
  {author} {\bibfnamefont {C.-C.}\ \bibnamefont {Lee}}, \bibinfo {author}
  {\bibfnamefont {S.-M.}\ \bibnamefont {Huang}}, \bibinfo {author}
  {\bibfnamefont {I.}~\bibnamefont {Belopolski}}, \bibinfo {author}
  {\bibfnamefont {D.~S.}\ \bibnamefont {Sanchez}}, \bibinfo {author}
  {\bibfnamefont {G.}~\bibnamefont {Bian}}, \bibinfo {author} {\bibfnamefont
  {N.}~\bibnamefont {Alidoust}}, \bibinfo {author} {\bibfnamefont {T.-R.}\
  \bibnamefont {Chang}}, \emph {et~al.},\ }\href
  {https://doi.org/10.1103/PhysRevLett.116.066601} {\bibfield  {journal}
  {\bibinfo  {journal} {Physical review letters}\ }\textbf {\bibinfo {volume}
  {116}},\ \bibinfo {pages} {066601} (\bibinfo {year} {2016})}\BibitemShut
  {NoStop}%
\bibitem [{\citenamefont {Sun}\ \emph {et~al.}(2015)\citenamefont {Sun},
  \citenamefont {Wu}, \citenamefont {Ali}, \citenamefont {Felser},\ and\
  \citenamefont {Yan}}]{sun2015prediction}%
  \BibitemOpen
  \bibfield  {author} {\bibinfo {author} {\bibfnamefont {Y.}~\bibnamefont
  {Sun}}, \bibinfo {author} {\bibfnamefont {S.-C.}\ \bibnamefont {Wu}},
  \bibinfo {author} {\bibfnamefont {M.~N.}\ \bibnamefont {Ali}}, \bibinfo
  {author} {\bibfnamefont {C.}~\bibnamefont {Felser}},\ and\ \bibinfo {author}
  {\bibfnamefont {B.}~\bibnamefont {Yan}},\ }\href
  {https://doi.org/10.1103/PhysRevB.92.161107} {\bibfield  {journal} {\bibinfo
  {journal} {Physical Review B}\ }\textbf {\bibinfo {volume} {92}},\ \bibinfo
  {pages} {161107} (\bibinfo {year} {2015})}\BibitemShut {NoStop}%
\bibitem [{\citenamefont {Alidoust}\ \emph {et~al.}(2017)\citenamefont
  {Alidoust}, \citenamefont {Halterman},\ and\ \citenamefont
  {Zyuzin}}]{alidoust}%
  \BibitemOpen
  \bibfield  {author} {\bibinfo {author} {\bibfnamefont {M.}~\bibnamefont
  {Alidoust}}, \bibinfo {author} {\bibfnamefont {K.}~\bibnamefont
  {Halterman}},\ and\ \bibinfo {author} {\bibfnamefont {A.}~\bibnamefont
  {Zyuzin}},\ }\href {https://doi.org/10.1103/PhysRevB.95.155124} {\bibfield
  {journal} {\bibinfo  {journal} {Physical Review B}\ }\textbf {\bibinfo
  {volume} {95}},\ \bibinfo {pages} {155124} (\bibinfo {year}
  {2017})}\BibitemShut {NoStop}%
\bibitem [{\citenamefont {Li}\ \emph {et~al.}(2021)\citenamefont {Li},
  \citenamefont {Deng}, \citenamefont {Fu}, \citenamefont {Li}, \citenamefont
  {Ma}, \citenamefont {Han}, \citenamefont {Zhou}, \citenamefont {Zhou},\ and\
  \citenamefont {Yao}}]{li2021type}%
  \BibitemOpen
  \bibfield  {author} {\bibinfo {author} {\bibfnamefont {X.-P.}\ \bibnamefont
  {Li}}, \bibinfo {author} {\bibfnamefont {K.}~\bibnamefont {Deng}}, \bibinfo
  {author} {\bibfnamefont {B.}~\bibnamefont {Fu}}, \bibinfo {author}
  {\bibfnamefont {Y.}~\bibnamefont {Li}}, \bibinfo {author} {\bibfnamefont
  {D.-S.}\ \bibnamefont {Ma}}, \bibinfo {author} {\bibfnamefont
  {J.}~\bibnamefont {Han}}, \bibinfo {author} {\bibfnamefont {J.}~\bibnamefont
  {Zhou}}, \bibinfo {author} {\bibfnamefont {S.}~\bibnamefont {Zhou}},\ and\
  \bibinfo {author} {\bibfnamefont {Y.}~\bibnamefont {Yao}},\ }\href
  {https://doi.org/10.1103/PhysRevB.103.L081402} {\bibfield  {journal}
  {\bibinfo  {journal} {Physical Review B}\ }\textbf {\bibinfo {volume}
  {103}},\ \bibinfo {pages} {L081402} (\bibinfo {year} {2021})}\BibitemShut
  {NoStop}%
\bibitem [{\citenamefont {Jin}\ \emph {et~al.}(2020{\natexlab{a}})\citenamefont
  {Jin}, \citenamefont {Zhang}, \citenamefont {Liu}, \citenamefont {Dai},
  \citenamefont {Wang},\ and\ \citenamefont {Liu}}]{jin2020fully}%
  \BibitemOpen
  \bibfield  {author} {\bibinfo {author} {\bibfnamefont {L.}~\bibnamefont
  {Jin}}, \bibinfo {author} {\bibfnamefont {X.}~\bibnamefont {Zhang}}, \bibinfo
  {author} {\bibfnamefont {Y.}~\bibnamefont {Liu}}, \bibinfo {author}
  {\bibfnamefont {X.}~\bibnamefont {Dai}}, \bibinfo {author} {\bibfnamefont
  {L.}~\bibnamefont {Wang}},\ and\ \bibinfo {author} {\bibfnamefont
  {G.}~\bibnamefont {Liu}},\ }\href
  {https://doi.org/10.1103/PhysRevB.102.195104} {\bibfield  {journal} {\bibinfo
   {journal} {Physical Review B}\ }\textbf {\bibinfo {volume} {102}},\ \bibinfo
  {pages} {195104} (\bibinfo {year} {2020}{\natexlab{a}})}\BibitemShut
  {NoStop}%
\bibitem [{\citenamefont {Shende}\ \emph
  {et~al.}(2023{\natexlab{b}})\citenamefont {Shende}, \citenamefont
  {Kumar~Gupta}, \citenamefont {Kore},\ and\ \citenamefont
  {Singh}}]{shende2023pressure}%
  \BibitemOpen
  \bibfield  {author} {\bibinfo {author} {\bibfnamefont {A.}~\bibnamefont
  {Shende}}, \bibinfo {author} {\bibfnamefont {S.}~\bibnamefont {Kumar~Gupta}},
  \bibinfo {author} {\bibfnamefont {A.}~\bibnamefont {Kore}},\ and\ \bibinfo
  {author} {\bibfnamefont {P.}~\bibnamefont {Singh}},\ }\bibfield  {journal}
  {\bibinfo  {journal} {Condensed Matter Physics}\ }\textbf {\bibinfo {volume}
  {26}},\ \href {https://doi.org/10.5488/CMP.26.23707} {10.5488/CMP.26.23707}
  (\bibinfo {year} {2023}{\natexlab{b}})\BibitemShut {NoStop}%
\bibitem [{\citenamefont {Cheung}\ \emph {et~al.}(2018)\citenamefont {Cheung},
  \citenamefont {Xiao}, \citenamefont {Hsu}, \citenamefont {Fuh}, \citenamefont
  {Lin},\ and\ \citenamefont {Chang}}]{cheung2018}%
  \BibitemOpen
  \bibfield  {author} {\bibinfo {author} {\bibfnamefont {C.-H.}\ \bibnamefont
  {Cheung}}, \bibinfo {author} {\bibfnamefont {R.}~\bibnamefont {Xiao}},
  \bibinfo {author} {\bibfnamefont {M.-C.}\ \bibnamefont {Hsu}}, \bibinfo
  {author} {\bibfnamefont {H.-R.}\ \bibnamefont {Fuh}}, \bibinfo {author}
  {\bibfnamefont {Y.-C.}\ \bibnamefont {Lin}},\ and\ \bibinfo {author}
  {\bibfnamefont {C.-R.}\ \bibnamefont {Chang}},\ }\href
  {https://doi.org/10.1088/1367-2630/aaf11d} {\bibfield  {journal} {\bibinfo
  {journal} {New Journal of Physics}\ }\textbf {\bibinfo {volume} {20}},\
  \bibinfo {pages} {123002} (\bibinfo {year} {2018})}\BibitemShut {NoStop}%
\bibitem [{\citenamefont {Barik}\ \emph {et~al.}(2018)\citenamefont {Barik},
  \citenamefont {Shinde},\ and\ \citenamefont {Singh}}]{barik2018}%
  \BibitemOpen
  \bibfield  {author} {\bibinfo {author} {\bibfnamefont {R.~K.}\ \bibnamefont
  {Barik}}, \bibinfo {author} {\bibfnamefont {R.}~\bibnamefont {Shinde}},\ and\
  \bibinfo {author} {\bibfnamefont {A.~K.}\ \bibnamefont {Singh}},\ }\href
  {https://doi.org/10.1088/1361-648X/aad8e1} {\bibfield  {journal} {\bibinfo
  {journal} {Journal of Physics: Condensed Matter}\ }\textbf {\bibinfo {volume}
  {30}},\ \bibinfo {pages} {375702} (\bibinfo {year} {2018})}\BibitemShut
  {NoStop}%
\bibitem [{\citenamefont {Lv}\ \emph {et~al.}(2017{\natexlab{a}})\citenamefont
  {Lv}, \citenamefont {Feng}, \citenamefont {Xu}, \citenamefont {Gao},
  \citenamefont {Ma}, \citenamefont {Kong}, \citenamefont {Richard},
  \citenamefont {Huang}, \citenamefont {Strocov}, \citenamefont {Fang} \emph
  {et~al.}}]{lv2017}%
  \BibitemOpen
  \bibfield  {author} {\bibinfo {author} {\bibfnamefont {B.}~\bibnamefont
  {Lv}}, \bibinfo {author} {\bibfnamefont {Z.-L.}\ \bibnamefont {Feng}},
  \bibinfo {author} {\bibfnamefont {Q.-N.}\ \bibnamefont {Xu}}, \bibinfo
  {author} {\bibfnamefont {X.}~\bibnamefont {Gao}}, \bibinfo {author}
  {\bibfnamefont {J.-Z.}\ \bibnamefont {Ma}}, \bibinfo {author} {\bibfnamefont
  {L.-Y.}\ \bibnamefont {Kong}}, \bibinfo {author} {\bibfnamefont
  {P.}~\bibnamefont {Richard}}, \bibinfo {author} {\bibfnamefont {Y.-B.}\
  \bibnamefont {Huang}}, \bibinfo {author} {\bibfnamefont {V.}~\bibnamefont
  {Strocov}}, \bibinfo {author} {\bibfnamefont {C.}~\bibnamefont {Fang}}, \emph
  {et~al.},\ }\href {https://doi.org/10.1038/nature22390} {\bibfield  {journal}
  {\bibinfo  {journal} {Nature}\ }\textbf {\bibinfo {volume} {546}},\ \bibinfo
  {pages} {627} (\bibinfo {year} {2017}{\natexlab{a}})}\BibitemShut {NoStop}%
\bibitem [{\citenamefont {Fang}\ \emph {et~al.}(2020)\citenamefont {Fang},
  \citenamefont {Gao}, \citenamefont {Venderbos},\ and\ \citenamefont
  {Rappe}}]{fang2020ideal}%
  \BibitemOpen
  \bibfield  {author} {\bibinfo {author} {\bibfnamefont {Z.}~\bibnamefont
  {Fang}}, \bibinfo {author} {\bibfnamefont {H.}~\bibnamefont {Gao}}, \bibinfo
  {author} {\bibfnamefont {J.~r.~W.}\ \bibnamefont {Venderbos}},\ and\ \bibinfo
  {author} {\bibfnamefont {A.~M.}\ \bibnamefont {Rappe}},\ }\href
  {https://doi.org/10.1103/PhysRevB.101.125202} {\bibfield  {journal} {\bibinfo
   {journal} {Physical Review B}\ }\textbf {\bibinfo {volume} {101}},\ \bibinfo
  {pages} {125202} (\bibinfo {year} {2020})}\BibitemShut {NoStop}%
\bibitem [{\citenamefont {Sun}\ \emph {et~al.}(2020)\citenamefont {Sun},
  \citenamefont {Hua}, \citenamefont {Liu}, \citenamefont {Liu}, \citenamefont
  {Ye}, \citenamefont {Qiao}, \citenamefont {Liu}, \citenamefont {Liu},
  \citenamefont {Guo}, \citenamefont {Lu} \emph {et~al.}}]{sun2020direct}%
  \BibitemOpen
  \bibfield  {author} {\bibinfo {author} {\bibfnamefont {Z.}~\bibnamefont
  {Sun}}, \bibinfo {author} {\bibfnamefont {C.}~\bibnamefont {Hua}}, \bibinfo
  {author} {\bibfnamefont {X.}~\bibnamefont {Liu}}, \bibinfo {author}
  {\bibfnamefont {Z.}~\bibnamefont {Liu}}, \bibinfo {author} {\bibfnamefont
  {M.}~\bibnamefont {Ye}}, \bibinfo {author} {\bibfnamefont {S.}~\bibnamefont
  {Qiao}}, \bibinfo {author} {\bibfnamefont {Z.}~\bibnamefont {Liu}}, \bibinfo
  {author} {\bibfnamefont {J.}~\bibnamefont {Liu}}, \bibinfo {author}
  {\bibfnamefont {Y.}~\bibnamefont {Guo}}, \bibinfo {author} {\bibfnamefont
  {Y.}~\bibnamefont {Lu}}, \emph {et~al.},\ }\href
  {https://doi.org/10.1103/PhysRevB.101.155114} {\bibfield  {journal} {\bibinfo
   {journal} {Physical Review B}\ }\textbf {\bibinfo {volume} {101}},\ \bibinfo
  {pages} {155114} (\bibinfo {year} {2020})}\BibitemShut {NoStop}%
\bibitem [{\citenamefont {Yang}\ \emph {et~al.}(2020)\citenamefont {Yang},
  \citenamefont {Cochran}, \citenamefont {Chapai}, \citenamefont {Tristant},
  \citenamefont {Yin}, \citenamefont {Belopolski}, \citenamefont {Cheng},
  \citenamefont {Multer}, \citenamefont {Zhang}, \citenamefont {Shumiya} \emph
  {et~al.}}]{yang2020observation}%
  \BibitemOpen
  \bibfield  {author} {\bibinfo {author} {\bibfnamefont {X.}~\bibnamefont
  {Yang}}, \bibinfo {author} {\bibfnamefont {T.~A.}\ \bibnamefont {Cochran}},
  \bibinfo {author} {\bibfnamefont {R.}~\bibnamefont {Chapai}}, \bibinfo
  {author} {\bibfnamefont {D.}~\bibnamefont {Tristant}}, \bibinfo {author}
  {\bibfnamefont {J.-X.}\ \bibnamefont {Yin}}, \bibinfo {author} {\bibfnamefont
  {I.}~\bibnamefont {Belopolski}}, \bibinfo {author} {\bibfnamefont
  {j.}~\bibnamefont {Cheng}, \bibfnamefont {Zui}}, \bibinfo {author}
  {\bibfnamefont {D.}~\bibnamefont {Multer}}, \bibinfo {author} {\bibfnamefont
  {S.~S.}\ \bibnamefont {Zhang}}, \bibinfo {author} {\bibfnamefont
  {N.}~\bibnamefont {Shumiya}}, \emph {et~al.},\ }\href
  {https://doi.org/10.1103/PhysRevB.101.201105} {\bibfield  {journal} {\bibinfo
   {journal} {Physical Review B}\ }\textbf {\bibinfo {volume} {101}},\ \bibinfo
  {pages} {201105} (\bibinfo {year} {2020})}\BibitemShut {NoStop}%
\bibitem [{\citenamefont {Thirupathaiah}\ \emph {et~al.}(2021)\citenamefont
  {Thirupathaiah}, \citenamefont {Kushnirenk}, \citenamefont {Koepernik},
  \citenamefont {Piening}, \citenamefont {B(\"u)chner}, \citenamefont
  {Aswartham}, \citenamefont {van~den Brink}, \citenamefont {Borisenko},\ and\
  \citenamefont {Fulga}}]{thiru}%
  \BibitemOpen
  \bibfield  {author} {\bibinfo {author} {\bibfnamefont {S.}~\bibnamefont
  {Thirupathaiah}}, \bibinfo {author} {\bibfnamefont {Y.}~\bibnamefont
  {Kushnirenk}}, \bibinfo {author} {\bibfnamefont {K.}~\bibnamefont
  {Koepernik}}, \bibinfo {author} {\bibfnamefont {B.~R.}\ \bibnamefont
  {Piening}}, \bibinfo {author} {\bibfnamefont {B.}~\bibnamefont
  {B(\"u)chner}}, \bibinfo {author} {\bibfnamefont {S.}~\bibnamefont
  {Aswartham}}, \bibinfo {author} {\bibfnamefont {J.}~\bibnamefont {van~den
  Brink}}, \bibinfo {author} {\bibfnamefont {S.}~\bibnamefont {Borisenko}},\
  and\ \bibinfo {author} {\bibfnamefont {I.~C.}\ \bibnamefont {Fulga}},\ }\href
  {https://doi.org/10.21468/SciPostPhys} {\bibfield  {journal} {\bibinfo
  {journal} {SciPost Physics}\ }\textbf {\bibinfo {volume} {10}},\ \bibinfo
  {pages} {004} (\bibinfo {year} {2021})}\BibitemShut {NoStop}%
\bibitem [{\citenamefont {Jin}\ \emph {et~al.}(2021)\citenamefont {Jin},
  \citenamefont {Liu}, \citenamefont {Zhang}, \citenamefont {Dai},\ and\
  \citenamefont {Liu}}]{jin2021}%
  \BibitemOpen
  \bibfield  {author} {\bibinfo {author} {\bibfnamefont {L.}~\bibnamefont
  {Jin}}, \bibinfo {author} {\bibfnamefont {Y.}~\bibnamefont {Liu}}, \bibinfo
  {author} {\bibfnamefont {X.}~\bibnamefont {Zhang}}, \bibinfo {author}
  {\bibfnamefont {X.}~\bibnamefont {Dai}},\ and\ \bibinfo {author}
  {\bibfnamefont {G.}~\bibnamefont {Liu}},\ }\href
  {https://doi.org/10.1103/PhysRevB.104.045111} {\bibfield  {journal} {\bibinfo
   {journal} {Physical Review B}\ }\textbf {\bibinfo {volume} {104}},\ \bibinfo
  {pages} {045111} (\bibinfo {year} {2021})}\BibitemShut {NoStop}%
\bibitem [{\citenamefont {Rong}\ \emph {et~al.}(2023)\citenamefont {Rong},
  \citenamefont {Huang}, \citenamefont {Zhang}, \citenamefont {Kumar},
  \citenamefont {Zhang}, \citenamefont {Zhang}, \citenamefont {Wang},
  \citenamefont {Hao}, \citenamefont {Cai}, \citenamefont {Wang} \emph
  {et~al.}}]{rong2023}%
  \BibitemOpen
  \bibfield  {author} {\bibinfo {author} {\bibfnamefont {H.}~\bibnamefont
  {Rong}}, \bibinfo {author} {\bibfnamefont {Z.}~\bibnamefont {Huang}},
  \bibinfo {author} {\bibfnamefont {X.}~\bibnamefont {Zhang}}, \bibinfo
  {author} {\bibfnamefont {S.}~\bibnamefont {Kumar}}, \bibinfo {author}
  {\bibfnamefont {F.}~\bibnamefont {Zhang}}, \bibinfo {author} {\bibfnamefont
  {C.}~\bibnamefont {Zhang}}, \bibinfo {author} {\bibfnamefont
  {Y.}~\bibnamefont {Wang}}, \bibinfo {author} {\bibfnamefont {Z.}~\bibnamefont
  {Hao}}, \bibinfo {author} {\bibfnamefont {Y.}~\bibnamefont {Cai}}, \bibinfo
  {author} {\bibfnamefont {L.}~\bibnamefont {Wang}}, \emph {et~al.},\ }\href
  {https://doi.org/10.1038/s41535-023-00565-8} {\bibfield  {journal} {\bibinfo
  {journal} {npj Quantum Materials}\ }\textbf {\bibinfo {volume} {8}},\
  \bibinfo {pages} {29} (\bibinfo {year} {2023})}\BibitemShut {NoStop}%
\bibitem [{\citenamefont {Guo}\ \emph {et~al.}(2021)\citenamefont {Guo},
  \citenamefont {Wei}, \citenamefont {Liu}, \citenamefont {Liu},\ and\
  \citenamefont {Lu}}]{guo2021eightfold}%
  \BibitemOpen
  \bibfield  {author} {\bibinfo {author} {\bibfnamefont {P.-J.}\ \bibnamefont
  {Guo}}, \bibinfo {author} {\bibfnamefont {Y.-W.}\ \bibnamefont {Wei}},
  \bibinfo {author} {\bibfnamefont {K.}~\bibnamefont {Liu}}, \bibinfo {author}
  {\bibfnamefont {Z.-X.}\ \bibnamefont {Liu}},\ and\ \bibinfo {author}
  {\bibfnamefont {Z.-Y.}\ \bibnamefont {Lu}},\ }\href
  {https://doi.org/10.1103/PhysRevLett.127.176401} {\bibfield  {journal}
  {\bibinfo  {journal} {Physical Review Letters}\ }\textbf {\bibinfo {volume}
  {127}},\ \bibinfo {pages} {176401} (\bibinfo {year} {2021})}\BibitemShut
  {NoStop}%
\bibitem [{\citenamefont {Lv}\ \emph {et~al.}(2017{\natexlab{b}})\citenamefont
  {Lv}, \citenamefont {Feng}, \citenamefont {Xu}, \citenamefont {Gao},
  \citenamefont {Ma}, \citenamefont {Kong}, \citenamefont {Richard},
  \citenamefont {Huang}, \citenamefont {Strocov}, \citenamefont {Fang} \emph
  {et~al.}}]{lv2017observation}%
  \BibitemOpen
  \bibfield  {author} {\bibinfo {author} {\bibfnamefont {B.}~\bibnamefont
  {Lv}}, \bibinfo {author} {\bibfnamefont {Z.-L.}\ \bibnamefont {Feng}},
  \bibinfo {author} {\bibfnamefont {Q.-N.}\ \bibnamefont {Xu}}, \bibinfo
  {author} {\bibfnamefont {X.}~\bibnamefont {Gao}}, \bibinfo {author}
  {\bibfnamefont {J.-Z.}\ \bibnamefont {Ma}}, \bibinfo {author} {\bibfnamefont
  {L.-Y.}\ \bibnamefont {Kong}}, \bibinfo {author} {\bibfnamefont
  {P.}~\bibnamefont {Richard}}, \bibinfo {author} {\bibfnamefont {Y.-B.}\
  \bibnamefont {Huang}}, \bibinfo {author} {\bibfnamefont {V.}~\bibnamefont
  {Strocov}}, \bibinfo {author} {\bibfnamefont {C.}~\bibnamefont {Fang}}, \emph
  {et~al.},\ }\href {https://doi.org/10.1038/nature22390} {\bibfield  {journal}
  {\bibinfo  {journal} {Nature}\ }\textbf {\bibinfo {volume} {546}},\ \bibinfo
  {pages} {627} (\bibinfo {year} {2017}{\natexlab{b}})}\BibitemShut {NoStop}%
\bibitem [{\citenamefont {He}\ \emph {et~al.}(2018)\citenamefont {He},
  \citenamefont {Kong}, \citenamefont {Wang},\ and\ \citenamefont
  {Kou}}]{he2018type}%
  \BibitemOpen
  \bibfield  {author} {\bibinfo {author} {\bibfnamefont {J.}~\bibnamefont
  {He}}, \bibinfo {author} {\bibfnamefont {X.}~\bibnamefont {Kong}}, \bibinfo
  {author} {\bibfnamefont {W.}~\bibnamefont {Wang}},\ and\ \bibinfo {author}
  {\bibfnamefont {S.-P.}\ \bibnamefont {Kou}},\ }\href
  {https://doi.org/10.1088/1367-2630/aabdf8} {\bibfield  {journal} {\bibinfo
  {journal} {New Journal of Physics}\ }\textbf {\bibinfo {volume} {20}},\
  \bibinfo {pages} {053019} (\bibinfo {year} {2018})}\BibitemShut {NoStop}%
\bibitem [{\citenamefont {Bian}\ \emph
  {et~al.}(2016{\natexlab{a}})\citenamefont {Bian}, \citenamefont {Chang},
  \citenamefont {Zheng}, \citenamefont {Velury}, \citenamefont {Xu},
  \citenamefont {Neupert}, \citenamefont {Chiu}, \citenamefont {Huang},
  \citenamefont {Sanchez}, \citenamefont {Belopolski} \emph
  {et~al.}}]{bian2016drumhead}%
  \BibitemOpen
  \bibfield  {author} {\bibinfo {author} {\bibfnamefont {G.}~\bibnamefont
  {Bian}}, \bibinfo {author} {\bibfnamefont {T.-R.}\ \bibnamefont {Chang}},
  \bibinfo {author} {\bibfnamefont {H.}~\bibnamefont {Zheng}}, \bibinfo
  {author} {\bibfnamefont {S.}~\bibnamefont {Velury}}, \bibinfo {author}
  {\bibfnamefont {S.-Y.}\ \bibnamefont {Xu}}, \bibinfo {author} {\bibfnamefont
  {T.}~\bibnamefont {Neupert}}, \bibinfo {author} {\bibfnamefont {C.-K.}\
  \bibnamefont {Chiu}}, \bibinfo {author} {\bibfnamefont {S.-M.}\ \bibnamefont
  {Huang}}, \bibinfo {author} {\bibfnamefont {D.~S.}\ \bibnamefont {Sanchez}},
  \bibinfo {author} {\bibfnamefont {I.}~\bibnamefont {Belopolski}}, \emph
  {et~al.},\ }\href {https://doi.org/10.1103/PhysRevB.93.121113} {\bibfield
  {journal} {\bibinfo  {journal} {Physical Review B}\ }\textbf {\bibinfo
  {volume} {93}},\ \bibinfo {pages} {121113} (\bibinfo {year}
  {2016}{\natexlab{a}})}\BibitemShut {NoStop}%
\bibitem [{\citenamefont {Yu}\ \emph {et~al.}(2015)\citenamefont {Yu},
  \citenamefont {Weng}, \citenamefont {Fang}, \citenamefont {Dai},\ and\
  \citenamefont {Hu}}]{yu2015topological}%
  \BibitemOpen
  \bibfield  {author} {\bibinfo {author} {\bibfnamefont {R.}~\bibnamefont
  {Yu}}, \bibinfo {author} {\bibfnamefont {H.}~\bibnamefont {Weng}}, \bibinfo
  {author} {\bibfnamefont {Z.}~\bibnamefont {Fang}}, \bibinfo {author}
  {\bibfnamefont {X.}~\bibnamefont {Dai}},\ and\ \bibinfo {author}
  {\bibfnamefont {X.}~\bibnamefont {Hu}},\ }\href
  {https://doi.org/10.1103/PhysRevLett.115.036807} {\bibfield  {journal}
  {\bibinfo  {journal} {Physical review letters}\ }\textbf {\bibinfo {volume}
  {115}},\ \bibinfo {pages} {036807} (\bibinfo {year} {2015})}\BibitemShut
  {NoStop}%
\bibitem [{\citenamefont {Fang}\ \emph {et~al.}(2015)\citenamefont {Fang},
  \citenamefont {Chen}, \citenamefont {Kee},\ and\ \citenamefont
  {Fu}}]{fang2015topological}%
  \BibitemOpen
  \bibfield  {author} {\bibinfo {author} {\bibfnamefont {C.}~\bibnamefont
  {Fang}}, \bibinfo {author} {\bibfnamefont {Y.}~\bibnamefont {Chen}}, \bibinfo
  {author} {\bibfnamefont {H.-Y.}\ \bibnamefont {Kee}},\ and\ \bibinfo {author}
  {\bibfnamefont {L.}~\bibnamefont {Fu}},\ }\href
  {https://doi.org/doi.org/10.1103/PhysRevB.92.081201} {\bibfield  {journal}
  {\bibinfo  {journal} {Physical Review B}\ }\textbf {\bibinfo {volume} {92}},\
  \bibinfo {pages} {081201} (\bibinfo {year} {2015})}\BibitemShut {NoStop}%
\bibitem [{\citenamefont {Chang}\ \emph {et~al.}(2019)\citenamefont {Chang},
  \citenamefont {Pletikosic}, \citenamefont {Kong}, \citenamefont {Bian},
  \citenamefont {Huang}, \citenamefont {Denlinger}, \citenamefont {Kushwaha},
  \citenamefont {Sinkovic}, \citenamefont {Jeng}, \citenamefont {Valla},
  \citenamefont {Xie},\ and\ \citenamefont {Cava}}]{201800897}%
  \BibitemOpen
  \bibfield  {author} {\bibinfo {author} {\bibfnamefont {T.-R.}\ \bibnamefont
  {Chang}}, \bibinfo {author} {\bibfnamefont {I.}~\bibnamefont {Pletikosic}},
  \bibinfo {author} {\bibfnamefont {T.}~\bibnamefont {Kong}}, \bibinfo {author}
  {\bibfnamefont {G.}~\bibnamefont {Bian}}, \bibinfo {author} {\bibfnamefont
  {A.}~\bibnamefont {Huang}}, \bibinfo {author} {\bibfnamefont
  {J.}~\bibnamefont {Denlinger}}, \bibinfo {author} {\bibfnamefont {S.~K.}\
  \bibnamefont {Kushwaha}}, \bibinfo {author} {\bibfnamefont {B.}~\bibnamefont
  {Sinkovic}}, \bibinfo {author} {\bibfnamefont {H.-T.}\ \bibnamefont {Jeng}},
  \bibinfo {author} {\bibfnamefont {T.}~\bibnamefont {Valla}}, \bibinfo
  {author} {\bibfnamefont {W.}~\bibnamefont {Xie}},\ and\ \bibinfo {author}
  {\bibfnamefont {R.~J.}\ \bibnamefont {Cava}},\ }\href
  {https://doi.org/https://doi.org/10.1002/advs.201800897} {\bibfield
  {journal} {\bibinfo  {journal} {Advanced Science}\ }\textbf {\bibinfo
  {volume} {6}},\ \bibinfo {pages} {1800897} (\bibinfo {year}
  {2019})}\BibitemShut {NoStop}%
\bibitem [{\citenamefont {Wang}\ \emph {et~al.}(2021)\citenamefont {Wang},
  \citenamefont {Qian}, \citenamefont {Yang}, \citenamefont {Chen},
  \citenamefont {Li}, \citenamefont {Tan}, \citenamefont {Cai}, \citenamefont
  {Zhao}, \citenamefont {Gao}, \citenamefont {Feng} \emph
  {et~al.}}]{wang2021spectroscopic}%
  \BibitemOpen
  \bibfield  {author} {\bibinfo {author} {\bibfnamefont {Y.}~\bibnamefont
  {Wang}}, \bibinfo {author} {\bibfnamefont {Y.}~\bibnamefont {Qian}}, \bibinfo
  {author} {\bibfnamefont {M.}~\bibnamefont {Yang}}, \bibinfo {author}
  {\bibfnamefont {H.}~\bibnamefont {Chen}}, \bibinfo {author} {\bibfnamefont
  {C.}~\bibnamefont {Li}}, \bibinfo {author} {\bibfnamefont {Z.}~\bibnamefont
  {Tan}}, \bibinfo {author} {\bibfnamefont {Y.}~\bibnamefont {Cai}}, \bibinfo
  {author} {\bibfnamefont {W.}~\bibnamefont {Zhao}}, \bibinfo {author}
  {\bibfnamefont {S.}~\bibnamefont {Gao}}, \bibinfo {author} {\bibfnamefont
  {Y.}~\bibnamefont {Feng}}, \emph {et~al.},\ }\href
  {https://doi.org/doi.org/10.1103/PhysRevB.103.125131} {\bibfield  {journal}
  {\bibinfo  {journal} {Physical Review B}\ }\textbf {\bibinfo {volume}
  {103}},\ \bibinfo {pages} {125131} (\bibinfo {year} {2021})}\BibitemShut
  {NoStop}%
\bibitem [{\citenamefont {Wang}\ \emph {et~al.}(2017)\citenamefont {Wang},
  \citenamefont {Ma}, \citenamefont {Emmanouilidou}, \citenamefont {Shen},
  \citenamefont {Hsu}, \citenamefont {Zhou}, \citenamefont {Zuo}, \citenamefont
  {Song}, \citenamefont {Xu}, \citenamefont {Wang} \emph
  {et~al.}}]{wang2017topological}%
  \BibitemOpen
  \bibfield  {author} {\bibinfo {author} {\bibfnamefont {X.-B.}\ \bibnamefont
  {Wang}}, \bibinfo {author} {\bibfnamefont {X.-M.}\ \bibnamefont {Ma}},
  \bibinfo {author} {\bibfnamefont {E.}~\bibnamefont {Emmanouilidou}}, \bibinfo
  {author} {\bibfnamefont {B.}~\bibnamefont {Shen}}, \bibinfo {author}
  {\bibfnamefont {C.-H.}\ \bibnamefont {Hsu}}, \bibinfo {author} {\bibfnamefont
  {C.-S.}\ \bibnamefont {Zhou}}, \bibinfo {author} {\bibfnamefont
  {Y.}~\bibnamefont {Zuo}}, \bibinfo {author} {\bibfnamefont {R.-R.}\
  \bibnamefont {Song}}, \bibinfo {author} {\bibfnamefont {S.-Y.}\ \bibnamefont
  {Xu}}, \bibinfo {author} {\bibfnamefont {G.}~\bibnamefont {Wang}}, \emph
  {et~al.},\ }\href {https://doi.org/doi.org/10.1103/PhysRevB.96.161112}
  {\bibfield  {journal} {\bibinfo  {journal} {Physical Review B}\ }\textbf
  {\bibinfo {volume} {96}},\ \bibinfo {pages} {161112} (\bibinfo {year}
  {2017})}\BibitemShut {NoStop}%
\bibitem [{\citenamefont {Takane}\ \emph {et~al.}(2018)\citenamefont {Takane},
  \citenamefont {Nakayama}, \citenamefont {Souma}, \citenamefont {Wada},
  \citenamefont {Okamoto}, \citenamefont {Takenaka}, \citenamefont {Yamakawa},
  \citenamefont {Yamakage}, \citenamefont {Mitsuhashi}, \citenamefont {Horiba}
  \emph {et~al.}}]{takane2018observation}%
  \BibitemOpen
  \bibfield  {author} {\bibinfo {author} {\bibfnamefont {D.}~\bibnamefont
  {Takane}}, \bibinfo {author} {\bibfnamefont {K.}~\bibnamefont {Nakayama}},
  \bibinfo {author} {\bibfnamefont {S.}~\bibnamefont {Souma}}, \bibinfo
  {author} {\bibfnamefont {T.}~\bibnamefont {Wada}}, \bibinfo {author}
  {\bibfnamefont {Y.}~\bibnamefont {Okamoto}}, \bibinfo {author} {\bibfnamefont
  {K.}~\bibnamefont {Takenaka}}, \bibinfo {author} {\bibfnamefont
  {Y.}~\bibnamefont {Yamakawa}}, \bibinfo {author} {\bibfnamefont
  {A.}~\bibnamefont {Yamakage}}, \bibinfo {author} {\bibfnamefont
  {T.}~\bibnamefont {Mitsuhashi}}, \bibinfo {author} {\bibfnamefont
  {K.}~\bibnamefont {Horiba}}, \emph {et~al.},\ }\href
  {https://doi.org/doi.org/10.1038/s41535-017-0074-z} {\bibfield  {journal}
  {\bibinfo  {journal} {npj Quantum Materials}\ }\textbf {\bibinfo {volume}
  {3}},\ \bibinfo {pages} {1} (\bibinfo {year} {2018})}\BibitemShut {NoStop}%
\bibitem [{\citenamefont {Xu}\ \emph {et~al.}(2018)\citenamefont {Xu},
  \citenamefont {Qian}, \citenamefont {Wu}, \citenamefont {Aut{\`e}s},
  \citenamefont {Matt}, \citenamefont {Lv}, \citenamefont {Yao}, \citenamefont
  {Strocov}, \citenamefont {Pomjakushina}, \citenamefont {Conder} \emph
  {et~al.}}]{xu2018trivial}%
  \BibitemOpen
  \bibfield  {author} {\bibinfo {author} {\bibfnamefont {N.}~\bibnamefont
  {Xu}}, \bibinfo {author} {\bibfnamefont {Y.}~\bibnamefont {Qian}}, \bibinfo
  {author} {\bibfnamefont {Q.}~\bibnamefont {Wu}}, \bibinfo {author}
  {\bibfnamefont {G.}~\bibnamefont {Aut{\`e}s}}, \bibinfo {author}
  {\bibfnamefont {C.~E.}\ \bibnamefont {Matt}}, \bibinfo {author}
  {\bibfnamefont {B.}~\bibnamefont {Lv}}, \bibinfo {author} {\bibfnamefont
  {M.}~\bibnamefont {Yao}}, \bibinfo {author} {\bibfnamefont {V.~N.}\
  \bibnamefont {Strocov}}, \bibinfo {author} {\bibfnamefont {E.}~\bibnamefont
  {Pomjakushina}}, \bibinfo {author} {\bibfnamefont {K.}~\bibnamefont
  {Conder}}, \emph {et~al.},\ }\href
  {https://doi.org/doi.org/10.1103/PhysRevB.97.161111} {\bibfield  {journal}
  {\bibinfo  {journal} {Physical Review B}\ }\textbf {\bibinfo {volume} {97}},\
  \bibinfo {pages} {161111} (\bibinfo {year} {2018})}\BibitemShut {NoStop}%
\bibitem [{\citenamefont {Lou}\ \emph {et~al.}(2018)\citenamefont {Lou},
  \citenamefont {Guo}, \citenamefont {Li}, \citenamefont {Wang}, \citenamefont
  {Liu}, \citenamefont {Sun}, \citenamefont {Li}, \citenamefont {Wu},
  \citenamefont {Wang}, \citenamefont {Sun} \emph
  {et~al.}}]{lou2018experimental}%
  \BibitemOpen
  \bibfield  {author} {\bibinfo {author} {\bibfnamefont {R.}~\bibnamefont
  {Lou}}, \bibinfo {author} {\bibfnamefont {P.}~\bibnamefont {Guo}}, \bibinfo
  {author} {\bibfnamefont {M.}~\bibnamefont {Li}}, \bibinfo {author}
  {\bibfnamefont {Q.}~\bibnamefont {Wang}}, \bibinfo {author} {\bibfnamefont
  {Z.}~\bibnamefont {Liu}}, \bibinfo {author} {\bibfnamefont {S.}~\bibnamefont
  {Sun}}, \bibinfo {author} {\bibfnamefont {C.}~\bibnamefont {Li}}, \bibinfo
  {author} {\bibfnamefont {X.}~\bibnamefont {Wu}}, \bibinfo {author}
  {\bibfnamefont {Z.}~\bibnamefont {Wang}}, \bibinfo {author} {\bibfnamefont
  {Z.}~\bibnamefont {Sun}}, \emph {et~al.},\ }\href
  {https://doi.org/doi.org/10.1038/s41535-018-0121-4} {\bibfield  {journal}
  {\bibinfo  {journal} {npj Quantum Materials}\ }\textbf {\bibinfo {volume}
  {3}},\ \bibinfo {pages} {43} (\bibinfo {year} {2018})}\BibitemShut {NoStop}%
\bibitem [{\citenamefont {Feng}\ \emph {et~al.}(2017)\citenamefont {Feng},
  \citenamefont {Fu}, \citenamefont {Kasamatsu}, \citenamefont {Ito},
  \citenamefont {Cheng}, \citenamefont {Liu}, \citenamefont {Feng},
  \citenamefont {Wu}, \citenamefont {Mahatha}, \citenamefont {Sheverdyaeva}
  \emph {et~al.}}]{feng2017experimental}%
  \BibitemOpen
  \bibfield  {author} {\bibinfo {author} {\bibfnamefont {B.}~\bibnamefont
  {Feng}}, \bibinfo {author} {\bibfnamefont {B.}~\bibnamefont {Fu}}, \bibinfo
  {author} {\bibfnamefont {S.}~\bibnamefont {Kasamatsu}}, \bibinfo {author}
  {\bibfnamefont {S.}~\bibnamefont {Ito}}, \bibinfo {author} {\bibfnamefont
  {P.}~\bibnamefont {Cheng}}, \bibinfo {author} {\bibfnamefont {C.-C.}\
  \bibnamefont {Liu}}, \bibinfo {author} {\bibfnamefont {Y.}~\bibnamefont
  {Feng}}, \bibinfo {author} {\bibfnamefont {S.}~\bibnamefont {Wu}}, \bibinfo
  {author} {\bibfnamefont {S.~K.}\ \bibnamefont {Mahatha}}, \bibinfo {author}
  {\bibfnamefont {P.}~\bibnamefont {Sheverdyaeva}}, \emph {et~al.},\ }\href
  {https://doi.org/10.1038/s41467-017-01108-z} {\bibfield  {journal} {\bibinfo
  {journal} {Nature communications}\ }\textbf {\bibinfo {volume} {8}},\
  \bibinfo {pages} {1007} (\bibinfo {year} {2017})}\BibitemShut {NoStop}%
\bibitem [{\citenamefont {Sato}\ \emph {et~al.}(2018)\citenamefont {Sato},
  \citenamefont {Wang}, \citenamefont {Nakayama}, \citenamefont {Souma},
  \citenamefont {Takane}, \citenamefont {Nakata}, \citenamefont {Iwasawa},
  \citenamefont {Cacho}, \citenamefont {Kim}, \citenamefont {Takahashi} \emph
  {et~al.}}]{sato2018observation}%
  \BibitemOpen
  \bibfield  {author} {\bibinfo {author} {\bibfnamefont {T.}~\bibnamefont
  {Sato}}, \bibinfo {author} {\bibfnamefont {Z.}~\bibnamefont {Wang}}, \bibinfo
  {author} {\bibfnamefont {K.}~\bibnamefont {Nakayama}}, \bibinfo {author}
  {\bibfnamefont {S.}~\bibnamefont {Souma}}, \bibinfo {author} {\bibfnamefont
  {D.}~\bibnamefont {Takane}}, \bibinfo {author} {\bibfnamefont
  {Y.}~\bibnamefont {Nakata}}, \bibinfo {author} {\bibfnamefont
  {H.}~\bibnamefont {Iwasawa}}, \bibinfo {author} {\bibfnamefont
  {C.}~\bibnamefont {Cacho}}, \bibinfo {author} {\bibfnamefont
  {T.}~\bibnamefont {Kim}}, \bibinfo {author} {\bibfnamefont {T.}~\bibnamefont
  {Takahashi}}, \emph {et~al.},\ }\href
  {https://doi.org/10.1103/PhysRevB.98.121111} {\bibfield  {journal} {\bibinfo
  {journal} {Physical Review B}\ }\textbf {\bibinfo {volume} {98}},\ \bibinfo
  {pages} {121111} (\bibinfo {year} {2018})}\BibitemShut {NoStop}%
\bibitem [{\citenamefont {Yi}\ \emph {et~al.}(2018)\citenamefont {Yi},
  \citenamefont {Lv}, \citenamefont {Wu}, \citenamefont {Fu}, \citenamefont
  {Gao}, \citenamefont {Yang}, \citenamefont {Peng}, \citenamefont {Li},
  \citenamefont {Huang}, \citenamefont {Richard} \emph
  {et~al.}}]{yi2018observation}%
  \BibitemOpen
  \bibfield  {author} {\bibinfo {author} {\bibfnamefont {C.-J.}\ \bibnamefont
  {Yi}}, \bibinfo {author} {\bibfnamefont {B.}~\bibnamefont {Lv}}, \bibinfo
  {author} {\bibfnamefont {Q.}~\bibnamefont {Wu}}, \bibinfo {author}
  {\bibfnamefont {B.-B.}\ \bibnamefont {Fu}}, \bibinfo {author} {\bibfnamefont
  {X.}~\bibnamefont {Gao}}, \bibinfo {author} {\bibfnamefont {M.}~\bibnamefont
  {Yang}}, \bibinfo {author} {\bibfnamefont {X.-L.}\ \bibnamefont {Peng}},
  \bibinfo {author} {\bibfnamefont {M.}~\bibnamefont {Li}}, \bibinfo {author}
  {\bibfnamefont {Y.-B.}\ \bibnamefont {Huang}}, \bibinfo {author}
  {\bibfnamefont {P.}~\bibnamefont {Richard}}, \emph {et~al.},\ }\href
  {https://doi.org/10.1103/PhysRevB.97.201107} {\bibfield  {journal} {\bibinfo
  {journal} {Physical Review B}\ }\textbf {\bibinfo {volume} {97}},\ \bibinfo
  {pages} {201107} (\bibinfo {year} {2018})}\BibitemShut {NoStop}%
\bibitem [{\citenamefont {Hosen}\ \emph {et~al.}(2017)\citenamefont {Hosen},
  \citenamefont {Dimitri}, \citenamefont {Belopolski}, \citenamefont
  {Maldonado}, \citenamefont {Sankar}, \citenamefont {Dhakal}, \citenamefont
  {Dhakal}, \citenamefont {Cole}, \citenamefont {Oppeneer}, \citenamefont
  {Kaczorowski} \emph {et~al.}}]{hosen2017tunability}%
  \BibitemOpen
  \bibfield  {author} {\bibinfo {author} {\bibfnamefont {M.~M.}\ \bibnamefont
  {Hosen}}, \bibinfo {author} {\bibfnamefont {K.}~\bibnamefont {Dimitri}},
  \bibinfo {author} {\bibfnamefont {I.}~\bibnamefont {Belopolski}}, \bibinfo
  {author} {\bibfnamefont {P.}~\bibnamefont {Maldonado}}, \bibinfo {author}
  {\bibfnamefont {R.}~\bibnamefont {Sankar}}, \bibinfo {author} {\bibfnamefont
  {N.}~\bibnamefont {Dhakal}}, \bibinfo {author} {\bibfnamefont
  {G.}~\bibnamefont {Dhakal}}, \bibinfo {author} {\bibfnamefont
  {T.}~\bibnamefont {Cole}}, \bibinfo {author} {\bibfnamefont {P.~M.}\
  \bibnamefont {Oppeneer}}, \bibinfo {author} {\bibfnamefont {D.}~\bibnamefont
  {Kaczorowski}}, \emph {et~al.},\ }\href
  {https://doi.org/10.1103/PhysRevB.95.161101} {\bibfield  {journal} {\bibinfo
  {journal} {Physical Review B}\ }\textbf {\bibinfo {volume} {95}},\ \bibinfo
  {pages} {161101} (\bibinfo {year} {2017})}\BibitemShut {NoStop}%
\bibitem [{\citenamefont {Bian}\ \emph
  {et~al.}(2016{\natexlab{b}})\citenamefont {Bian}, \citenamefont {Chang},
  \citenamefont {Sankar}, \citenamefont {Xu}, \citenamefont {Zheng},
  \citenamefont {Neupert}, \citenamefont {Chiu}, \citenamefont {Huang},
  \citenamefont {Chang}, \citenamefont {Belopolski} \emph
  {et~al.}}]{bian2016topological}%
  \BibitemOpen
  \bibfield  {author} {\bibinfo {author} {\bibfnamefont {G.}~\bibnamefont
  {Bian}}, \bibinfo {author} {\bibfnamefont {T.-R.}\ \bibnamefont {Chang}},
  \bibinfo {author} {\bibfnamefont {R.}~\bibnamefont {Sankar}}, \bibinfo
  {author} {\bibfnamefont {S.-Y.}\ \bibnamefont {Xu}}, \bibinfo {author}
  {\bibfnamefont {H.}~\bibnamefont {Zheng}}, \bibinfo {author} {\bibfnamefont
  {T.}~\bibnamefont {Neupert}}, \bibinfo {author} {\bibfnamefont {C.-K.}\
  \bibnamefont {Chiu}}, \bibinfo {author} {\bibfnamefont {S.-M.}\ \bibnamefont
  {Huang}}, \bibinfo {author} {\bibfnamefont {G.}~\bibnamefont {Chang}},
  \bibinfo {author} {\bibfnamefont {I.}~\bibnamefont {Belopolski}}, \emph
  {et~al.},\ }\href {https://doi.org/10.1038/ncomms10556} {\bibfield  {journal}
  {\bibinfo  {journal} {Nature communications}\ }\textbf {\bibinfo {volume}
  {7}},\ \bibinfo {pages} {1} (\bibinfo {year}
  {2016}{\natexlab{b}})}\BibitemShut {NoStop}%
\bibitem [{\citenamefont {Chen}\ \emph
  {et~al.}(2017{\natexlab{a}})\citenamefont {Chen}, \citenamefont {Xu},
  \citenamefont {Jiang}, \citenamefont {Wu}, \citenamefont {Qi}, \citenamefont
  {Yang}, \citenamefont {Wang}, \citenamefont {Sun}, \citenamefont {Schroter},
  \citenamefont {Yang} \emph {et~al.}}]{chen2017dirac}%
  \BibitemOpen
  \bibfield  {author} {\bibinfo {author} {\bibfnamefont {C.}~\bibnamefont
  {Chen}}, \bibinfo {author} {\bibfnamefont {X.}~\bibnamefont {Xu}}, \bibinfo
  {author} {\bibfnamefont {J.}~\bibnamefont {Jiang}}, \bibinfo {author}
  {\bibfnamefont {S.-C.}\ \bibnamefont {Wu}}, \bibinfo {author} {\bibfnamefont
  {Y.}~\bibnamefont {Qi}}, \bibinfo {author} {\bibfnamefont {L.}~\bibnamefont
  {Yang}}, \bibinfo {author} {\bibfnamefont {M.}~\bibnamefont {Wang}}, \bibinfo
  {author} {\bibfnamefont {Y.}~\bibnamefont {Sun}}, \bibinfo {author}
  {\bibfnamefont {N.}~\bibnamefont {Schroter}}, \bibinfo {author}
  {\bibfnamefont {H.}~\bibnamefont {Yang}}, \emph {et~al.},\ }\href
  {https://doi.org/10.1103/PhysRevB.95.125126} {\bibfield  {journal} {\bibinfo
  {journal} {Physical Review B}\ }\textbf {\bibinfo {volume} {95}},\ \bibinfo
  {pages} {125126} (\bibinfo {year} {2017}{\natexlab{a}})}\BibitemShut
  {NoStop}%
\bibitem [{\citenamefont {Hu}\ \emph {et~al.}(2016)\citenamefont {Hu},
  \citenamefont {Tang}, \citenamefont {Liu}, \citenamefont {Liu}, \citenamefont
  {Zhu}, \citenamefont {Graf}, \citenamefont {Myhro}, \citenamefont {Tran},
  \citenamefont {Lau}, \citenamefont {Wei} \emph {et~al.}}]{hu2016evidence}%
  \BibitemOpen
  \bibfield  {author} {\bibinfo {author} {\bibfnamefont {J.}~\bibnamefont
  {Hu}}, \bibinfo {author} {\bibfnamefont {Z.}~\bibnamefont {Tang}}, \bibinfo
  {author} {\bibfnamefont {J.}~\bibnamefont {Liu}}, \bibinfo {author}
  {\bibfnamefont {X.}~\bibnamefont {Liu}}, \bibinfo {author} {\bibfnamefont
  {Y.}~\bibnamefont {Zhu}}, \bibinfo {author} {\bibfnamefont {D.}~\bibnamefont
  {Graf}}, \bibinfo {author} {\bibfnamefont {K.}~\bibnamefont {Myhro}},
  \bibinfo {author} {\bibfnamefont {S.}~\bibnamefont {Tran}}, \bibinfo {author}
  {\bibfnamefont {C.~N.}\ \bibnamefont {Lau}}, \bibinfo {author} {\bibfnamefont
  {J.}~\bibnamefont {Wei}}, \emph {et~al.},\ }\href
  {https://doi.org/10.1103/PhysRevLett.117.016602} {\bibfield  {journal}
  {\bibinfo  {journal} {Physical review letters}\ }\textbf {\bibinfo {volume}
  {117}},\ \bibinfo {pages} {016602} (\bibinfo {year} {2016})}\BibitemShut
  {NoStop}%
\bibitem [{\citenamefont {Liu}\ \emph {et~al.}(2018)\citenamefont {Liu},
  \citenamefont {Lou}, \citenamefont {Guo}, \citenamefont {Wang}, \citenamefont
  {Sun}, \citenamefont {Li}, \citenamefont {Thirupathaiah}, \citenamefont
  {Fedorov}, \citenamefont {Shen}, \citenamefont {Liu} \emph
  {et~al.}}]{liu2018experimental}%
  \BibitemOpen
  \bibfield  {author} {\bibinfo {author} {\bibfnamefont {Z.}~\bibnamefont
  {Liu}}, \bibinfo {author} {\bibfnamefont {R.}~\bibnamefont {Lou}}, \bibinfo
  {author} {\bibfnamefont {P.}~\bibnamefont {Guo}}, \bibinfo {author}
  {\bibfnamefont {Q.}~\bibnamefont {Wang}}, \bibinfo {author} {\bibfnamefont
  {S.}~\bibnamefont {Sun}}, \bibinfo {author} {\bibfnamefont {C.}~\bibnamefont
  {Li}}, \bibinfo {author} {\bibfnamefont {S.}~\bibnamefont {Thirupathaiah}},
  \bibinfo {author} {\bibfnamefont {A.}~\bibnamefont {Fedorov}}, \bibinfo
  {author} {\bibfnamefont {D.}~\bibnamefont {Shen}}, \bibinfo {author}
  {\bibfnamefont {K.}~\bibnamefont {Liu}}, \emph {et~al.},\ }\href
  {https://doi.org/10.1103/PhysRevX.8.031044} {\bibfield  {journal} {\bibinfo
  {journal} {Physical Review X}\ }\textbf {\bibinfo {volume} {8}},\ \bibinfo
  {pages} {031044} (\bibinfo {year} {2018})}\BibitemShut {NoStop}%
\bibitem [{\citenamefont {Hirayama}\ \emph {et~al.}(2017)\citenamefont
  {Hirayama}, \citenamefont {Okugawa}, \citenamefont {Miyake},\ and\
  \citenamefont {Murakami}}]{hirayama2017topological}%
  \BibitemOpen
  \bibfield  {author} {\bibinfo {author} {\bibfnamefont {M.}~\bibnamefont
  {Hirayama}}, \bibinfo {author} {\bibfnamefont {R.}~\bibnamefont {Okugawa}},
  \bibinfo {author} {\bibfnamefont {T.}~\bibnamefont {Miyake}},\ and\ \bibinfo
  {author} {\bibfnamefont {S.}~\bibnamefont {Murakami}},\ }\href
  {https://doi.org/doi.org/10.1038/ncomms14022} {\bibfield  {journal} {\bibinfo
   {journal} {Nature communications}\ }\textbf {\bibinfo {volume} {8}},\
  \bibinfo {pages} {14022} (\bibinfo {year} {2017})}\BibitemShut {NoStop}%
\bibitem [{\citenamefont {Cheng}\ \emph {et~al.}(2019)\citenamefont {Cheng},
  \citenamefont {Zhang}, \citenamefont {Sun}, \citenamefont {Li}, \citenamefont
  {Yuan}, \citenamefont {Wang}, \citenamefont {Cao}, \citenamefont {Shao},
  \citenamefont {Bian}, \citenamefont {Zhang} \emph
  {et~al.}}]{cheng2019visualizing}%
  \BibitemOpen
  \bibfield  {author} {\bibinfo {author} {\bibfnamefont {Z.}~\bibnamefont
  {Cheng}}, \bibinfo {author} {\bibfnamefont {Z.}~\bibnamefont {Zhang}},
  \bibinfo {author} {\bibfnamefont {H.}~\bibnamefont {Sun}}, \bibinfo {author}
  {\bibfnamefont {S.}~\bibnamefont {Li}}, \bibinfo {author} {\bibfnamefont
  {H.}~\bibnamefont {Yuan}}, \bibinfo {author} {\bibfnamefont {Z.}~\bibnamefont
  {Wang}}, \bibinfo {author} {\bibfnamefont {Y.}~\bibnamefont {Cao}}, \bibinfo
  {author} {\bibfnamefont {Z.}~\bibnamefont {Shao}}, \bibinfo {author}
  {\bibfnamefont {Q.}~\bibnamefont {Bian}}, \bibinfo {author} {\bibfnamefont
  {X.}~\bibnamefont {Zhang}}, \emph {et~al.},\ }\bibfield  {journal} {\bibinfo
  {journal} {APL Materials}\ }\textbf {\bibinfo {volume} {7}},\ \href
  {https://doi.org/10.1063/1.5084090} {10.1063/1.5084090} (\bibinfo {year}
  {2019})\BibitemShut {NoStop}%
\bibitem [{\citenamefont {Jin}\ \emph {et~al.}(2020{\natexlab{b}})\citenamefont
  {Jin}, \citenamefont {Zhang}, \citenamefont {He}, \citenamefont {Meng},
  \citenamefont {Dai},\ and\ \citenamefont {Liu}}]{jin2020ferromagnetic}%
  \BibitemOpen
  \bibfield  {author} {\bibinfo {author} {\bibfnamefont {L.}~\bibnamefont
  {Jin}}, \bibinfo {author} {\bibfnamefont {X.}~\bibnamefont {Zhang}}, \bibinfo
  {author} {\bibfnamefont {T.}~\bibnamefont {He}}, \bibinfo {author}
  {\bibfnamefont {W.}~\bibnamefont {Meng}}, \bibinfo {author} {\bibfnamefont
  {X.}~\bibnamefont {Dai}},\ and\ \bibinfo {author} {\bibfnamefont
  {G.}~\bibnamefont {Liu}},\ }\href
  {https://doi.org/10.1016/j.apsusc.2020.146376} {\bibfield  {journal}
  {\bibinfo  {journal} {Applied Surface Science}\ }\textbf {\bibinfo {volume}
  {520}},\ \bibinfo {pages} {146376} (\bibinfo {year}
  {2020}{\natexlab{b}})}\BibitemShut {NoStop}%
\bibitem [{\citenamefont {Cui}\ \emph {et~al.}(2020)\citenamefont {Cui},
  \citenamefont {Song}, \citenamefont {Cai}, \citenamefont {Cui}, \citenamefont
  {Liu},\ and\ \citenamefont {Zhao}}]{cui2020three}%
  \BibitemOpen
  \bibfield  {author} {\bibinfo {author} {\bibfnamefont {L.}~\bibnamefont
  {Cui}}, \bibinfo {author} {\bibfnamefont {T.}~\bibnamefont {Song}}, \bibinfo
  {author} {\bibfnamefont {J.}~\bibnamefont {Cai}}, \bibinfo {author}
  {\bibfnamefont {X.}~\bibnamefont {Cui}}, \bibinfo {author} {\bibfnamefont
  {Z.}~\bibnamefont {Liu}},\ and\ \bibinfo {author} {\bibfnamefont
  {J.}~\bibnamefont {Zhao}},\ }\href
  {https://doi.org/10.1103/PhysRevB.102.155133} {\bibfield  {journal} {\bibinfo
   {journal} {Physical Review B}\ }\textbf {\bibinfo {volume} {102}},\ \bibinfo
  {pages} {155133} (\bibinfo {year} {2020})}\BibitemShut {NoStop}%
\bibitem [{\citenamefont {Jin}\ \emph {et~al.}(2020{\natexlab{c}})\citenamefont
  {Jin}, \citenamefont {Zhang}, \citenamefont {Liu}, \citenamefont {Dai},
  \citenamefont {Shen}, \citenamefont {Wang},\ and\ \citenamefont
  {Liu}}]{jin2020two}%
  \BibitemOpen
  \bibfield  {author} {\bibinfo {author} {\bibfnamefont {L.}~\bibnamefont
  {Jin}}, \bibinfo {author} {\bibfnamefont {X.}~\bibnamefont {Zhang}}, \bibinfo
  {author} {\bibfnamefont {Y.}~\bibnamefont {Liu}}, \bibinfo {author}
  {\bibfnamefont {X.}~\bibnamefont {Dai}}, \bibinfo {author} {\bibfnamefont
  {X.}~\bibnamefont {Shen}}, \bibinfo {author} {\bibfnamefont {L.}~\bibnamefont
  {Wang}},\ and\ \bibinfo {author} {\bibfnamefont {G.}~\bibnamefont {Liu}},\
  }\href {https://doi.org/10.1103/PhysRevB.102.125118} {\bibfield  {journal}
  {\bibinfo  {journal} {Physical Review B}\ }\textbf {\bibinfo {volume}
  {102}},\ \bibinfo {pages} {125118} (\bibinfo {year}
  {2020}{\natexlab{c}})}\BibitemShut {NoStop}%
\bibitem [{\citenamefont {Huang}\ \emph {et~al.}(2016)\citenamefont {Huang},
  \citenamefont {Liu}, \citenamefont {Vanderbilt},\ and\ \citenamefont
  {Duan}}]{huang2016topological}%
  \BibitemOpen
  \bibfield  {author} {\bibinfo {author} {\bibfnamefont {H.}~\bibnamefont
  {Huang}}, \bibinfo {author} {\bibfnamefont {J.}~\bibnamefont {Liu}}, \bibinfo
  {author} {\bibfnamefont {D.}~\bibnamefont {Vanderbilt}},\ and\ \bibinfo
  {author} {\bibfnamefont {W.}~\bibnamefont {Duan}},\ }\href
  {https://doi.org/10.1103/PhysRevB.93.201114} {\bibfield  {journal} {\bibinfo
  {journal} {Physical Review B}\ }\textbf {\bibinfo {volume} {93}},\ \bibinfo
  {pages} {201114} (\bibinfo {year} {2016})}\BibitemShut {NoStop}%
\bibitem [{\citenamefont {Suzumura}\ and\ \citenamefont
  {Yamakage}(2018)}]{suzumura2018berry}%
  \BibitemOpen
  \bibfield  {author} {\bibinfo {author} {\bibfnamefont {Y.}~\bibnamefont
  {Suzumura}}\ and\ \bibinfo {author} {\bibfnamefont {A.}~\bibnamefont
  {Yamakage}},\ }\href {https://doi.org/10.7566/JPSJ.87.093704} {\bibfield
  {journal} {\bibinfo  {journal} {Journal of the Physical Society of Japan}\
  }\textbf {\bibinfo {volume} {87}},\ \bibinfo {pages} {093704} (\bibinfo
  {year} {2018})}\BibitemShut {NoStop}%
\bibitem [{\citenamefont {Yu}\ \emph {et~al.}(2017)\citenamefont {Yu},
  \citenamefont {Fang}, \citenamefont {Dai},\ and\ \citenamefont
  {Weng}}]{yu2017topological}%
  \BibitemOpen
  \bibfield  {author} {\bibinfo {author} {\bibfnamefont {R.}~\bibnamefont
  {Yu}}, \bibinfo {author} {\bibfnamefont {Z.}~\bibnamefont {Fang}}, \bibinfo
  {author} {\bibfnamefont {X.}~\bibnamefont {Dai}},\ and\ \bibinfo {author}
  {\bibfnamefont {H.}~\bibnamefont {Weng}},\ }\href
  {https://doi.org/10.1007/s11467-016-0630-1} {\bibfield  {journal} {\bibinfo
  {journal} {Frontiers of Physics}\ }\textbf {\bibinfo {volume} {12}},\
  \bibinfo {pages} {1} (\bibinfo {year} {2017})}\BibitemShut {NoStop}%
\bibitem [{\citenamefont {Gao}\ \emph {et~al.}(2019)\citenamefont {Gao},
  \citenamefont {Venderbos}, \citenamefont {Kim},\ and\ \citenamefont
  {Rappe}}]{gao2019topological}%
  \BibitemOpen
  \bibfield  {author} {\bibinfo {author} {\bibfnamefont {H.}~\bibnamefont
  {Gao}}, \bibinfo {author} {\bibfnamefont {J.~r.~W.}\ \bibnamefont
  {Venderbos}}, \bibinfo {author} {\bibfnamefont {Y.}~\bibnamefont {Kim}},\
  and\ \bibinfo {author} {\bibfnamefont {A.~M.}\ \bibnamefont {Rappe}},\ }\href
  {https://doi.org/10.1146/annurev-matsci-070218-010049} {\bibfield  {journal}
  {\bibinfo  {journal} {Annual Review of Materials Research}\ }\textbf
  {\bibinfo {volume} {49}},\ \bibinfo {pages} {153} (\bibinfo {year}
  {2019})}\BibitemShut {NoStop}%
\bibitem [{\citenamefont {Li}\ \emph {et~al.}(2017)\citenamefont {Li},
  \citenamefont {Yu}, \citenamefont {Liu}, \citenamefont {Guan}, \citenamefont
  {Wang}, \citenamefont {Zhang}, \citenamefont {Yao},\ and\ \citenamefont
  {Yang}}]{li2017type}%
  \BibitemOpen
  \bibfield  {author} {\bibinfo {author} {\bibfnamefont {S.}~\bibnamefont
  {Li}}, \bibinfo {author} {\bibfnamefont {Z.-M.}\ \bibnamefont {Yu}}, \bibinfo
  {author} {\bibfnamefont {Y.}~\bibnamefont {Liu}}, \bibinfo {author}
  {\bibfnamefont {S.}~\bibnamefont {Guan}}, \bibinfo {author} {\bibfnamefont
  {S.-S.}\ \bibnamefont {Wang}}, \bibinfo {author} {\bibfnamefont
  {X.}~\bibnamefont {Zhang}}, \bibinfo {author} {\bibfnamefont
  {Y.}~\bibnamefont {Yao}},\ and\ \bibinfo {author} {\bibfnamefont {S.~A.}\
  \bibnamefont {Yang}},\ }\href {https://doi.org/10.1103/PhysRevB.96.081106}
  {\bibfield  {journal} {\bibinfo  {journal} {Physical Review B}\ }\textbf
  {\bibinfo {volume} {96}},\ \bibinfo {pages} {081106} (\bibinfo {year}
  {2017})}\BibitemShut {NoStop}%
\bibitem [{\citenamefont {Xu}\ \emph {et~al.}(2020{\natexlab{a}})\citenamefont
  {Xu}, \citenamefont {Zhang}, \citenamefont {Meng}, \citenamefont {He},
  \citenamefont {Liu}, \citenamefont {Dai}, \citenamefont {Zhang},\ and\
  \citenamefont {Liu}}]{xu2020centrosymmetric}%
  \BibitemOpen
  \bibfield  {author} {\bibinfo {author} {\bibfnamefont {L.}~\bibnamefont
  {Xu}}, \bibinfo {author} {\bibfnamefont {X.}~\bibnamefont {Zhang}}, \bibinfo
  {author} {\bibfnamefont {W.}~\bibnamefont {Meng}}, \bibinfo {author}
  {\bibfnamefont {T.}~\bibnamefont {He}}, \bibinfo {author} {\bibfnamefont
  {Y.}~\bibnamefont {Liu}}, \bibinfo {author} {\bibfnamefont {X.}~\bibnamefont
  {Dai}}, \bibinfo {author} {\bibfnamefont {Y.}~\bibnamefont {Zhang}},\ and\
  \bibinfo {author} {\bibfnamefont {G.}~\bibnamefont {Liu}},\ }\href
  {https://doi.org/10.1039/D0TC03600E} {\bibfield  {journal} {\bibinfo
  {journal} {Journal of Materials Chemistry C}\ }\textbf {\bibinfo {volume}
  {8}},\ \bibinfo {pages} {14109} (\bibinfo {year}
  {2020}{\natexlab{a}})}\BibitemShut {NoStop}%
\bibitem [{\citenamefont {Wang}\ \emph {et~al.}(2018)\citenamefont {Wang},
  \citenamefont {Gao}, \citenamefont {Lu}, \citenamefont {Xie}, \citenamefont
  {Ge}, \citenamefont {Zhao}, \citenamefont {Zhang},\ and\ \citenamefont
  {Liu}}]{wang2018type}%
  \BibitemOpen
  \bibfield  {author} {\bibinfo {author} {\bibfnamefont {B.}~\bibnamefont
  {Wang}}, \bibinfo {author} {\bibfnamefont {H.}~\bibnamefont {Gao}}, \bibinfo
  {author} {\bibfnamefont {Q.}~\bibnamefont {Lu}}, \bibinfo {author}
  {\bibfnamefont {W.}~\bibnamefont {Xie}}, \bibinfo {author} {\bibfnamefont
  {Y.}~\bibnamefont {Ge}}, \bibinfo {author} {\bibfnamefont {Y.-H.}\
  \bibnamefont {Zhao}}, \bibinfo {author} {\bibfnamefont {K.}~\bibnamefont
  {Zhang}},\ and\ \bibinfo {author} {\bibfnamefont {Y.}~\bibnamefont {Liu}},\
  }\href {https://doi.org/10.1103/PhysRevB.98.115164} {\bibfield  {journal}
  {\bibinfo  {journal} {Physical Review B}\ }\textbf {\bibinfo {volume} {98}},\
  \bibinfo {pages} {115164} (\bibinfo {year} {2018})}\BibitemShut {NoStop}%
\bibitem [{\citenamefont {Bzdu{\v{s}}ek}\ \emph {et~al.}(2016)\citenamefont
  {Bzdu{\v{s}}ek}, \citenamefont {Wu}, \citenamefont {R{\"u}egg}, \citenamefont
  {Sigrist},\ and\ \citenamefont {Soluyanov}}]{bzduvsek2016nodal}%
  \BibitemOpen
  \bibfield  {author} {\bibinfo {author} {\bibfnamefont {T.}~\bibnamefont
  {Bzdu{\v{s}}ek}}, \bibinfo {author} {\bibfnamefont {Q.}~\bibnamefont {Wu}},
  \bibinfo {author} {\bibfnamefont {A.}~\bibnamefont {R{\"u}egg}}, \bibinfo
  {author} {\bibfnamefont {M.}~\bibnamefont {Sigrist}},\ and\ \bibinfo {author}
  {\bibfnamefont {A.~A.}\ \bibnamefont {Soluyanov}},\ }\href@noop {} {\bibfield
   {journal} {\bibinfo  {journal} {Nature}\ }\textbf {\bibinfo {volume}
  {538}},\ \bibinfo {pages} {75} (\bibinfo {year} {2016})}\BibitemShut
  {NoStop}%
\bibitem [{\citenamefont {Song}\ \emph {et~al.}(2022)\citenamefont {Song},
  \citenamefont {Jin}, \citenamefont {Song}, \citenamefont {Rong},
  \citenamefont {Zhu}, \citenamefont {Liang}, \citenamefont {Cui},
  \citenamefont {Sun}, \citenamefont {Zhao}, \citenamefont {Shi}, \citenamefont
  {Zhang}, \citenamefont {Liu},\ and\ \citenamefont
  {Zhou}}]{song2022spectroscopic}%
  \BibitemOpen
  \bibfield  {author} {\bibinfo {author} {\bibfnamefont {C.}~\bibnamefont
  {Song}}, \bibinfo {author} {\bibfnamefont {L.}~\bibnamefont {Jin}}, \bibinfo
  {author} {\bibfnamefont {P.}~\bibnamefont {Song}}, \bibinfo {author}
  {\bibfnamefont {H.}~\bibnamefont {Rong}}, \bibinfo {author} {\bibfnamefont
  {W.}~\bibnamefont {Zhu}}, \bibinfo {author} {\bibfnamefont {B.}~\bibnamefont
  {Liang}}, \bibinfo {author} {\bibfnamefont {S.}~\bibnamefont {Cui}}, \bibinfo
  {author} {\bibfnamefont {Z.}~\bibnamefont {Sun}}, \bibinfo {author}
  {\bibfnamefont {L.}~\bibnamefont {Zhao}}, \bibinfo {author} {\bibfnamefont
  {Y.}~\bibnamefont {Shi}}, \bibinfo {author} {\bibfnamefont {X.}~\bibnamefont
  {Zhang}}, \bibinfo {author} {\bibfnamefont {G.}~\bibnamefont {Liu}},\ and\
  \bibinfo {author} {\bibfnamefont {X.~J.}\ \bibnamefont {Zhou}},\ }\href
  {https://doi.org/10.1103/PhysRevB.105.L161104} {\bibfield  {journal}
  {\bibinfo  {journal} {Phys. Rev. B}\ }\textbf {\bibinfo {volume} {105}},\
  \bibinfo {pages} {L161104} (\bibinfo {year} {2022})}\BibitemShut {NoStop}%
\bibitem [{\citenamefont {Bi}\ \emph {et~al.}(2017)\citenamefont {Bi},
  \citenamefont {Yan}, \citenamefont {Lu},\ and\ \citenamefont
  {Wang}}]{PhysRevB.96.201305}%
  \BibitemOpen
  \bibfield  {author} {\bibinfo {author} {\bibfnamefont {R.}~\bibnamefont
  {Bi}}, \bibinfo {author} {\bibfnamefont {Z.}~\bibnamefont {Yan}}, \bibinfo
  {author} {\bibfnamefont {L.}~\bibnamefont {Lu}},\ and\ \bibinfo {author}
  {\bibfnamefont {Z.}~\bibnamefont {Wang}},\ }\href
  {https://doi.org/10.1103/PhysRevB.96.201305} {\bibfield  {journal} {\bibinfo
  {journal} {Phys. Rev. B}\ }\textbf {\bibinfo {volume} {96}},\ \bibinfo
  {pages} {201305} (\bibinfo {year} {2017})}\BibitemShut {NoStop}%
\bibitem [{\citenamefont {Chen}\ \emph
  {et~al.}(2017{\natexlab{b}})\citenamefont {Chen}, \citenamefont {Lu},\ and\
  \citenamefont {Hou}}]{PhysRevB.96.041102}%
  \BibitemOpen
  \bibfield  {author} {\bibinfo {author} {\bibfnamefont {W.}~\bibnamefont
  {Chen}}, \bibinfo {author} {\bibfnamefont {H.-Z.}\ \bibnamefont {Lu}},\ and\
  \bibinfo {author} {\bibfnamefont {J.-M.}\ \bibnamefont {Hou}},\ }\href
  {https://doi.org/10.1103/PhysRevB.96.041102} {\bibfield  {journal} {\bibinfo
  {journal} {Phys. Rev. B}\ }\textbf {\bibinfo {volume} {96}},\ \bibinfo
  {pages} {041102} (\bibinfo {year} {2017}{\natexlab{b}})}\BibitemShut
  {NoStop}%
\bibitem [{\citenamefont {Chang}\ and\ \citenamefont
  {Yee}(2017)}]{PhysRevB.96.081114}%
  \BibitemOpen
  \bibfield  {author} {\bibinfo {author} {\bibfnamefont {P.-Y.}\ \bibnamefont
  {Chang}}\ and\ \bibinfo {author} {\bibfnamefont {C.-H.}\ \bibnamefont
  {Yee}},\ }\href {https://doi.org/10.1103/PhysRevB.96.081114} {\bibfield
  {journal} {\bibinfo  {journal} {Phys. Rev. B}\ }\textbf {\bibinfo {volume}
  {96}},\ \bibinfo {pages} {081114} (\bibinfo {year} {2017})}\BibitemShut
  {NoStop}%
\bibitem [{\citenamefont {Ji}\ \emph {et~al.}(2023)\citenamefont {Ji},
  \citenamefont {Li}, \citenamefont {Tang}, \citenamefont {Zhou}, \citenamefont
  {Wang}, \citenamefont {Li}, \citenamefont {Li},\ and\ \citenamefont
  {Yao}}]{ji2023observation}%
  \BibitemOpen
  \bibfield  {author} {\bibinfo {author} {\bibfnamefont {C.-Y.}\ \bibnamefont
  {Ji}}, \bibinfo {author} {\bibfnamefont {X.-P.}\ \bibnamefont {Li}}, \bibinfo
  {author} {\bibfnamefont {Z.}~\bibnamefont {Tang}}, \bibinfo {author}
  {\bibfnamefont {D.}~\bibnamefont {Zhou}}, \bibinfo {author} {\bibfnamefont
  {Y.}~\bibnamefont {Wang}}, \bibinfo {author} {\bibfnamefont {F.}~\bibnamefont
  {Li}}, \bibinfo {author} {\bibfnamefont {J.}~\bibnamefont {Li}},\ and\
  \bibinfo {author} {\bibfnamefont {Y.}~\bibnamefont {Yao}},\ }\bibfield
  {journal} {\bibinfo  {journal} {arXiv preprint arXiv:2307.03370}\ }\href
  {https://doi.org/10.48550/arXiv.2307.03370} {10.48550/arXiv.2307.03370}
  (\bibinfo {year} {2023})\BibitemShut {NoStop}%
\bibitem [{\citenamefont {Kim}\ \emph {et~al.}(2015)\citenamefont {Kim},
  \citenamefont {Wieder}, \citenamefont {Kane},\ and\ \citenamefont
  {Rappe}}]{kim2015dirac}%
  \BibitemOpen
  \bibfield  {author} {\bibinfo {author} {\bibfnamefont {Y.}~\bibnamefont
  {Kim}}, \bibinfo {author} {\bibfnamefont {B.~J.}\ \bibnamefont {Wieder}},
  \bibinfo {author} {\bibfnamefont {C.}~\bibnamefont {Kane}},\ and\ \bibinfo
  {author} {\bibfnamefont {A.~M.}\ \bibnamefont {Rappe}},\ }\href
  {https://doi.org/10.1103/PhysRevLett.115.036806} {\bibfield  {journal}
  {\bibinfo  {journal} {Physical review letters}\ }\textbf {\bibinfo {volume}
  {115}},\ \bibinfo {pages} {036806} (\bibinfo {year} {2015})}\BibitemShut
  {NoStop}%
\bibitem [{\citenamefont {Zhao}\ \emph {et~al.}(2016)\citenamefont {Zhao},
  \citenamefont {Yu}, \citenamefont {Weng},\ and\ \citenamefont
  {Fang}}]{zhao2016topological}%
  \BibitemOpen
  \bibfield  {author} {\bibinfo {author} {\bibfnamefont {J.}~\bibnamefont
  {Zhao}}, \bibinfo {author} {\bibfnamefont {R.}~\bibnamefont {Yu}}, \bibinfo
  {author} {\bibfnamefont {H.}~\bibnamefont {Weng}},\ and\ \bibinfo {author}
  {\bibfnamefont {Z.}~\bibnamefont {Fang}},\ }\href
  {https://doi.org/10.1103/PhysRevB.94.195104} {\bibfield  {journal} {\bibinfo
  {journal} {Physical Review B}\ }\textbf {\bibinfo {volume} {94}},\ \bibinfo
  {pages} {195104} (\bibinfo {year} {2016})}\BibitemShut {NoStop}%
\bibitem [{\citenamefont {Xie}\ \emph {et~al.}()\citenamefont {Xie},
  \citenamefont {Schoop}, \citenamefont {Seibel}, \citenamefont {Gibson},
  \citenamefont {Xie},\ and\ \citenamefont {Cava}}]{xie3apl}%
  \BibitemOpen
  \bibfield  {author} {\bibinfo {author} {\bibfnamefont {L.}~\bibnamefont
  {Xie}}, \bibinfo {author} {\bibfnamefont {L.}~\bibnamefont {Schoop}},
  \bibinfo {author} {\bibfnamefont {E.}~\bibnamefont {Seibel}}, \bibinfo
  {author} {\bibfnamefont {Q.}~\bibnamefont {Gibson}}, \bibinfo {author}
  {\bibfnamefont {W.}~\bibnamefont {Xie}},\ and\ \bibinfo {author}
  {\bibfnamefont {R.}~\bibnamefont {Cava}},\ }\bibfield  {journal} {\bibinfo
  {journal} {APL materials}\ }\href {https://doi.org/10.1063/1.4926545}
  {10.1063/1.4926545}\BibitemShut {NoStop}%
\bibitem [{\citenamefont {Okamoto}\ \emph {et~al.}(2016)\citenamefont
  {Okamoto}, \citenamefont {Inohara}, \citenamefont {Yamakage}, \citenamefont
  {Yamakawa},\ and\ \citenamefont {Takenaka}}]{okamoto2016low}%
  \BibitemOpen
  \bibfield  {author} {\bibinfo {author} {\bibfnamefont {Y.}~\bibnamefont
  {Okamoto}}, \bibinfo {author} {\bibfnamefont {T.}~\bibnamefont {Inohara}},
  \bibinfo {author} {\bibfnamefont {A.}~\bibnamefont {Yamakage}}, \bibinfo
  {author} {\bibfnamefont {Y.}~\bibnamefont {Yamakawa}},\ and\ \bibinfo
  {author} {\bibfnamefont {K.}~\bibnamefont {Takenaka}},\ }\href
  {https://doi.org/10.7566/JPSJ.85.123701} {\bibfield  {journal} {\bibinfo
  {journal} {Journal of the Physical Society of Japan}\ }\textbf {\bibinfo
  {volume} {85}},\ \bibinfo {pages} {123701} (\bibinfo {year}
  {2016})}\BibitemShut {NoStop}%
\bibitem [{\citenamefont {Zhang}\ \emph {et~al.}(2017)\citenamefont {Zhang},
  \citenamefont {Jin}, \citenamefont {Dai},\ and\ \citenamefont
  {Liu}}]{zhang2017topological}%
  \BibitemOpen
  \bibfield  {author} {\bibinfo {author} {\bibfnamefont {X.}~\bibnamefont
  {Zhang}}, \bibinfo {author} {\bibfnamefont {L.}~\bibnamefont {Jin}}, \bibinfo
  {author} {\bibfnamefont {X.}~\bibnamefont {Dai}},\ and\ \bibinfo {author}
  {\bibfnamefont {G.}~\bibnamefont {Liu}},\ }\href
  {https://doi.org/10.1021/acs.jpclett.7b02129} {\bibfield  {journal} {\bibinfo
   {journal} {The journal of physical chemistry letters}\ }\textbf {\bibinfo
  {volume} {8}},\ \bibinfo {pages} {4814} (\bibinfo {year} {2017})}\BibitemShut
  {NoStop}%
\bibitem [{\citenamefont {Rosmus}\ \emph {et~al.}(2022)\citenamefont {Rosmus},
  \citenamefont {Olszowska}, \citenamefont {Bukowski}, \citenamefont
  {Starowicz}, \citenamefont {Piekarz},\ and\ \citenamefont
  {Ptok}}]{rosmus2022electronic}%
  \BibitemOpen
  \bibfield  {author} {\bibinfo {author} {\bibfnamefont {M.}~\bibnamefont
  {Rosmus}}, \bibinfo {author} {\bibfnamefont {N.}~\bibnamefont {Olszowska}},
  \bibinfo {author} {\bibfnamefont {Z.}~\bibnamefont {Bukowski}}, \bibinfo
  {author} {\bibfnamefont {P.}~\bibnamefont {Starowicz}}, \bibinfo {author}
  {\bibfnamefont {P.}~\bibnamefont {Piekarz}},\ and\ \bibinfo {author}
  {\bibfnamefont {A.}~\bibnamefont {Ptok}},\ }\href
  {https://doi.org/10.3390/ma15207168} {\bibfield  {journal} {\bibinfo
  {journal} {Materials}\ }\textbf {\bibinfo {volume} {15}},\ \bibinfo {pages}
  {7168} (\bibinfo {year} {2022})}\BibitemShut {NoStop}%
\bibitem [{\citenamefont {Shi}\ \emph {et~al.}(2016)\citenamefont {Shi},
  \citenamefont {Richard}, \citenamefont {Wang}, \citenamefont {Liu},
  \citenamefont {Matt}, \citenamefont {Xu}, \citenamefont {Dhaka},
  \citenamefont {Ristic}, \citenamefont {Qian}, \citenamefont {Yang},
  \citenamefont {Petrovic}, \citenamefont {Shi},\ and\ \citenamefont
  {Ding}}]{observation}%
  \BibitemOpen
  \bibfield  {author} {\bibinfo {author} {\bibfnamefont {X.}~\bibnamefont
  {Shi}}, \bibinfo {author} {\bibfnamefont {P.}~\bibnamefont {Richard}},
  \bibinfo {author} {\bibfnamefont {K.}~\bibnamefont {Wang}}, \bibinfo {author}
  {\bibfnamefont {M.}~\bibnamefont {Liu}}, \bibinfo {author} {\bibfnamefont
  {C.~E.}\ \bibnamefont {Matt}}, \bibinfo {author} {\bibfnamefont
  {N.}~\bibnamefont {Xu}}, \bibinfo {author} {\bibfnamefont {R.~S.}\
  \bibnamefont {Dhaka}}, \bibinfo {author} {\bibfnamefont {Z.}~\bibnamefont
  {Ristic}}, \bibinfo {author} {\bibfnamefont {T.}~\bibnamefont {Qian}},
  \bibinfo {author} {\bibfnamefont {Y.-F.}\ \bibnamefont {Yang}}, \bibinfo
  {author} {\bibfnamefont {C.}~\bibnamefont {Petrovic}}, \bibinfo {author}
  {\bibfnamefont {M.}~\bibnamefont {Shi}},\ and\ \bibinfo {author}
  {\bibfnamefont {H.}~\bibnamefont {Ding}},\ }\href
  {https://doi.org/10.1103/PhysRevB.93.081105} {\bibfield  {journal} {\bibinfo
  {journal} {Phys. Rev. B}\ }\textbf {\bibinfo {volume} {93}},\ \bibinfo
  {pages} {081105} (\bibinfo {year} {2016})}\BibitemShut {NoStop}%
\bibitem [{\citenamefont {Zhan}\ \emph {et~al.}(2023)\citenamefont {Zhan},
  \citenamefont {Li}, \citenamefont {Shi}, \citenamefont {Chen},\ and\
  \citenamefont {Sun}}]{zhan2023coexistence}%
  \BibitemOpen
  \bibfield  {author} {\bibinfo {author} {\bibfnamefont {J.}~\bibnamefont
  {Zhan}}, \bibinfo {author} {\bibfnamefont {J.}~\bibnamefont {Li}}, \bibinfo
  {author} {\bibfnamefont {W.}~\bibnamefont {Shi}}, \bibinfo {author}
  {\bibfnamefont {X.-Q.}\ \bibnamefont {Chen}},\ and\ \bibinfo {author}
  {\bibfnamefont {Y.}~\bibnamefont {Sun}},\ }\href
  {https://doi.org/10.1103/PhysRevB.107.224402} {\bibfield  {journal} {\bibinfo
   {journal} {Physical Review B}\ }\textbf {\bibinfo {volume} {107}},\ \bibinfo
  {pages} {224402} (\bibinfo {year} {2023})}\BibitemShut {NoStop}%
\bibitem [{\citenamefont {Zhao}\ \emph {et~al.}(2021)\citenamefont {Zhao},
  \citenamefont {Guo}, \citenamefont {Ma},\ and\ \citenamefont
  {Lu}}]{zhao2021coexistence}%
  \BibitemOpen
  \bibfield  {author} {\bibinfo {author} {\bibfnamefont {X.}~\bibnamefont
  {Zhao}}, \bibinfo {author} {\bibfnamefont {P.-j.}\ \bibnamefont {Guo}},
  \bibinfo {author} {\bibfnamefont {F.}~\bibnamefont {Ma}},\ and\ \bibinfo
  {author} {\bibfnamefont {Z.-Y.}\ \bibnamefont {Lu}},\ }\href
  {https://doi.org/10.1103/PhysRevB.103.085138} {\bibfield  {journal} {\bibinfo
   {journal} {Physical Review B}\ }\textbf {\bibinfo {volume} {103}},\ \bibinfo
  {pages} {085138} (\bibinfo {year} {2021})}\BibitemShut {NoStop}%
\bibitem [{\citenamefont {Sun}\ \emph {et~al.}(2017)\citenamefont {Sun},
  \citenamefont {Zhang},\ and\ \citenamefont {Chang}}]{sun2017coexistence}%
  \BibitemOpen
  \bibfield  {author} {\bibinfo {author} {\bibfnamefont {J.-P.}\ \bibnamefont
  {Sun}}, \bibinfo {author} {\bibfnamefont {D.}~\bibnamefont {Zhang}},\ and\
  \bibinfo {author} {\bibfnamefont {K.}~\bibnamefont {Chang}},\ }\href
  {https://doi.org/10.1103/PhysRevB.96.045121} {\bibfield  {journal} {\bibinfo
  {journal} {Physical Review B}\ }\textbf {\bibinfo {volume} {96}},\ \bibinfo
  {pages} {045121} (\bibinfo {year} {2017})}\BibitemShut {NoStop}%
\bibitem [{\citenamefont {Li}\ \emph {et~al.}(2020)\citenamefont {Li},
  \citenamefont {Xia}, \citenamefont {Khenata},\ and\ \citenamefont
  {Kuang}}]{li2020insight}%
  \BibitemOpen
  \bibfield  {author} {\bibinfo {author} {\bibfnamefont {Y.}~\bibnamefont
  {Li}}, \bibinfo {author} {\bibfnamefont {J.}~\bibnamefont {Xia}}, \bibinfo
  {author} {\bibfnamefont {R.}~\bibnamefont {Khenata}},\ and\ \bibinfo {author}
  {\bibfnamefont {M.}~\bibnamefont {Kuang}},\ }\href
  {https://doi.org/10.3390/ma13173841} {\bibfield  {journal} {\bibinfo
  {journal} {Materials}\ }\textbf {\bibinfo {volume} {13}},\ \bibinfo {pages}
  {3841} (\bibinfo {year} {2020})}\BibitemShut {NoStop}%
\bibitem [{\citenamefont {Xu}\ \emph {et~al.}(2020{\natexlab{b}})\citenamefont
  {Xu}, \citenamefont {Xi},\ and\ \citenamefont {Gao}}]{xu2020hexagonal}%
  \BibitemOpen
  \bibfield  {author} {\bibinfo {author} {\bibfnamefont {H.}~\bibnamefont
  {Xu}}, \bibinfo {author} {\bibfnamefont {H.}~\bibnamefont {Xi}},\ and\
  \bibinfo {author} {\bibfnamefont {Y.-C.}\ \bibnamefont {Gao}},\ }\href
  {https://doi.org/10.3389/fchem.2020.608398} {\bibfield  {journal} {\bibinfo
  {journal} {Frontiers in Chemistry}\ }\textbf {\bibinfo {volume} {8}},\
  \bibinfo {pages} {608398} (\bibinfo {year} {2020}{\natexlab{b}})}\BibitemShut
  {NoStop}%
\bibitem [{\citenamefont {Blaha}\ \emph {et~al.}(2001)\citenamefont {Blaha},
  \citenamefont {Schwarz}, \citenamefont {Madsen}, \citenamefont {Kvasnicka},
  \citenamefont {Luitz} \emph {et~al.}}]{blaha2001wien2k}%
  \BibitemOpen
  \bibfield  {author} {\bibinfo {author} {\bibfnamefont {P.}~\bibnamefont
  {Blaha}}, \bibinfo {author} {\bibfnamefont {K.}~\bibnamefont {Schwarz}},
  \bibinfo {author} {\bibfnamefont {G.~K.}\ \bibnamefont {Madsen}}, \bibinfo
  {author} {\bibfnamefont {D.}~\bibnamefont {Kvasnicka}}, \bibinfo {author}
  {\bibfnamefont {J.}~\bibnamefont {Luitz}}, \emph {et~al.},\ }\href
  {http://www.wien2k.at/reg_user/textbooks/usersguide.pdf} {\bibfield
  {journal} {\bibinfo  {journal} {An augmented plane wave+ local orbitals
  program for calculating crystal properties}\ }\textbf {\bibinfo {volume}
  {60}} (\bibinfo {year} {2001})}\BibitemShut {NoStop}%
\bibitem [{\citenamefont {Mostofi}\ \emph {et~al.}(2008)\citenamefont
  {Mostofi}, \citenamefont {Yates}, \citenamefont {Lee}, \citenamefont {Souza},
  \citenamefont {Vanderbilt},\ and\ \citenamefont
  {Marzari}}]{mostofi2008wannier90}%
  \BibitemOpen
  \bibfield  {author} {\bibinfo {author} {\bibfnamefont {A.~A.}\ \bibnamefont
  {Mostofi}}, \bibinfo {author} {\bibfnamefont {J.~R.}\ \bibnamefont {Yates}},
  \bibinfo {author} {\bibfnamefont {Y.-S.}\ \bibnamefont {Lee}}, \bibinfo
  {author} {\bibfnamefont {I.}~\bibnamefont {Souza}}, \bibinfo {author}
  {\bibfnamefont {D.}~\bibnamefont {Vanderbilt}},\ and\ \bibinfo {author}
  {\bibfnamefont {N.}~\bibnamefont {Marzari}},\ }\href
  {https://doi.org/10.1016/j.cpc.2007.11.016} {\bibfield  {journal} {\bibinfo
  {journal} {Computer physics communications}\ }\textbf {\bibinfo {volume}
  {178}},\ \bibinfo {pages} {685} (\bibinfo {year} {2008})}\BibitemShut
  {NoStop}%
\bibitem [{\citenamefont {Wu}\ \emph {et~al.}(2018)\citenamefont {Wu},
  \citenamefont {Zhang}, \citenamefont {Song}, \citenamefont {Troyer},\ and\
  \citenamefont {Soluyanov}}]{wu2018wanniertools}%
  \BibitemOpen
  \bibfield  {author} {\bibinfo {author} {\bibfnamefont {Q.}~\bibnamefont
  {Wu}}, \bibinfo {author} {\bibfnamefont {S.}~\bibnamefont {Zhang}}, \bibinfo
  {author} {\bibfnamefont {H.-F.}\ \bibnamefont {Song}}, \bibinfo {author}
  {\bibfnamefont {M.}~\bibnamefont {Troyer}},\ and\ \bibinfo {author}
  {\bibfnamefont {A.~A.}\ \bibnamefont {Soluyanov}},\ }\href
  {https://doi.org/10.1016/j.cpc.2017.09.033} {\bibfield  {journal} {\bibinfo
  {journal} {Computer Physics Communications}\ }\textbf {\bibinfo {volume}
  {224}},\ \bibinfo {pages} {405} (\bibinfo {year} {2018})}\BibitemShut
  {NoStop}%
\bibitem [{\citenamefont {Togo}\ \emph {et~al.}(2023)\citenamefont {Togo},
  \citenamefont {Chaput}, \citenamefont {Tadano},\ and\ \citenamefont
  {Tanaka}}]{phonopy}%
  \BibitemOpen
  \bibfield  {author} {\bibinfo {author} {\bibfnamefont {A.}~\bibnamefont
  {Togo}}, \bibinfo {author} {\bibfnamefont {L.}~\bibnamefont {Chaput}},
  \bibinfo {author} {\bibfnamefont {T.}~\bibnamefont {Tadano}},\ and\ \bibinfo
  {author} {\bibfnamefont {I.}~\bibnamefont {Tanaka}},\ }\href
  {https://doi.org/10.1088/1361-648X/acd831} {\bibfield  {journal} {\bibinfo
  {journal} {J. Phys. Condens. Matter}\ }\textbf {\bibinfo {volume} {35}},\
  \bibinfo {pages} {353001} (\bibinfo {year} {2023})}\BibitemShut {NoStop}%
\bibitem [{\citenamefont {Togo}(2023)}]{phonopy1}%
  \BibitemOpen
  \bibfield  {author} {\bibinfo {author} {\bibfnamefont {A.}~\bibnamefont
  {Togo}},\ }\href {https://doi.org/10.7566/JPSJ.92.012001} {\bibfield
  {journal} {\bibinfo  {journal} {J. Phys. Soc. Jpn.}\ }\textbf {\bibinfo
  {volume} {92}},\ \bibinfo {pages} {012001} (\bibinfo {year}
  {2023})}\BibitemShut {NoStop}%
\bibitem [{sup()}]{supplementary2024}%
  \BibitemOpen
  \href@noop {} {\bibinfo  {journal} {Refer supplementary material for detailed
  analysis}\ }\BibitemShut {NoStop}%
\end{thebibliography}%


\begin{thebibliography}{6}%
\makeatletter
\providecommand \@ifxundefined [1]{%
 \@ifx{#1\undefined}
}%
\providecommand \@ifnum [1]{%
 \ifnum #1\expandafter \@firstoftwo
 \else \expandafter \@secondoftwo
 \fi
}%
\providecommand \@ifx [1]{%
 \ifx #1\expandafter \@firstoftwo
 \else \expandafter \@secondoftwo
 \fi
}%
\providecommand \natexlab [1]{#1}%
\providecommand \enquote  [1]{``#1''}%
\providecommand \bibnamefont  [1]{#1}%
\providecommand \bibfnamefont [1]{#1}%
\providecommand \citenamefont [1]{#1}%
\providecommand \href@noop [0]{\@secondoftwo}%
\providecommand \href [0]{\begingroup \@sanitize@url \@href}%
\providecommand \@href[1]{\@@startlink{#1}\@@href}%
\providecommand \@@href[1]{\endgroup#1\@@endlink}%
\providecommand \@sanitize@url [0]{\catcode `\\12\catcode `\$12\catcode
  `\&12\catcode `\#12\catcode `\^12\catcode `\_12\catcode `\%12\relax}%
\providecommand \@@startlink[1]{}%
\providecommand \@@endlink[0]{}%
\providecommand \url  [0]{\begingroup\@sanitize@url \@url }%
\providecommand \@url [1]{\endgroup\@href {#1}{\urlprefix }}%
\providecommand \urlprefix  [0]{URL }%
\providecommand \Eprint [0]{\href }%
\providecommand \doibase [0]{http://dx.doi.org/}%
\providecommand \selectlanguage [0]{\@gobble}%
\providecommand \bibinfo  [0]{\@secondoftwo}%
\providecommand \bibfield  [0]{\@secondoftwo}%
\providecommand \translation [1]{[#1]}%
\providecommand \BibitemOpen [0]{}%
\providecommand \bibitemStop [0]{}%
\providecommand \bibitemNoStop [0]{.\EOS\space}%
\providecommand \EOS [0]{\spacefactor3000\relax}%
\providecommand \BibitemShut  [1]{\csname bibitem#1\endcsname}%
\let\auto@bib@innerbib\@empty
\bibitem [{\citenamefont {Togo}\ \emph {et~al.}(2023)\citenamefont {Togo},
  \citenamefont {Chaput}, \citenamefont {Tadano},\ and\ \citenamefont
  {Tanaka}}]{phonopy}%
  \BibitemOpen
  \bibfield  {author} {\bibinfo {author} {\bibfnamefont {A.}~\bibnamefont
  {Togo}}, \bibinfo {author} {\bibfnamefont {L.}~\bibnamefont {Chaput}},
  \bibinfo {author} {\bibfnamefont {T.}~\bibnamefont {Tadano}}, \ and\ \bibinfo
  {author} {\bibfnamefont {I.}~\bibnamefont {Tanaka}},\ }\bibfield  {title}
  {\enquote {\bibinfo {title} {Implementation strategies in phonopy and
  phono3py},}\ }\href {\doibase 10.1088/1361-648X/acd831} {\bibfield  {journal}
  {\bibinfo  {journal} {J. Phys. Condens. Matter}\ }\textbf {\bibinfo {volume}
  {35}},\ \bibinfo {pages} {353001} (\bibinfo {year} {2023})}\BibitemShut
  {NoStop}%
\bibitem [{\citenamefont {Togo}(2023)}]{phonopy1}%
  \BibitemOpen
  \bibfield  {author} {\bibinfo {author} {\bibfnamefont {A.}~\bibnamefont
  {Togo}},\ }\bibfield  {title} {\enquote {\bibinfo {title} {First-principles
  phonon calculations with phonopy and phono3py},}\ }\href {\doibase
  10.7566/JPSJ.92.012001} {\bibfield  {journal} {\bibinfo  {journal} {J. Phys.
  Soc. Jpn.}\ }\textbf {\bibinfo {volume} {92}},\ \bibinfo {pages} {012001}
  (\bibinfo {year} {2023})}\BibitemShut {NoStop}%
\bibitem [{\citenamefont {M{\"u}ller}\ \emph {et~al.}(2020)\citenamefont
  {M{\"u}ller}, \citenamefont {Khouri}, \citenamefont {Van~Delft},
  \citenamefont {Pezzini}, \citenamefont {Hsu}, \citenamefont {Ayres},
  \citenamefont {Breitkreiz}, \citenamefont {Schoop}, \citenamefont
  {Carrington}, \citenamefont {Hussey} \emph
  {et~al.}}]{muller2020determination}%
  \BibitemOpen
  \bibfield  {author} {\bibinfo {author} {\bibfnamefont {C.}~\bibnamefont
  {M{\"u}ller}}, \bibinfo {author} {\bibfnamefont {T.}~\bibnamefont {Khouri}},
  \bibinfo {author} {\bibfnamefont {M.}~\bibnamefont {Van~Delft}}, \bibinfo
  {author} {\bibfnamefont {S.}~\bibnamefont {Pezzini}}, \bibinfo {author}
  {\bibfnamefont {Y.-T.}\ \bibnamefont {Hsu}}, \bibinfo {author} {\bibfnamefont
  {J.}~\bibnamefont {Ayres}}, \bibinfo {author} {\bibfnamefont
  {M.}~\bibnamefont {Breitkreiz}}, \bibinfo {author} {\bibfnamefont
  {L.}~\bibnamefont {Schoop}}, \bibinfo {author} {\bibfnamefont
  {A.}~\bibnamefont {Carrington}}, \bibinfo {author} {\bibfnamefont
  {N.}~\bibnamefont {Hussey}},  \emph {et~al.},\ }\bibfield  {title} {\enquote
  {\bibinfo {title} {Determination of the fermi surface and field-induced
  quasiparticle tunneling around the dirac nodal loop in zrsis},}\ }\href
  {\doibase 10.1103/PhysRevResearch.2.023217} {\bibfield  {journal} {\bibinfo
  {journal} {Physical Review Research}\ }\textbf {\bibinfo {volume} {2}},\
  \bibinfo {pages} {023217} (\bibinfo {year} {2020})}\BibitemShut {NoStop}%
\bibitem [{\citenamefont {Lou}\ \emph {et~al.}(2018)\citenamefont {Lou},
  \citenamefont {Guo}, \citenamefont {Li}, \citenamefont {Wang}, \citenamefont
  {Liu}, \citenamefont {Sun}, \citenamefont {Li}, \citenamefont {Wu},
  \citenamefont {Wang}, \citenamefont {Sun} \emph
  {et~al.}}]{lou2018experimental}%
  \BibitemOpen
  \bibfield  {author} {\bibinfo {author} {\bibfnamefont {R.}~\bibnamefont
  {Lou}}, \bibinfo {author} {\bibfnamefont {P.}~\bibnamefont {Guo}}, \bibinfo
  {author} {\bibfnamefont {M.}~\bibnamefont {Li}}, \bibinfo {author}
  {\bibfnamefont {Q.}~\bibnamefont {Wang}}, \bibinfo {author} {\bibfnamefont
  {Z.}~\bibnamefont {Liu}}, \bibinfo {author} {\bibfnamefont {S.}~\bibnamefont
  {Sun}}, \bibinfo {author} {\bibfnamefont {C.}~\bibnamefont {Li}}, \bibinfo
  {author} {\bibfnamefont {X.}~\bibnamefont {Wu}}, \bibinfo {author}
  {\bibfnamefont {Z.}~\bibnamefont {Wang}}, \bibinfo {author} {\bibfnamefont
  {Z.}~\bibnamefont {Sun}},  \emph {et~al.},\ }\bibfield  {title} {\enquote
  {\bibinfo {title} {Experimental observation of bulk nodal lines and
  electronic surface states in zrb2},}\ }\href {\doibase
  doi.org/10.1038/s41535-018-0121-4} {\bibfield  {journal} {\bibinfo  {journal}
  {npj Quantum Materials}\ }\textbf {\bibinfo {volume} {3}},\ \bibinfo {pages}
  {43} (\bibinfo {year} {2018})}\BibitemShut {NoStop}%
\bibitem [{\citenamefont {Wang}\ \emph {et~al.}(2024)\citenamefont {Wang},
  \citenamefont {Nepal},\ and\ \citenamefont {Canfield}}]{wang2024origin}%
  \BibitemOpen
  \bibfield  {author} {\bibinfo {author} {\bibfnamefont {L.-L.}\ \bibnamefont
  {Wang}}, \bibinfo {author} {\bibfnamefont {N.~K.}\ \bibnamefont {Nepal}}, \
  and\ \bibinfo {author} {\bibfnamefont {P.~C.}\ \bibnamefont {Canfield}},\
  }\bibfield  {title} {\enquote {\bibinfo {title} {Origin of charge density
  wave in topological semimetals sral4 and eual4},}\ }\href {\doibase
  10.1038/s42005-024-01600-1} {\bibfield  {journal} {\bibinfo  {journal}
  {Communications Physics}\ }\textbf {\bibinfo {volume} {7}},\ \bibinfo {pages}
  {111} (\bibinfo {year} {2024})}\BibitemShut {NoStop}%
\bibitem [{\citenamefont {Fu}\ \emph {et~al.}(2019)\citenamefont {Fu},
  \citenamefont {Yi}, \citenamefont {Zhang}, \citenamefont {Caputo},
  \citenamefont {Ma}, \citenamefont {Gao}, \citenamefont {Lv}, \citenamefont
  {Kong}, \citenamefont {Huang}, \citenamefont {Richard} \emph
  {et~al.}}]{fu2019dirac}%
  \BibitemOpen
  \bibfield  {author} {\bibinfo {author} {\bibfnamefont {B.-B.}\ \bibnamefont
  {Fu}}, \bibinfo {author} {\bibfnamefont {C.-J.}\ \bibnamefont {Yi}}, \bibinfo
  {author} {\bibfnamefont {T.-T.}\ \bibnamefont {Zhang}}, \bibinfo {author}
  {\bibfnamefont {M.}~\bibnamefont {Caputo}}, \bibinfo {author} {\bibfnamefont
  {J.-Z.}\ \bibnamefont {Ma}}, \bibinfo {author} {\bibfnamefont
  {X.}~\bibnamefont {Gao}}, \bibinfo {author} {\bibfnamefont {B.}~\bibnamefont
  {Lv}}, \bibinfo {author} {\bibfnamefont {L.-Y.}\ \bibnamefont {Kong}},
  \bibinfo {author} {\bibfnamefont {Y.-B.}\ \bibnamefont {Huang}}, \bibinfo
  {author} {\bibfnamefont {P.}~\bibnamefont {Richard}},  \emph {et~al.},\
  }\bibfield  {title} {\enquote {\bibinfo {title} {Dirac nodal surfaces and
  nodal lines in zrsis},}\ }\href {\doibase 10.1126/sciadv.aau6459} {\bibfield
  {journal} {\bibinfo  {journal} {Science advances}\ }\textbf {\bibinfo
  {volume} {5}},\ \bibinfo {pages} {eaau6459} (\bibinfo {year}
  {2019})}\BibitemShut {NoStop}%
\end{thebibliography}%


\end{document}



\title{Supplementary Information for "Coexistence of Topological Dirac and Dirac Nodal line semimetal in SrCaP belonging to Nodal line semimetal family SrCaX(X= Bi, Sb, As, P)"}

\author{Shivendra Kumar Gupta}
\email{shivendrakumarg900@gmail.com}
\affiliation{Department of Physics, Visvesvaraya National Institute of Technology, Nagpur, 440010, India \\}
\author{Ashish Kore}%
\affiliation{Physics Division, National Center for Theoretical Sciences, National Taiwan University, Taipei 106319, Taiwan \\}
\author{Saurabh Kumar Sen}%
 
\author{Poorva Singh}
\email{poorvasingh@phy.vnit.ac.in}
\affiliation{Department of Physics, Visvesvaraya National Institute of Technology, Nagpur, 440010, India \\}
\date{\today}

\maketitle

\onecolumngrid



\begin{figure}[hbtp]
 \includegraphics[width=1 \textwidth]{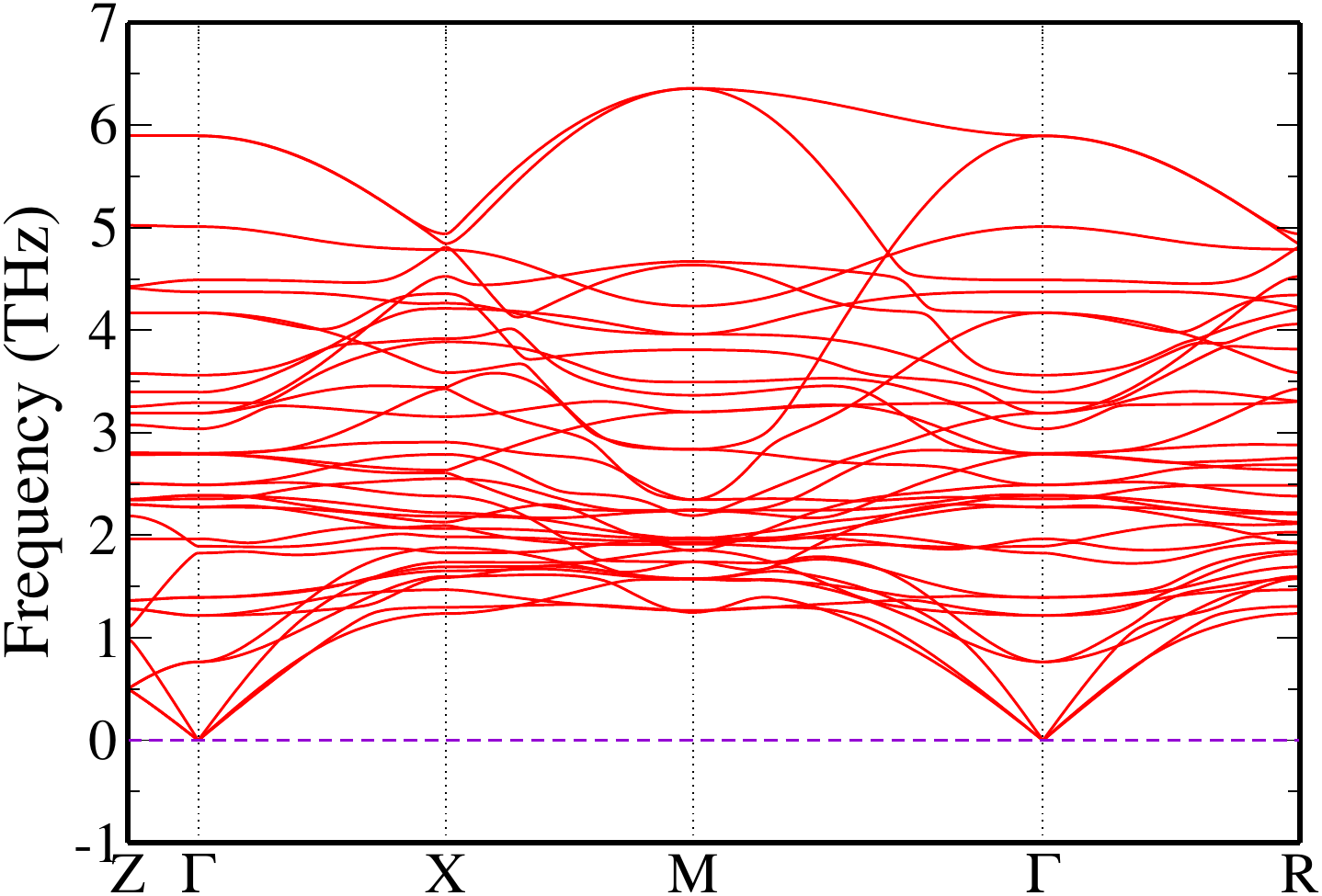}
 \caption{\label{fig:epsart} Phonon band structure of SrCaBi} 
 \label{fig:fragmentation2} 
 \end{figure}

 To confirm the dynamical stability, the phonon band structure for SrCaBi has been calculated  as shown in Figure \ref{fig:fragmentation2}, using phonopy code \cite{phonopy,phonopy1}. $3\times3\times1$ supercell has been used to incorporate the force constants. The phonon band structure for SrCaBi has no imaginary phonon frequency indicating that it is dynamically stable.  
 \linebreak
 \begin{figure}[hbtp]
 \includegraphics[width=18cm,height=8.8cm]{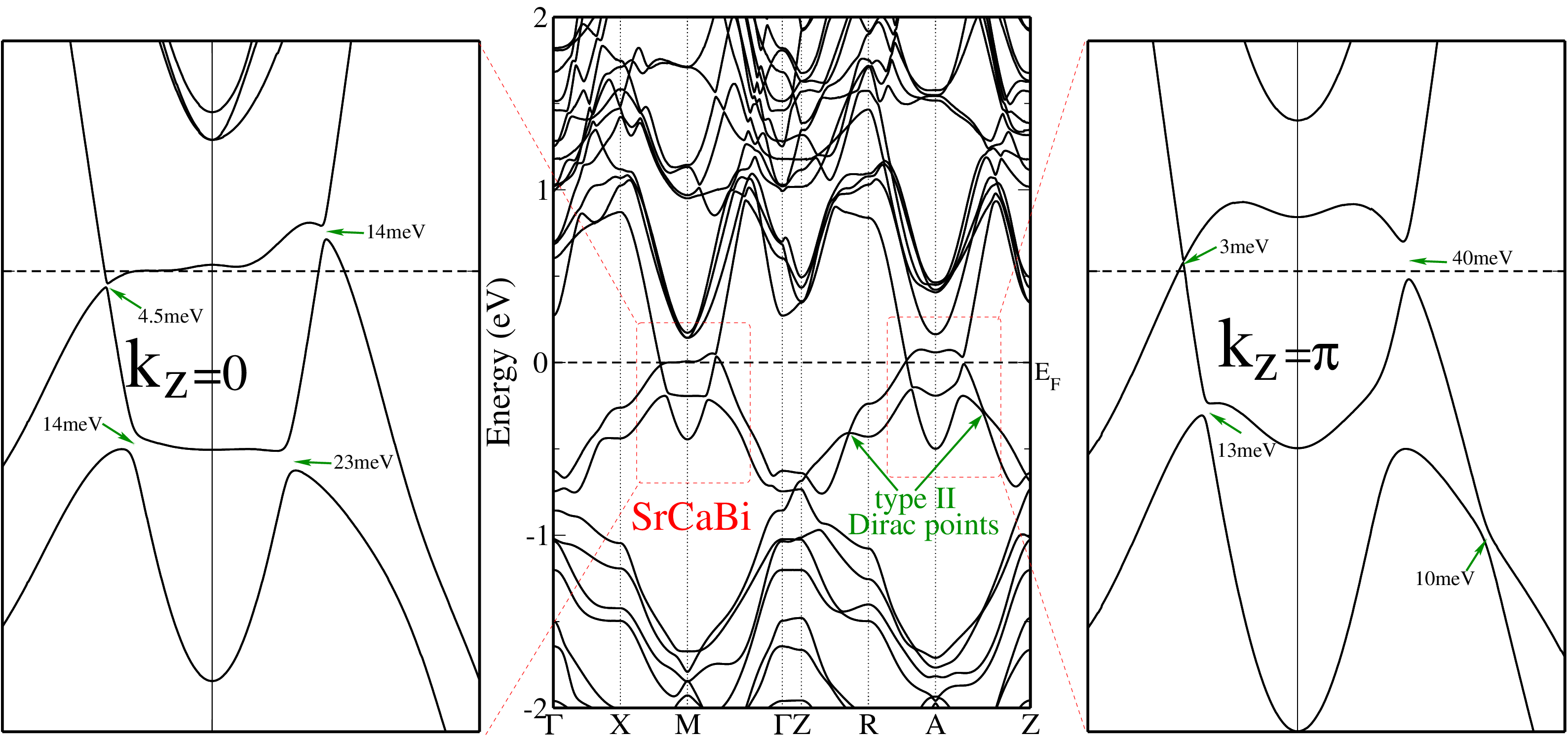}
 \caption{SOC induced electronic band structure of SrCaBi. The crossing points near the Fermi level are shown in zoomed version in left(for $k_z =0$) and right(for $k_z =\pi$) side of the figure.}
 \label{fig:frozen1} 
 \end{figure}

\vspace{5cm}
 \begin{figure}[hbtp]
 \includegraphics[width=18cm,height=8.8cm]{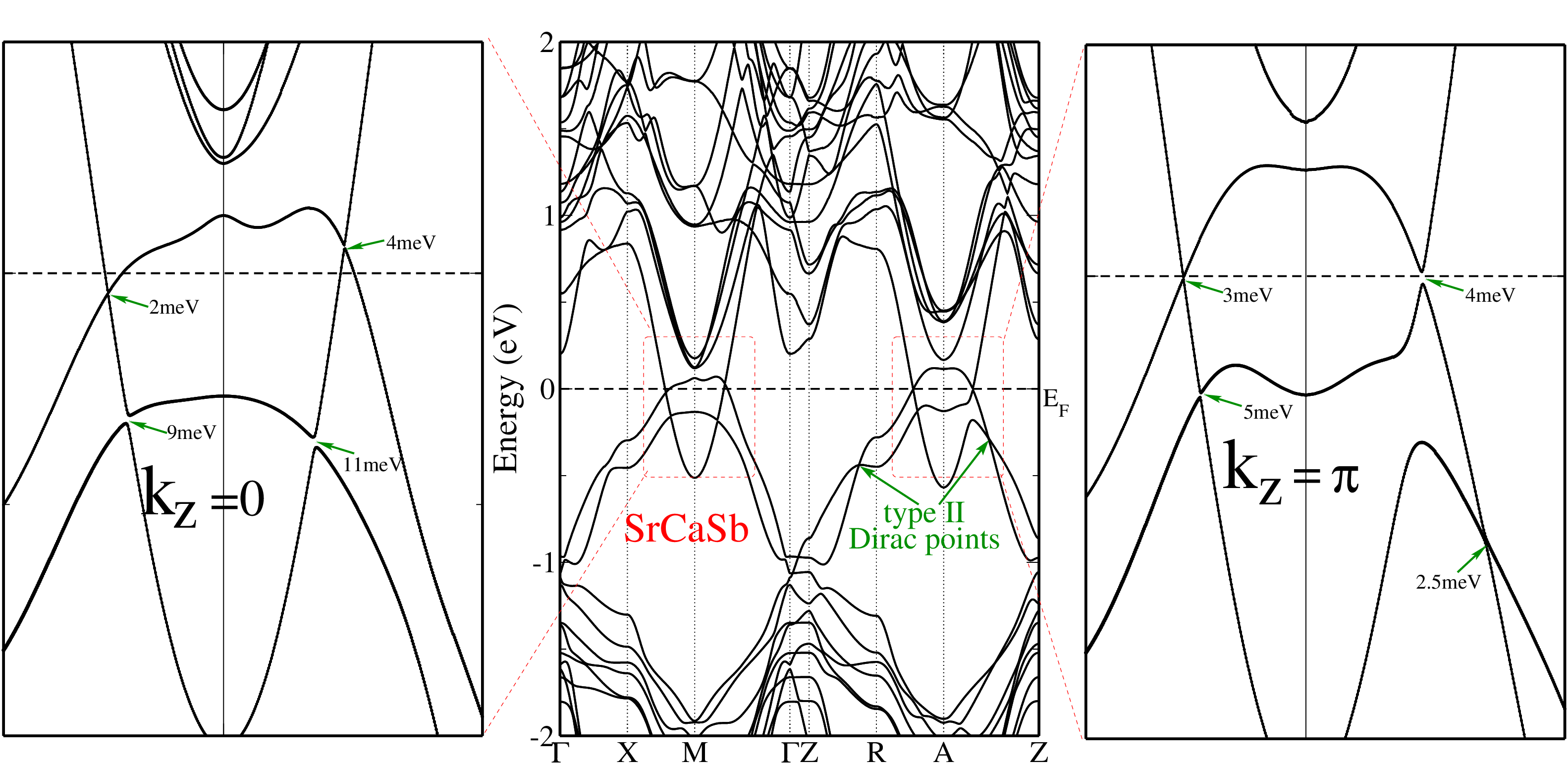}
 \caption{SOC induced electronic band structure of SrCaSb. The crossing points near the Fermi level are shown in zoomed version in left(for $k_z =0$) and right(for $k_z =\pi$) side of the figure.}
 \label{fig:frozen2} 
 \end{figure}
 \begin{figure}[hbtp]
\includegraphics[width=18cm,height=8.8cm]{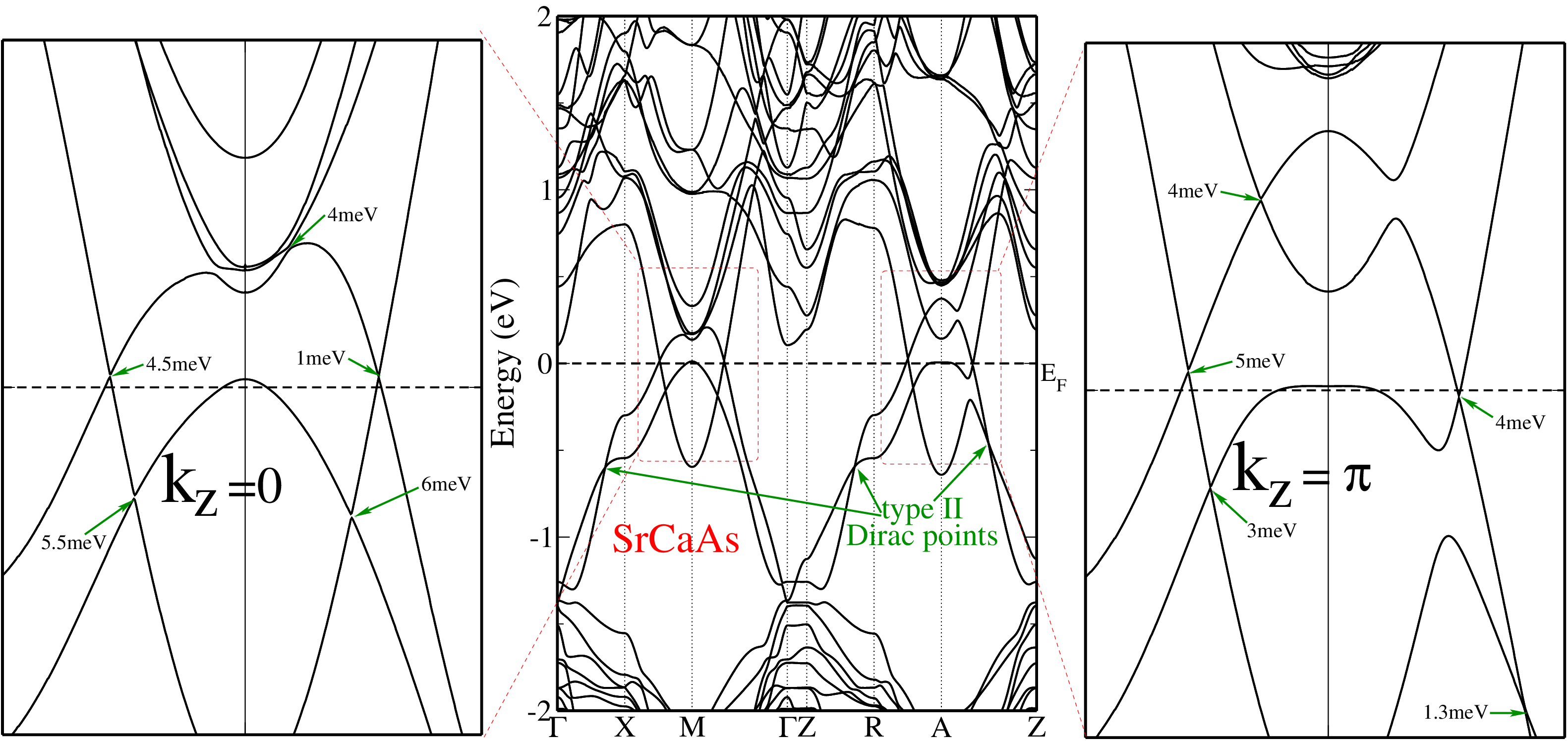}
 \caption{SOC induced electronic band structure of SrCaAs. The crossing points near the Fermi level are shown in zoomed version in left(for $k_z =0$) and right(for $k_z =\pi$) side of the figure.}
 \label{fig:frozen3} 
 \end{figure}
\vspace{5cm}
 \begin{figure}[hbtp]
\includegraphics[width=18cm,height=8.8cm]{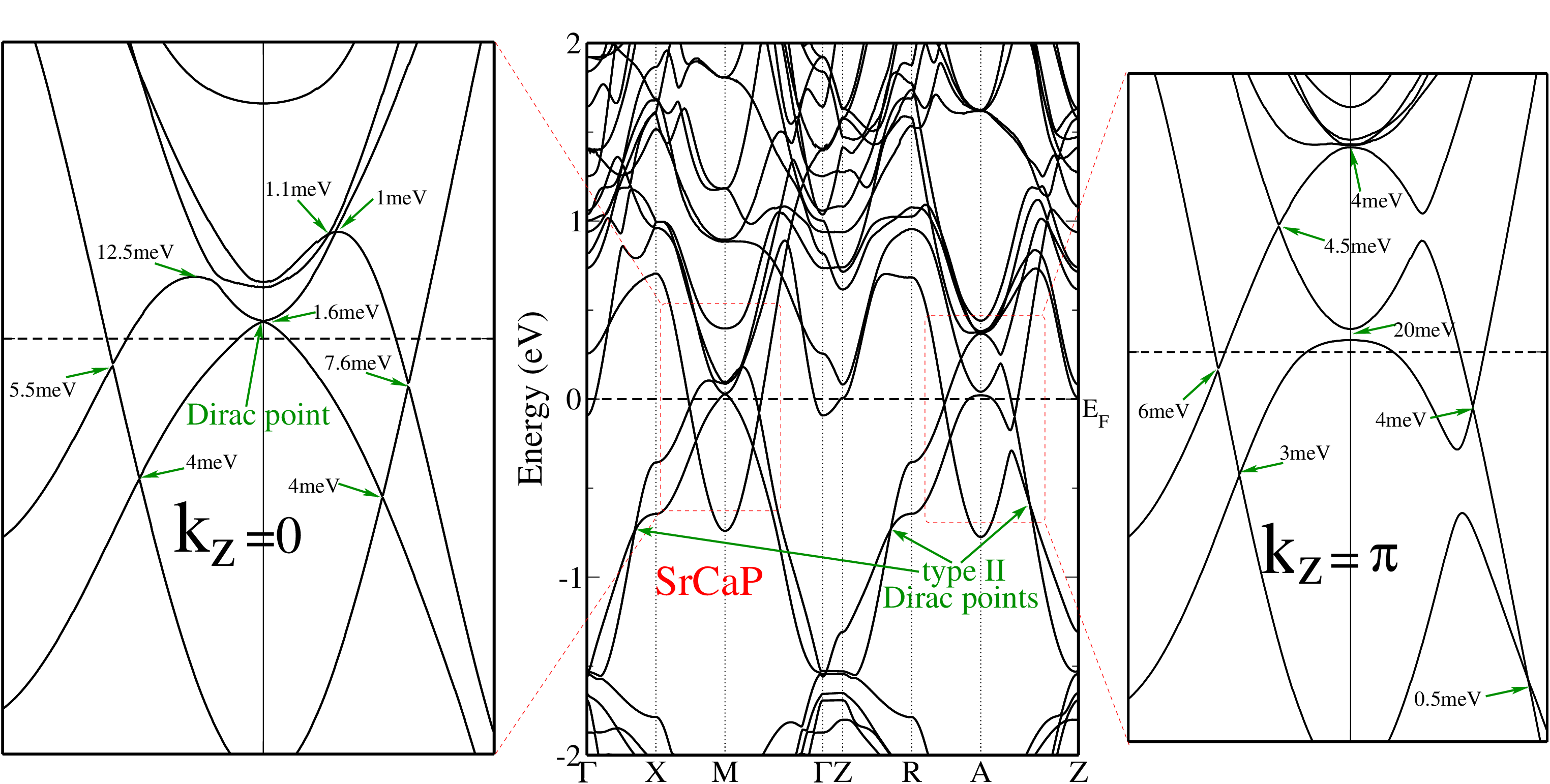}
 \caption{SOC induced electronic band structure of SrCaP. The crossing points near the Fermi level are shown in zoomed version in left(for $k_z =0$) and right(for $k_z =\pi$) side of the figure.}
 \label{fig:frozen4} 
 \end{figure}
\begin{figure}[hbtp]
 \includegraphics[width=1 \textwidth]{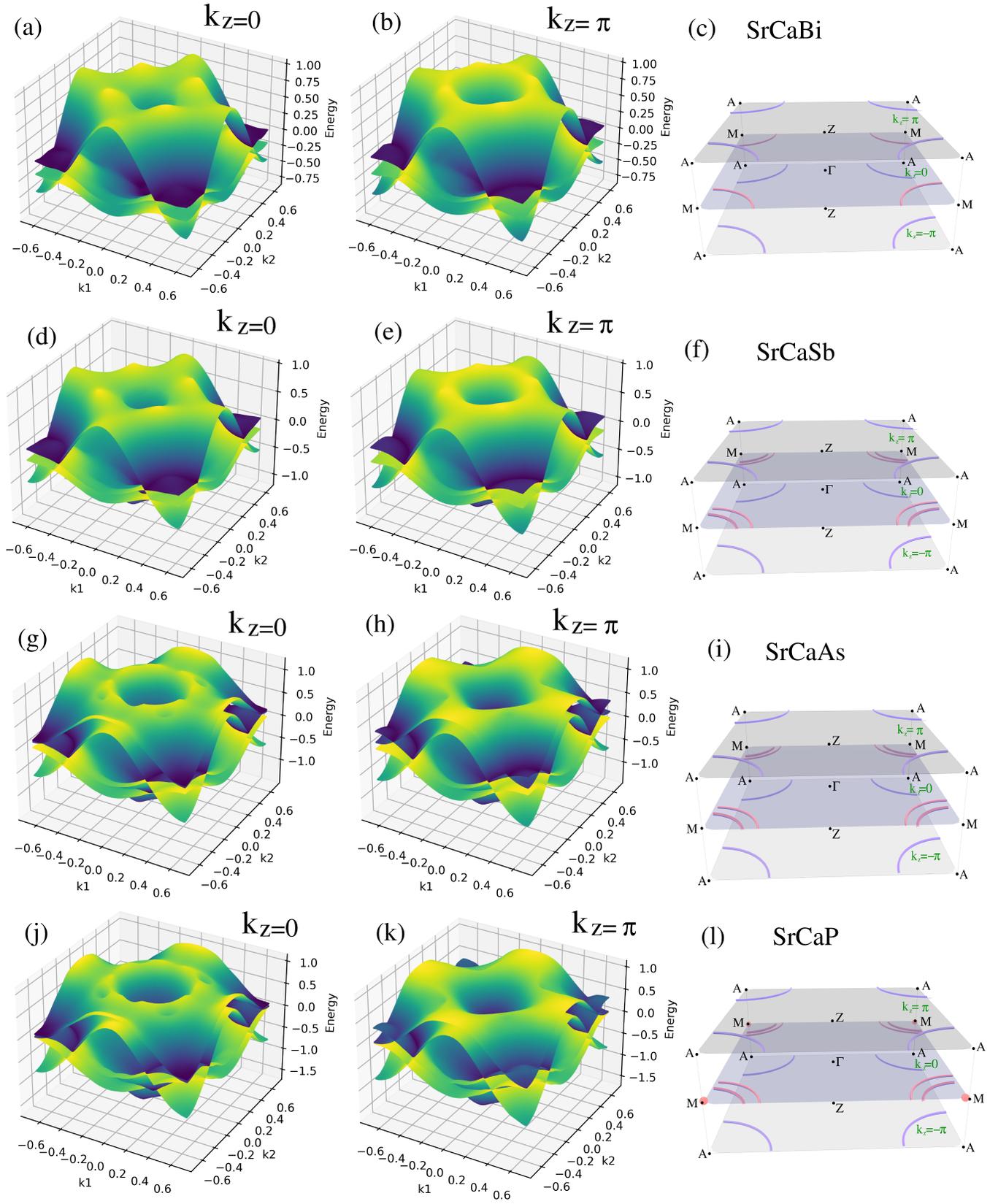}
 \caption{\label{fig:epsart} 3D band structure of (a) SrCaBi $k_z =0$ (b) SrCaBi $k_z =\pi$ (c) nodal loop in first BZ in SrCaBi. 3D band structure of (d) SrCaSb $k_z =0$ (e) SrCaSb $k_z =\pi$ (f) nodal loop in first BZ of SrCaSb. 3D band structure of (g) SrCaAs $k_z =0$ (h) SrCaAs $k_z =\pi$ (i) nodal loop in first BZ in SrCaAs. 3D band structure of (g) SrCaP $k_z =0$ (h) SrCaP $k_z =\pi$ (i) nodal loop in first BZ in SrCaP. }
 \label{fig:S6} 
 \end{figure}
 \begin{figure}[h]
 \includegraphics[width=.63 \textwidth]{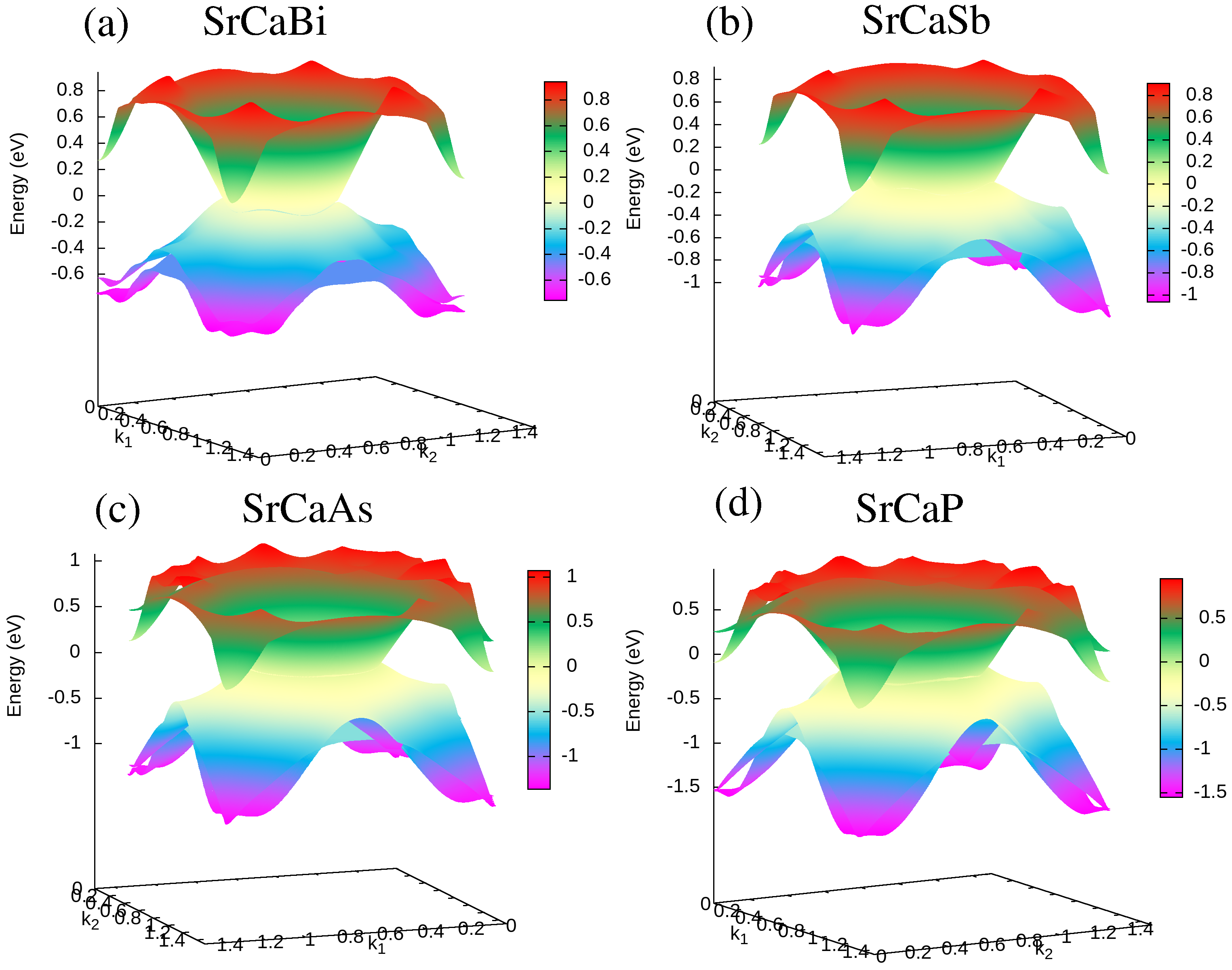}
 \caption{\label{fig:epsart} 3D band structure of (a) SrCaBi (b) SrCaSb (c) SrCaAs (d) SrCaP at  $k_z =0$ centered at M point near Fermi level }
 \label{fig:fragmentation} 
 \end{figure}
\vspace{5cm}
 \begin{figure}[h]
 \includegraphics[width=0.6 \textwidth]{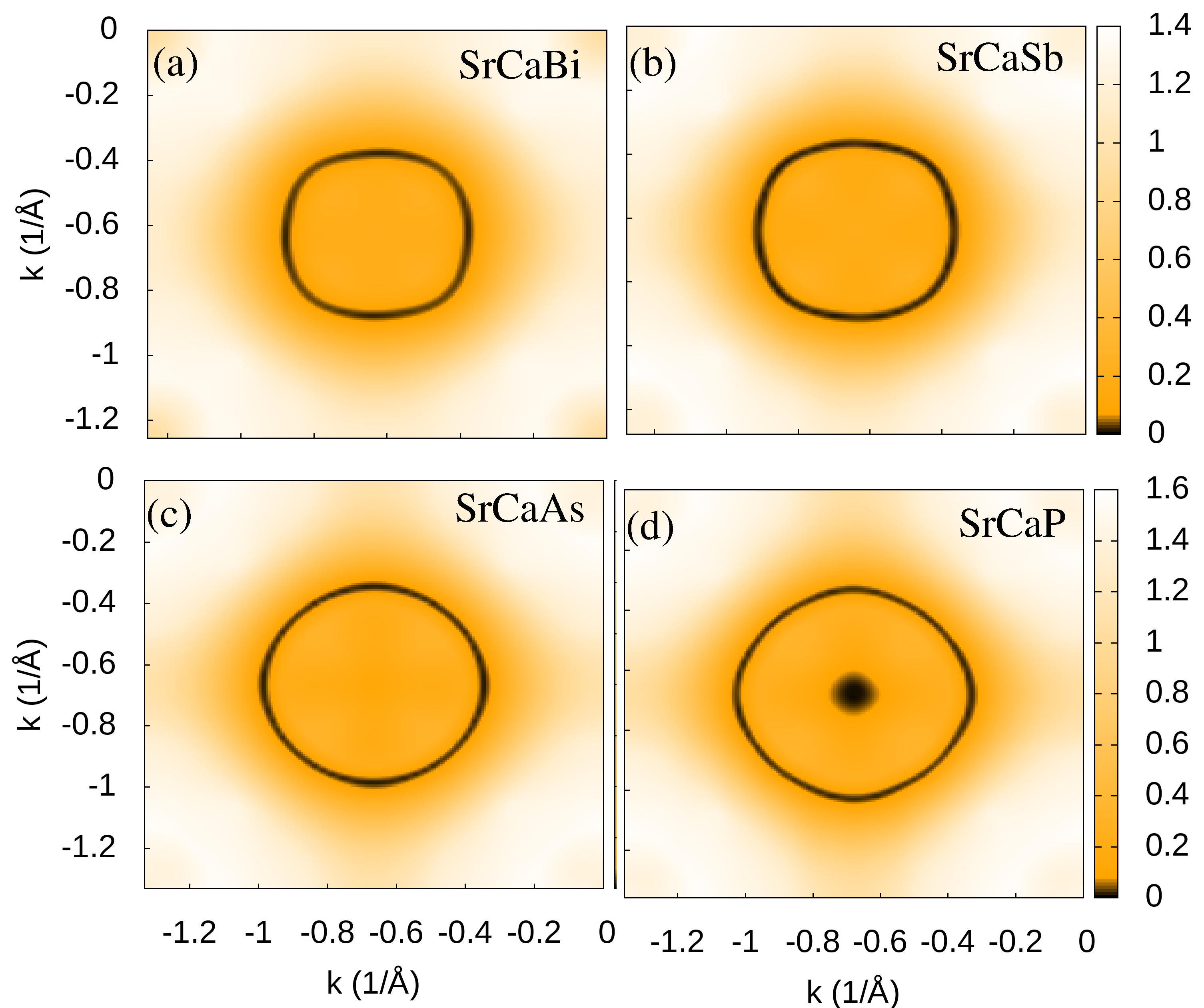}
 \caption{\label{fig:epsart} Gap plane showing the band closing of two bands centered at M point at Fermi level for (a) SrCaBi (b) SrCaSb (c) SrCaAs (d) SrCaP.}
 \label{fig:fragmentation} 
 \end{figure}

 \begin{figure}[hbtp]
 \includegraphics[width=0.58 \textwidth]{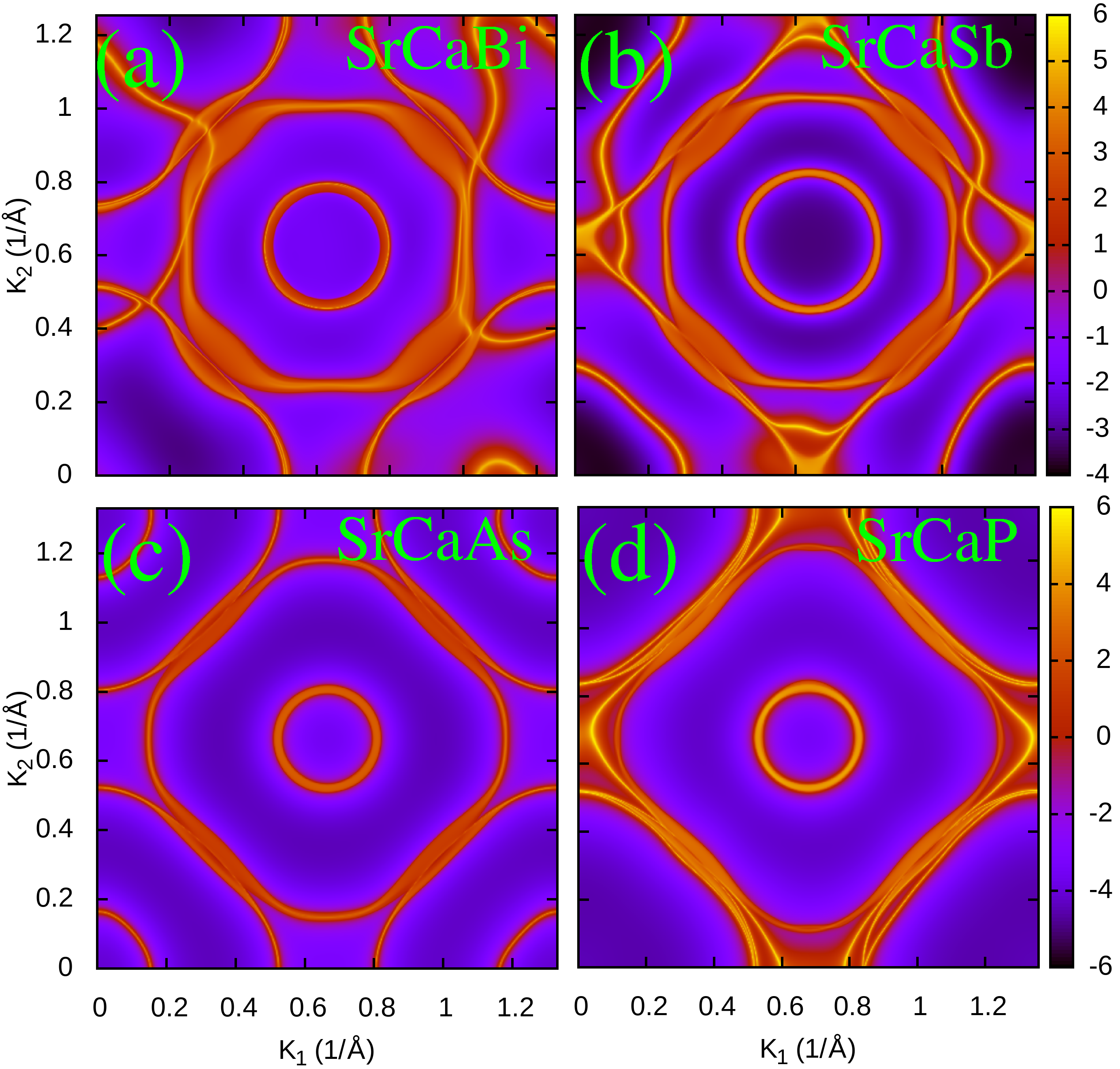}
 \caption{\label{fig:epsart} Fermi arc at energy around the type II crossing point for (a) SrCaBi (b) SrCaSb (c) SrCaAs (d) SrCaP.}
 \label{fig:fragmentation} 
 \end{figure}

 \begin{figure}[hbtp]
 \includegraphics[width=0.58 \textwidth]{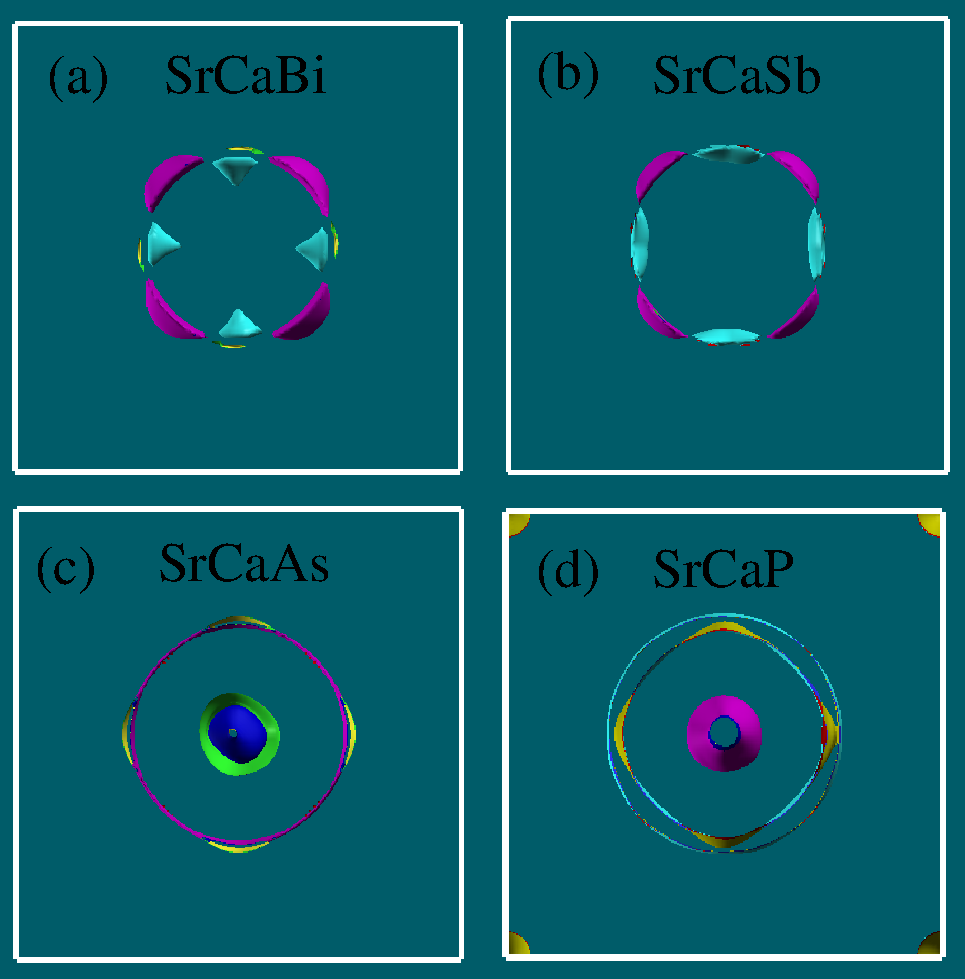}
 \caption{\label{fig:epsart} Fermi surface for (a) SrCaBi (b) SrCaSb (c) SrCaAs (d) SrCaP }
 \label{fig:fragmentation1} 
 \end{figure}
\vspace{5cm}

Figure \ref{fig:S6} confirmed that the band crossing of the nodal line for SrCaX occurs around the M and A points. To understand the topology of the Fermi surface around these points, we plotted the top view of the 3D Fermi surface for Nodal Line Semimetals (NLSMs) SrCaX, as depicted in Figure S10. This plot aligns with the corresponding bulk band structures shown in Figures \ref{fig:frozen1}, \ref{fig:frozen2}, \ref{fig:frozen3} and \ref{fig:frozen4} respectively.
Notably, in SrCaBi and SrCaSb, there is a discontinuity in their Fermi surfaces due to band crossings above and below the Fermi level, altering the shape of the electron-hole pocket. Although SrCaBi and SrCaSb have the same number of crossings, the hole pocket in SrCaBi evolves from a cone shape to an ellipsoidal shape in SrCaSb. In SrCaAs, a crossing above the Fermi level disrupts the continuity of its Fermi surface.
In contrast, SrCaP exhibits crossings entirely below or above the Fermi level, resulting in a continuous Fermi surface. Additionally, in both SrCaAs and SrCaP, a central hole pocket emerges, taking a circular form due to valence band crossings around the high-symmetry point M\cite{muller2020determination, lou2018experimental, wang2024origin, fu2019dirac}.

\bibliography{biblo}